\documentclass[prb,twocolumn,preprintnumbers ,amsmath, amssymb, superscriptaddress, aps,floatfix]{revtex4-2}

\usepackage{color}
\usepackage{amsmath,amssymb}
\usepackage{pifont}
\usepackage{amssymb}  
\usepackage{bbold}
\usepackage{float}
\usepackage{tikz}
\usepackage{xcolor}
\usepackage{makecell}
\usepackage{pifont}   
\usepackage{graphicx} 
\usepackage{subfig}
\usepackage{dcolumn}  
\usepackage{bm}       
\usepackage{multirow} 
\usepackage[colorlinks]{hyperref}
\usepackage{placeins}

\newcommand{\ket}[1]{\left|{#1}\right\rangle}

\newcommand{\eqfitpage}[1]{\resizebox{\linewidth}{!}{$#1$}}

\renewcommand{\vec}[1]{\boldsymbol{#1}}
\renewcommand{\Omega}{\omega}
\definecolor{AAAMTc1}{RGB}{238, 99, 99}
\definecolor{AAAMTc2}{RGB}{128, 0, 0}
\definecolor{AAAMTc3}{RGB}{139, 58, 58}
\definecolor{AAAMTc4}{RGB}{238, 0, 0 }
\definecolor{AAATTc1}{RGB}{255,140,0}
\definecolor{AAATTc2}{RGB}{205,92,92}
\definecolor{AAATTc3}{RGB}{128,0,0 }
\definecolor{AAATTc4}{RGB}{178,34,34}
\definecolor{AAATTc5}{RGB}{240,128,128}
\definecolor{AAATTc6}{RGB}{255,165,0}
\definecolor{AAATTc7}{RGB}{255,99,71}
\definecolor{AAATTc8}{RGB}{233,150,122}
\definecolor{AAATTc9}{RGB}{220,20,60}
\definecolor{AAATTc10}{RGB}{189,183,107}
\definecolor{AAATTc11}{RGB}{165,42,42}
\definecolor{AAATTc12}{RGB}{255,69,0}
\definecolor{AAATTc13}{RGB}{255,0,0}
\definecolor{AAATTc14}{RGB}{250,128,114}
\definecolor{AAATTc15}{RGB}{255,127,80}
\definecolor{ABCTTc1}{RGB}{100,149,237}
\definecolor{ABCTTc2}{RGB}{0,191,255}
\definecolor{ABCTTc3}{RGB}{135,206,235}
\definecolor{ABCTTc4}{RGB}{47,79,79}
\definecolor{ABCTTc5}{RGB}{65,105,225}
\definecolor{ABCTTc6}{RGB}{0,0,128}
\definecolor{ABCTTc7}{RGB}{147,112,219}
\definecolor{ABCTTc8}{RGB}{0,0,255}
\definecolor{ABCTTflipc1}{RGB}{220,20,60}
\definecolor{ABCTTflipc2}{RGB}{255,99,71}
\definecolor{ABCTTflipc3}{RGB}{255,127,80}
\definecolor{ABCTTflipc4}{RGB}{255,0,0}
\definecolor{ABCTTflipc5}{RGB}{178,34,34}
\definecolor{ABCTTflipc6}{RGB}{233,150,122}
\definecolor{ABCTTflipc7}{RGB}{250,128,114}
\definecolor{ABCMTc1}{RGB}{70,130,180}
\definecolor{ABCMTc2}{RGB}{30,144,255}
\definecolor{ABCMTc3}{RGB}{0,0,255}
\definecolor{ABCMTc4}{RGB}{0,0,205}
\definecolor{ABCMTc5}{RGB}{0,191,255}

\begin{document}
\title{Floquet engineering and non-equilibrium topological maps in twisted trilayer graphene}
\author{I. A. Assi}
\affiliation{Department of Physics and Physical Oceanography, Memorial University of Newfoundland, St. John's, Newfoundland \& Labrador, A1B 3X7, Canada}
\author{J. P. F. LeBlanc}
\affiliation{Department of Physics and Physical Oceanography, Memorial University of Newfoundland, St. John's, Newfoundland \& Labrador, A1B 3X7, Canada}
\author{Martin Rodriguez-Vega}
\affiliation{Theoretical Division, Los Alamos National Laboratory, Los Alamos, New Mexico 87545, USA}
\author{Hocine Bahlouli}
\affiliation{Department of Physics, King Fahd University of Petroleum and Minerals, 31261 Dhahran, Saudi Arabia}
\author{Michael Vogl}
\affiliation{Department of Physics, King Fahd University of Petroleum and Minerals, 31261 Dhahran, Saudi Arabia}
\date{\today}

\begin{abstract}
   Motivated by the recent experimental realization of twisted trilayer graphene and the observed superconductivity that is associated with its flat bands at specific angles, we study trilayer graphene under the influence of different forms of light in the non-interacting limit. Specifically, we study four different types of stacking configurations with a single twisted layer. In all four cases, we study the impact of circularly polarized light and longitudinal light coming from a waveguide. We derive effective time-independent Floquet Hamiltonians and review light-induced changes to the band structure. For circularly polarized light, we find band flattening effects as well as band gap openings.  We emphasize that there is a rich band topology, which we summarize in Chern number maps that are different for all four studied lattice configurations. The case of a so-called ABC stacking with top layer twist is especially rich and shows a different phase diagram depending on the handedness of the circularly polarized light. Consequently, we propose an experiment where this difference in typologies could be captured via optical conductivity measurements. In contrast for the case of longitudinal light that is coming from a waveguide, we find that the band structure is very closely related to the equilibrium one but the magic angles can be tuned in-situ by varying the intensity of the incident beam of light.
\end{abstract}
\maketitle

\section{Introduction}

Graphene \cite{Novoselov666}, a single layer of carbon atoms arranged in a hexagonal lattice structure, is an amazing material with some highly unusual physical and electronic properties such as good conduction of electricity and heat almost without scattering \cite{RevModPhys.81.109}. Electrons in graphene behave as massless relativistic fermions at low energies which led to much excitement. However, some aspects of this feature, such as Klein tunneling, which enables charge carriers to tunnel through very high electrostatic barriers \cite{Klein1929, Katsnelson_2006}, are detrimental to the realization of electronic applications. Hence, a lot of efforts have focused on modifying the band structure of graphene \cite{Zhang2009,RevModPhys.81.109}, which is hoped to allow for better control of graphene's exciting features. 

One approach to modifying graphene's band structure that is exceptionally versatile, is to stack multiple graphene layers on top of each other and introduce a relative twist angle between them. Such a twist manifests itself in the appearance of moiré patterns, which can be visualized using scanning tunneling microscope (STM) techniques \cite{STMTTG}. These structures have been observed in many samples constructed using mechanical exfoliation techniques, which makes it possible to obtain flakes with the desired number of layers \cite{Pong_2005}. In samples created this way one commonly finds rotation of the top layer with respect to the lower ones. 
One of the most exciting features of twisted graphene bilayer was noticed in 2011 by Bistritzer \& MacDonald \cite{Bistritzer12233}. Particularly, they found that for a twist angle $\theta\approx 1.05^\circ$ - the magic angle - the lowest energy moiré bands become flat. This finding led to predictions that the electron-electron interactions at this special magic angle play an important role and may give rise to strong correlation effects such as superfluidity \cite{Eisenstein2004, Min_2008},  magnetism \cite{PhysRevLett.88.127202}, or other types of ordered states.

In this context, the discovery of superconductivity near the magic angle in twisted bilayer graphene by Cao et. al \cite{Cao2018} has led to much excitement about the potential occurrence of a similar effect in a growing number of twisted moire materials.  Since then, it has been experimentally found that twisted materials can host a variety of strongly correlated states \cite{Cao2020,Shen2020,Liu2020,PhysRevB.99.235417,PhysRevB.99.235406,Lee2019,doi:10.1021/acs.nanolett.9b05117, C9NR10830K, PhysRevB.99.075127,kerelsky2019moireless,rubioverdu2020universal,halbertal2020moire,christos2021correlated,Cao2018,Cao2018sc,Codecidoeaaw9770,wong2019cascade,Lu2019efetov,Sharpe605, Seo_2019} ranging from correlated insulator \cite{christos2021correlated,Cao2018,Cao2018sc,Codecidoeaaw9770,wong2019cascade,Lu2019efetov} to ferromagnetic behaviour \cite{Sharpe605, Seo_2019}.

A second common approach at modifying graphene's bandstructure, topology and transport is to subject it to different forms of light such as circularly polarized light \cite{Oka_2019, rudner2020_review,Giovannini_2019,McIver2020,oka2009,lindner2011,rechtsman2013,PhysRevB.103.195422,PhysRevB.93.115420}. For this it is important to be able to solve the time-dependent Floquet eigenvalue problem. For instance it can be useful to make use of effective time-independent Floquet Hamiltonians, Floquet perturbation theories and various other techniques that allow for both perturbative and non-perturbative descriptions \cite{MoireFloquetRev,blanes2009,rahav2003,rahav2003b,Bukov_2015,Eckardt_2015,Feldm1984,Magnus1954,PhysRevB.95.014112,PhysRevX.4.031027,PhysRevLett.115.075301,PhysRevB.93.144307,PhysRevB.94.235419,PhysRevLett.116.125301,Vogl_2019AnalogHJ,verdeny2013,PhysRevLett.110.200403,Vogl2020_effham,vogl2019,Rodriguez_Vega_2018,PhysRevLett.121.036402,Martiskainen2015,rigolin2008,weinberg2015,PhysRevB.100.041103,PhysRevB.96.155438,Kennes-Klinovaja-2019,PhysRevB.101.155417}.

More recently, both moir\'e and Floquet approaches have been combined, and a wealth of additional features such as modified band structures, light induced flat bands \cite{katz2019floquet,PhysRevB.101.241408,vogl2020effective} and various topological phases have been theoretically predicted \cite{Topp_2019,PhysRevB.103.014310,rodriguezvega_2020a,PhysRevB.103.195146,Li_2020}. 

In this work, we have been motivated by the recent interest in twisted trilayer graphene (TTG), which was sparked by the discovery of superconductivity \cite{Park2021}. However, instead of focusing on the impact of interactions in this material, which have already been studied in \cite{christos2021correlated,lake2021reentrant,ramires2021emulating,chou2021correlationinduced}, we will focus on the interplay of twisted trilayer graphene under the action of different forms of light. This interaction can lead to effective bands with distinct properties from their equilibrium counterparts such as topological Chern phases similar to what has been found in graphene \cite{oka2009} as we will see later.

The rest of this work is structured as follows. In Sec. \ref{sec:model} we describe twisted trilayer graphene in equilibrium,  introduce  the model, and review some of its properties. In Sec.
\ref{sec:III}, we describe the different forms of light that are used in this work, and describe how they enter the Hamiltonian description. We also discuss the numerical approach to our theoretical model. In Sec. \ref{sec:circ_pol_lighte}, we focus on circularly polarized light where we first analyze the band structures for the different TTGs. We then derive effective time-independent Hamiltonians that allow for less computationally costly treatment of light-driven TTG. These effective Hamiltonians are then used to study  the band topology and suggest an experimental setup to test some of our predictions. Finally, in section \ref{sec:waveguide} we consider longitudinal light coming from a waveguide.  Here, we focus exclusively on changes to band structure and the effective Hamiltonian. We find that this type of light  makes it possible to tune the magic angles, where flat bands appear. The effective Hamiltonian allows us to gain insight into the mechanism behind this observation. Lastly, in Sec. \ref{sec:conclusion} we present our conclusion.

\section{Setup of equilibrium model and equilibrium results}
\label{sec:model}

The system we will study in this work is twisted trilayer graphene in its various stacking configurations subjected to different forms of light. Here, we briefly review some of its equilibrium properties.

\begin{figure}[!htbp]
	\begin{center}
	\subfloat[]{\includegraphics[width = 0.5\linewidth]{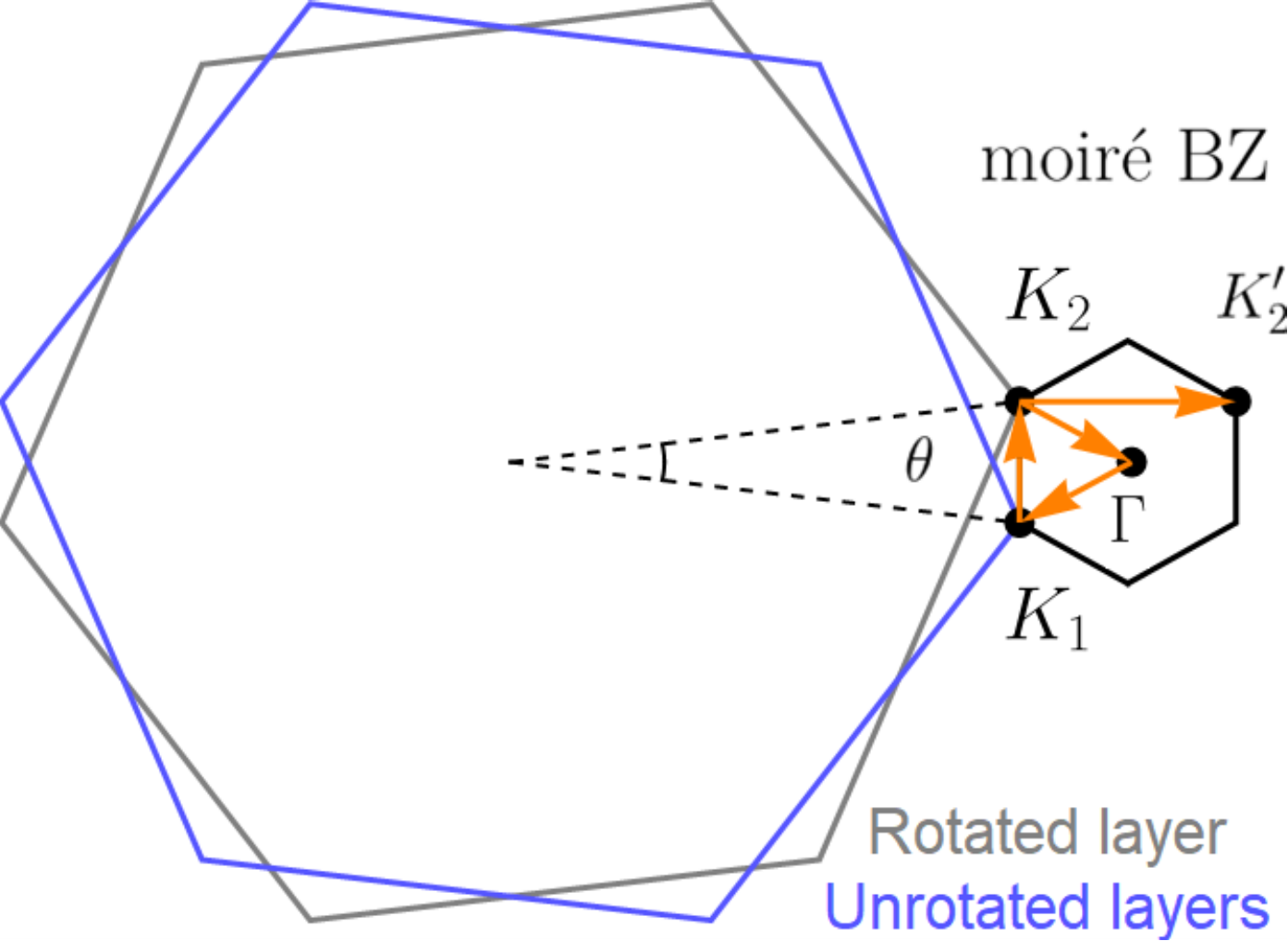}}
	\subfloat[]{\includegraphics[width = 0.5\linewidth]{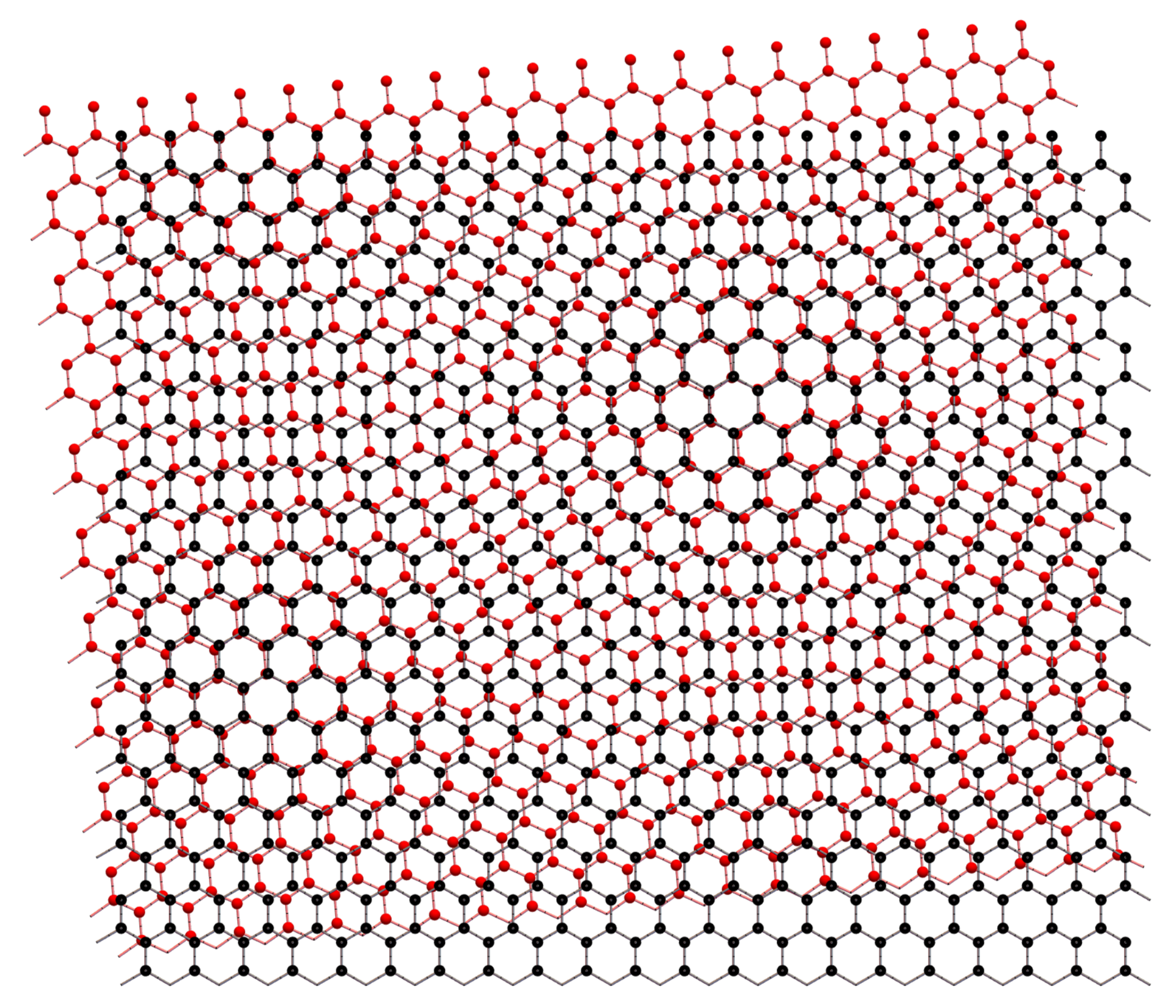}}
	\caption{(a) The moiré Brillouin zone for a trilayer graphene (TTG) system with only one layer twisted. (b) Schematic plot of the TTG system for the AAA stacking with middle layer twist showing the moiré pattern.  }
	\label{fig:latticeBZ}
	\end{center}
\end{figure}

Trilayer graphene is formed by three layers of graphene stacked on top of one another. If two sheets are stacked in such a way that all the carbon atoms of one layer are exactly on top of an atom in the layer below this is called AA stacking. The case where only half of the atoms in the top layer have an atom exactly below it is referred to as either AB or BA stacking (there are two possible orderings). For trilayer graphene there can be various possible configurations such as AAA, ABA and ABC stacking. To obtain twisted trilayer graphene each of these stackings can then have either the top or middle layer rotated with respect to the other layers, which will lead to a moiré pattern that is associated with a smaller moiré Brillouin zone (MBZ), as we show in Fig. \ref{fig:latticeBZ}.

The static Hamiltonian for twisted trilayer graphene we will work with is based on the Hamiltonian introduced in earlier works \cite{multilayerTG,LiWuMcDonald}. It is given as
\begin{equation}
\eqfitpage{
    H(\vec{x},\vec{k})=\begin{pmatrix}
    h_{1}(\theta_{1},\vec{k}-\vec{\kappa}_{1})&T_{12}(\vec{x})&0\\
    T_{12}^{\dagger}(\vec{x})&h_{1}(\theta_{2},\vec{k}-\vec{\kappa}_{2})&T_{23}(\vec{x})\\
    0&T_{23}^{\dagger}(\vec{x})&h_{3}(\theta_{3},\vec{k}-\vec{\kappa}_{3})
    \end{pmatrix}},
    \label{eq:macdHam_mod}
\end{equation}

where 
\begin{equation}
    h_{\ell}(\theta_{\ell},\vec{k})=\gamma\left(\begin{array}{ccc}
    0&f\left(R(\theta_{\ell})\vec{k}\right)\\
    f^{*}\left(R(\theta_{\ell})\vec{k}\right)&0\\
    \end{array}\right),
\end{equation}
is the single layer graphene Hamiltonian with hoppings between graphene's sublattices given as $f(\vec{k})=e^{-i2a_{0}k_{y}/3}+2e^{ia_{0}k_{y}/3}\sin(a_0k_x/\sqrt{3}-\pi/6)$, $R(\theta_{\ell})$ is the rotation matrix in the layer plane, and $\gamma=\hbar v_{F}/a_0=2.364$ eV is the strength of the interlayer hoping, where $v_F=10^6$ m/s is the Fermi velocity and $a_0=0.246 {\rm A}^\circ$ is the lattice constant \cite{McCann2012}. Here, we used the bounded tight binding form of $f(\vec k)$ for single layer graphene rather than a linearized dispersion used in Ref. \cite{LiWuMcDonald} because bounded Hamiltonians are more well-behaved for the purposes of Floquet theory. 

The Hamiltonian above can be used to model the various possible configurations of twisted trilayer graphene we will consider. Particularly at twist angle $\theta$, a top layer twist (TLT) can be modelled with parameter choices $\theta_1=-\theta_2=-\theta_3=\theta/2$, $\vec{\kappa}_{1}=\vec{\kappa}_{-}$ and $\vec{\kappa}_{2,3}=\vec{\kappa}_{+}$ where $\vec{\kappa}_{\pm}=\frac{k_\theta}{2}\left(-\sqrt{3},\pm1\right)$ and $k_\theta=8\pi\sin(\theta/2)/3a_0$. Similarly, for a middle layer twist (MLT), one would have to set $\theta_1=-\theta_2=\theta_3=\theta/2$, $\vec{\kappa}_{1,3}=\vec{\kappa}_{-}$ and $\vec{\kappa}_{2}=\vec{\kappa}_{+}$. 

The $\vec{\kappa}_{\pm}$ shifts were introduced into the Hamiltonian through a unitary transformation that ensures that the momenta in  all three layers are measured with respect to the $\Gamma$ point in the moiré Brillouin zone seen in Fig. \ref{fig:latticeBZ}. This transformation is also the reason why our $T(\vec x)$ matrices, where $x$ is the spatial coordinate, at first glance seem to differ by the choice of $\vec q$ vectors from those used in \cite{LiWuMcDonald} as we will see below. The advantage of this approach is that an expansion of the Hamiltonian in terms of plane waves can be done in a conventional way (without the need to introduce additional phases for the different layer components of the wavefunction). In our case, the $T$ matrices for a middle twist are given as $T_{12}(\vec{x})=\sum_{\ell=1}^{3}e^{-i\vec{q}_{\ell}\cdot\vec{x}}T_{\ell}$,  $T_{23}(\vec{x})=\sum_{\ell=1}^{3}e^{+i\vec{q}_{\ell}\cdot\vec{x}}T_{\ell}$. The case of a top twist differs in that $T_{23}(\vec{x})=\sum_{\ell=1}^{3}T_{\ell}$. In addition, we have $\vec{q}_{1,2}=\frac{\sqrt{3}k_\theta}{2}(\pm 1,\sqrt{3})$ and $\vec{q}_{3}=(0,0)$. Finally, the matrices $T_\ell$ are defined as follows \cite{LiWuMcDonald}
\begin{equation}
    T_\ell^{AB}=\left[T_\ell^{BA}\right]^{\dagger}=\left(\begin{array}{cc}
     w_0 e^{i\frac{2\ell\pi}{3}} & w_1\\
     w_1 e^{-i\frac{2\ell\pi}{3}} & w_0 e^{i\frac{2\ell\pi}{3}}\\
    \end{array}\right),
\end{equation}
\begin{equation}
    T_\ell^{AA}=\left(\begin{array}{cc}
     w_0 & w_1 e^{-i\frac{2\ell\pi}{3}}\\
     w_1 e^{i\frac{2\ell\pi}{3}} & w_0\\
    \end{array}\right)
\end{equation}
where the superscripts refer to the type of stacking we have which can be $AA$, $AB$, or $BA$ stacking. We choose the following tunneling parameters $w_1=110$ meV and $w_0\approx0.8w_1$ so that they are close to those in twisted bilayer graphene where distortions in a relaxed lattice can be modelled this way \cite{katz2019floquet,PhysRevB.101.241408}, which is expected to happen for twisted trilayer graphene if we neglect next nearest layer interactions.

\begin{figure}[!htbp]
\centering
    \includegraphics[width=0.47\linewidth]{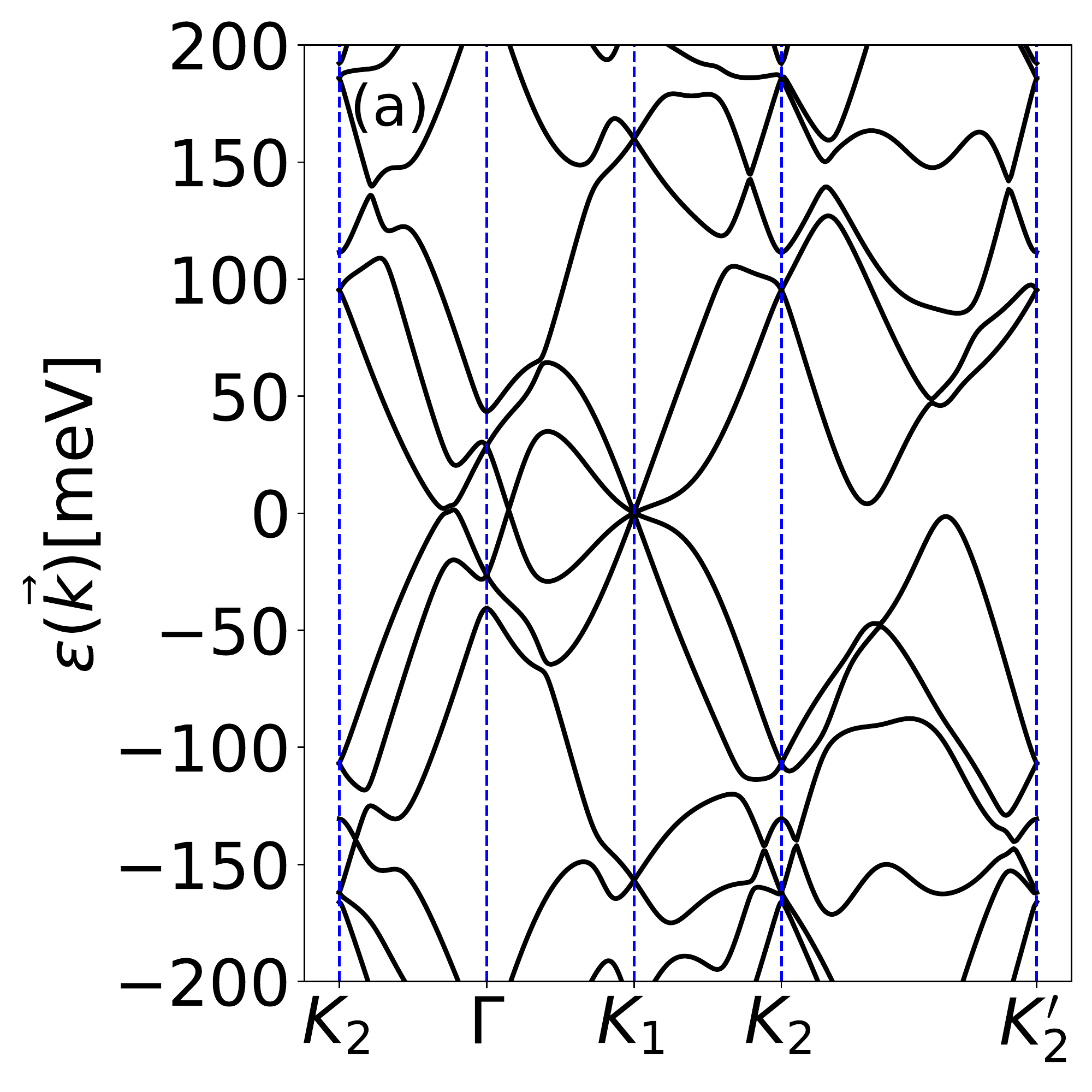}\hfil
    \includegraphics[width=0.47\linewidth]{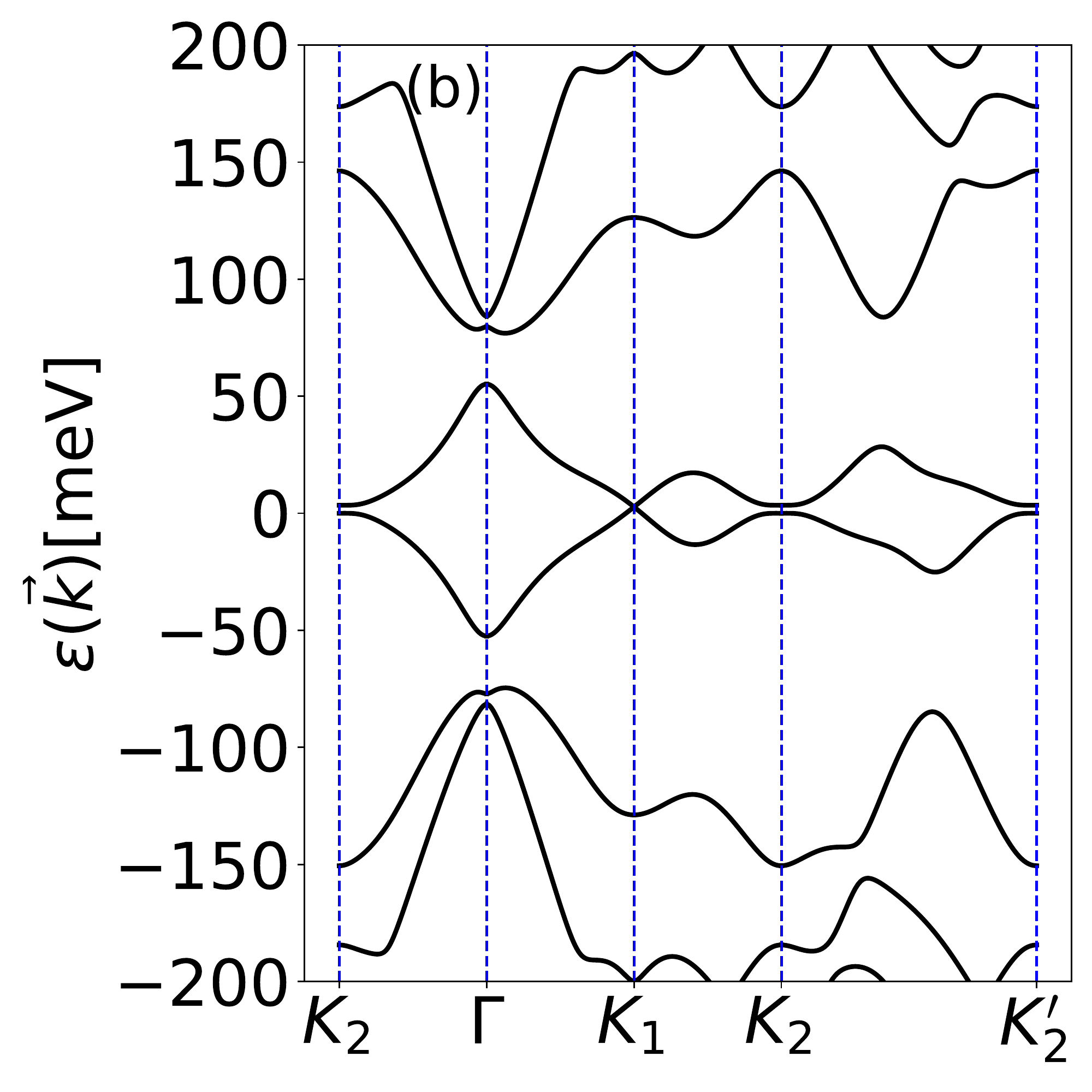}\par\medskip
    \includegraphics[width=0.47\linewidth]{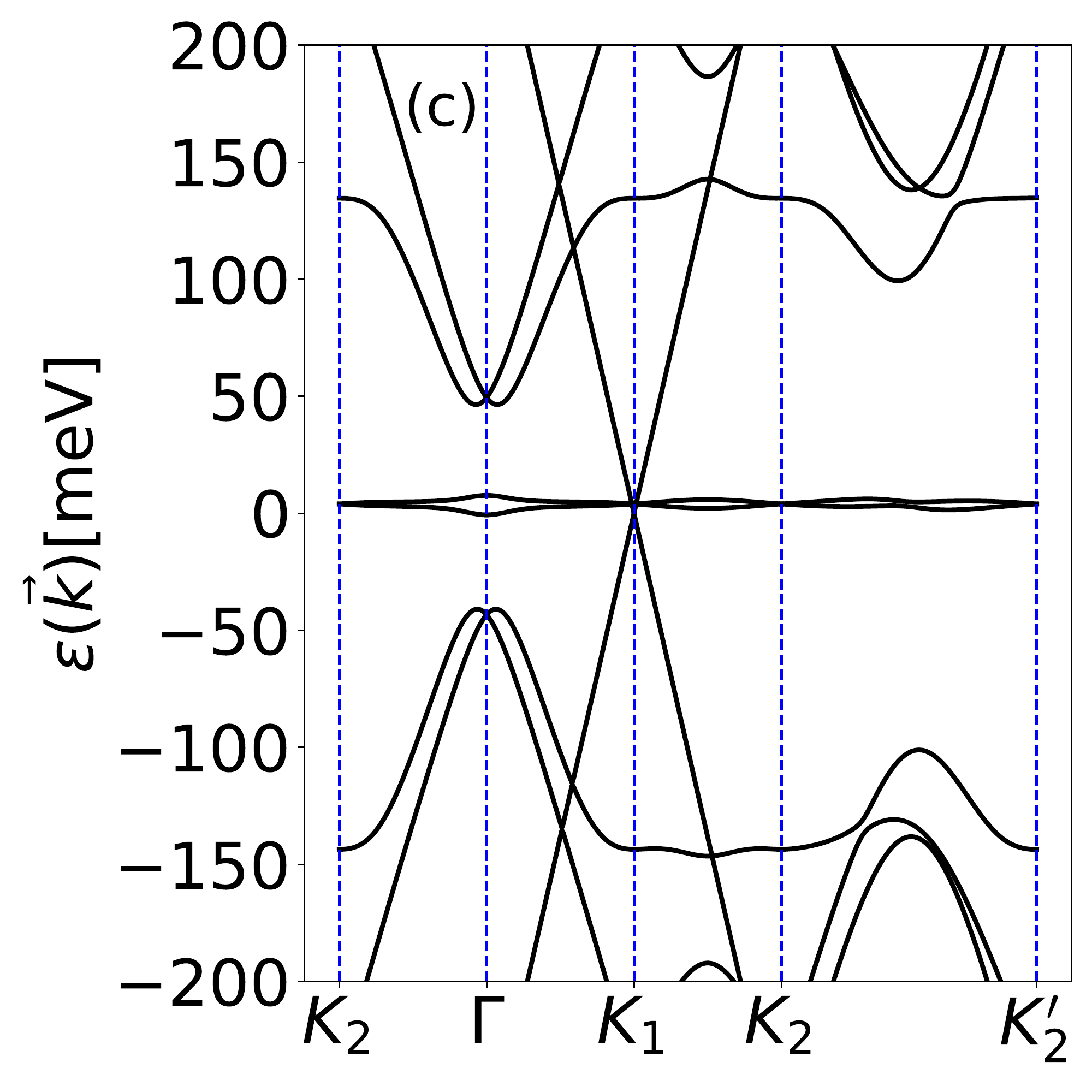}\hfil
    \includegraphics[width=0.47\linewidth]{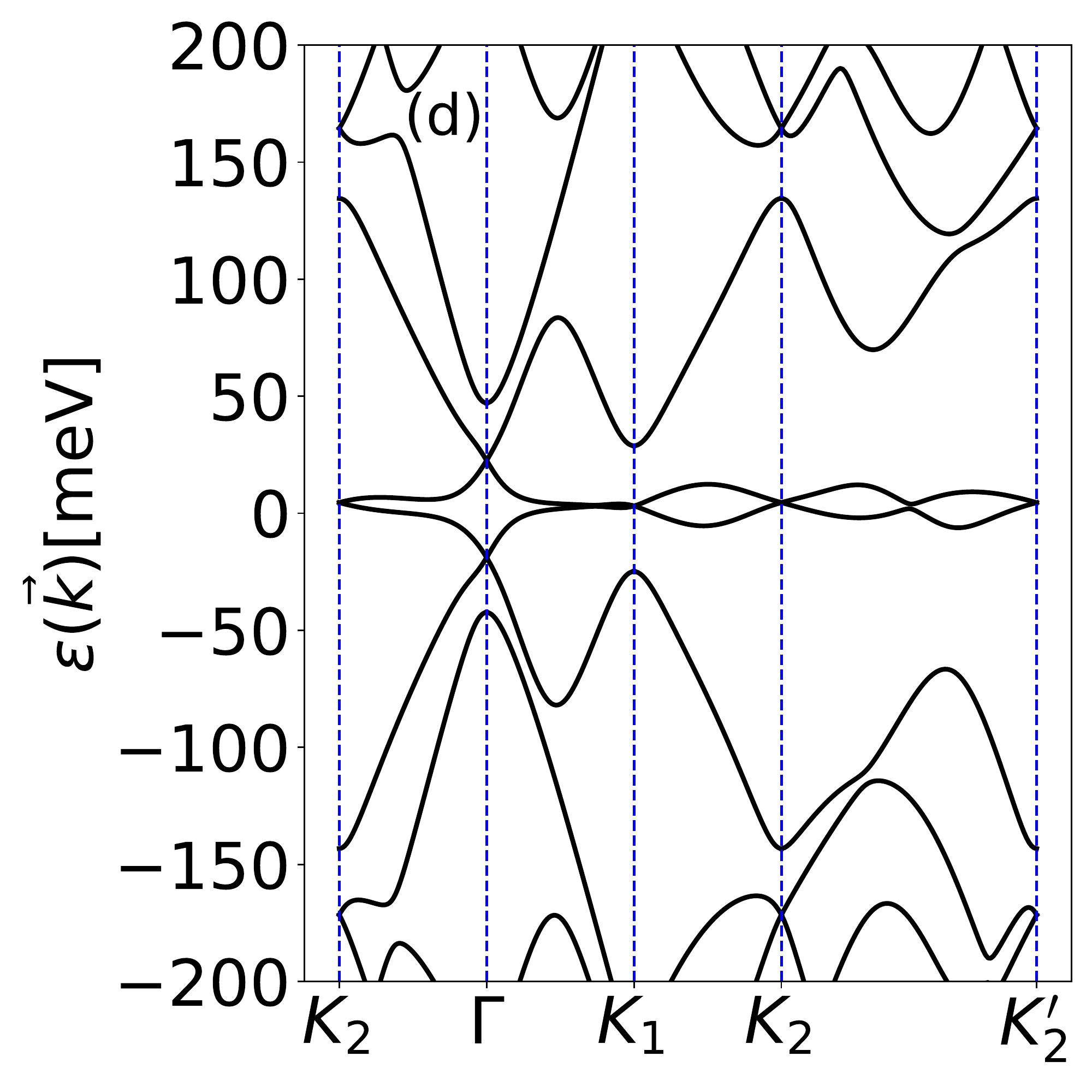}
\caption{(Color online)  Band structure of the TTG with equilibrium Hamiltonian (Eq.~\ref{eq:macdHam_mod}) top layer twisted  (top row) and middle layer twisted (bottom row) with the parameters $w_0=0.8w_1$, $w_1=110$ meV and $\theta=1.6^\circ$, and  $\gamma=2.364$ eV. (a,c) Starting from AAA stacking and (b,d) starting from ABC stacking.}
\label{fig:static}
\end{figure}

To provide a reference for our discussion of the non-equilibrium case, we remind the reader of some equilibrium properties of twisted trilayer graphene. We plotted the band structure for the AAA and ABC stacked twisted trilayer graphene with top and middle layers twisted as shown in Fig. \ref{fig:static}. We have omitted the ABA stacking case because we find that top layer twisted  ABA TTG has a band structure that for small twist angles is equivalent to the top layer twisted ABC case. Similarly, the middle layer twisted ABA case has a band structure that is equivalent to middle layer twisted AAA TTG . This phenomenon is similar to twisted bilayer graphene where for small twist angles it does not matter whether one started from AA stacking or AB stacking. This property is also preserved once we introduce circularly polarized light and longitudinal light coming from a waveguide.

\section{Non-equilibrium system and theoretical approach}
\label{sec:III}

\begin{figure}[!htb]
	\begin{center}
		\includegraphics[width=1\linewidth]{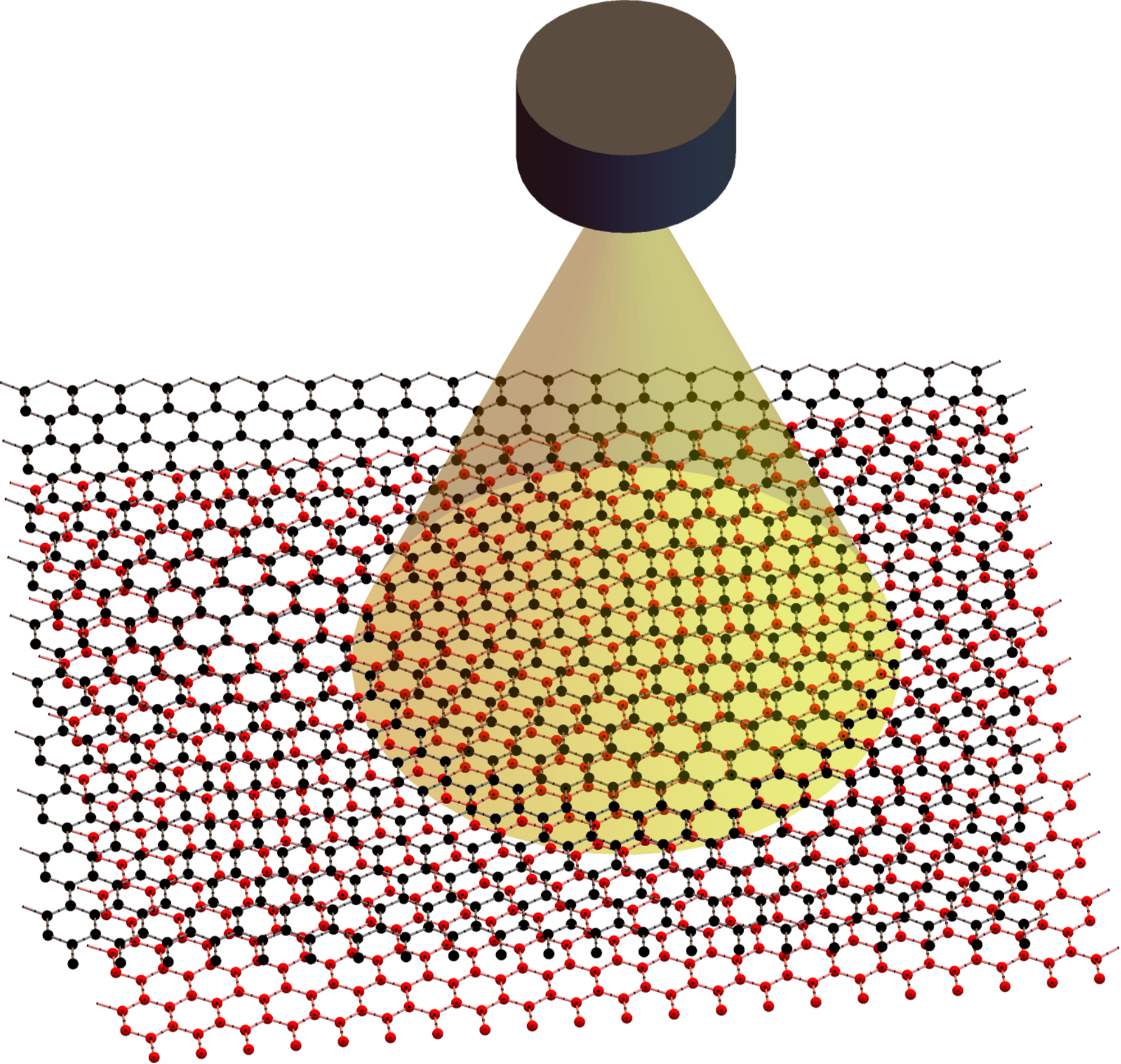}
		\caption{Light source applied on TTG system with AAA middle layer twisted.}
	\label{fig:light}
	\end{center}
\end{figure}

In this section we discuss the effect of shining different types of light on TTG samples (Fig.~\ref{fig:light}) and how to include this effect in our theoretical model. The first light source we consider in this work is a circularly polarized light. If this light is applied perpendicular to the graphene layers, at frequency $\omega$ and driving strength $A$, then we include its effect in a semi-classical fashion - assuming large photon numbers-  by making use of the minimal substitution prescription $k_x\to k_x-A\cos(\omega t)$ and $k_y\to k_y-A\sin(\omega t)$ \cite{TBG2020}. Thus, we have a time-periodic Hamiltonian satisfying $H(\vec{x},\vec{k},t)=H(\vec{x},\vec{k},t+2\pi/\omega)$. We should mention that small deviations from normal incidence leads to small corrections in the high-frequency limit, which is the case in this work, and thus it can be neglected. For a study of oblique incidence light in bilayer graphene, see \cite{Kumar_2021}.

The second type of light we will be considering in this work is longitudinal light coming from a waveguide. Here, the boundary conditions of a waveguide allow for light with longitudinal components to exist, which is not possible in vacuum \cite{PhysRevB.101.241408}. The semi-classical Peirls substitution teaches us how to include a vector potential in a tight binding model \cite{PhysRevB.101.241408}. Since our model can be derived from a tight binding model, it becomes clear that the effect of this type of light is included in the Hamiltonian via the substitution $w_0\to w_0e^{-ia_{AA}A\cos(\omega t)}$ and $w_1\to w_1e^{-ia_{AB}A\cos(\omega t)}$, where $a_{AA}=0.36 $ nm and $a_{AB}=0.34$ nm are interlayer distances in AA and AB regions of the twisted materials \cite{MoireFloquetRev}. The effect of this type of light therefore is to turn interlayer hoppings time-dependent.  

In both cases light leads to a periodically time-dependent Hamiltonian.  It is therefore necessary to introduce tools that allow us to properly treat the periodic time dependence. For this purpose, we note that the solution of the time-dependent Schrodinger equation can be written in Bloch form $\psi(t)=e^{-i\varepsilon t}u(t)$, for some periodic function $u(t+T)=u(t)$ and quasi-energy $\varepsilon$, this gives the following equation for $u(t)$ \cite{MoireFloquetRev}
 
\begin{equation}
\label{eqn:timeDepSchr}
    (H-i\partial_t)u(t)=\varepsilon u(t)
\end{equation}
Because $u(t)$ is periodic we can expand the Hamiltonian in the Fourier basis $\ket{n}=e^{in\omega t}/\sqrt{2\pi/\omega}$ and find
\begin{equation}
\sum_{\ell}\left[H^{(j-\ell)}+\delta_{\ell,j}\ell\omega\right]u_{\ell}=\varepsilon u_j,
\label{eq:Floquet_schroedinger}
\end{equation}
where $H^{(M)}=T^{-1}\int_{0}^{T}dt e^{-iM\omega t}H$ and $T=2\pi/\omega$ \cite{MoireFloquetRev}. This equation is effectively time-independent and can therefore be treated using equilibrium techniques. Clearly, this comes at the price, however, that the resulting matrix needs to be truncated to finite order since it can be really large. It is therefore computationally expensive and thus approximations can be convenient. In later sections we will consider two types of approximations, which allow us to work with much smaller matrices that will help speed up computation times \cite{TBG2020}. 

\section{Circularly polarized light}
\label{sec:circ_pol_lighte}

\subsection{Numerical band structure results}

In this section we study the effects that circularly polarized light has on the band structure of the various twisted trilayer graphene systems. To do this we evaluate the Floquet-Schr\"odinger equation \eqref{eq:Floquet_schroedinger} numerically.  

\begin{figure}[!htbp]
\centering
    \includegraphics[height=0.47\linewidth,width=0.5\linewidth]{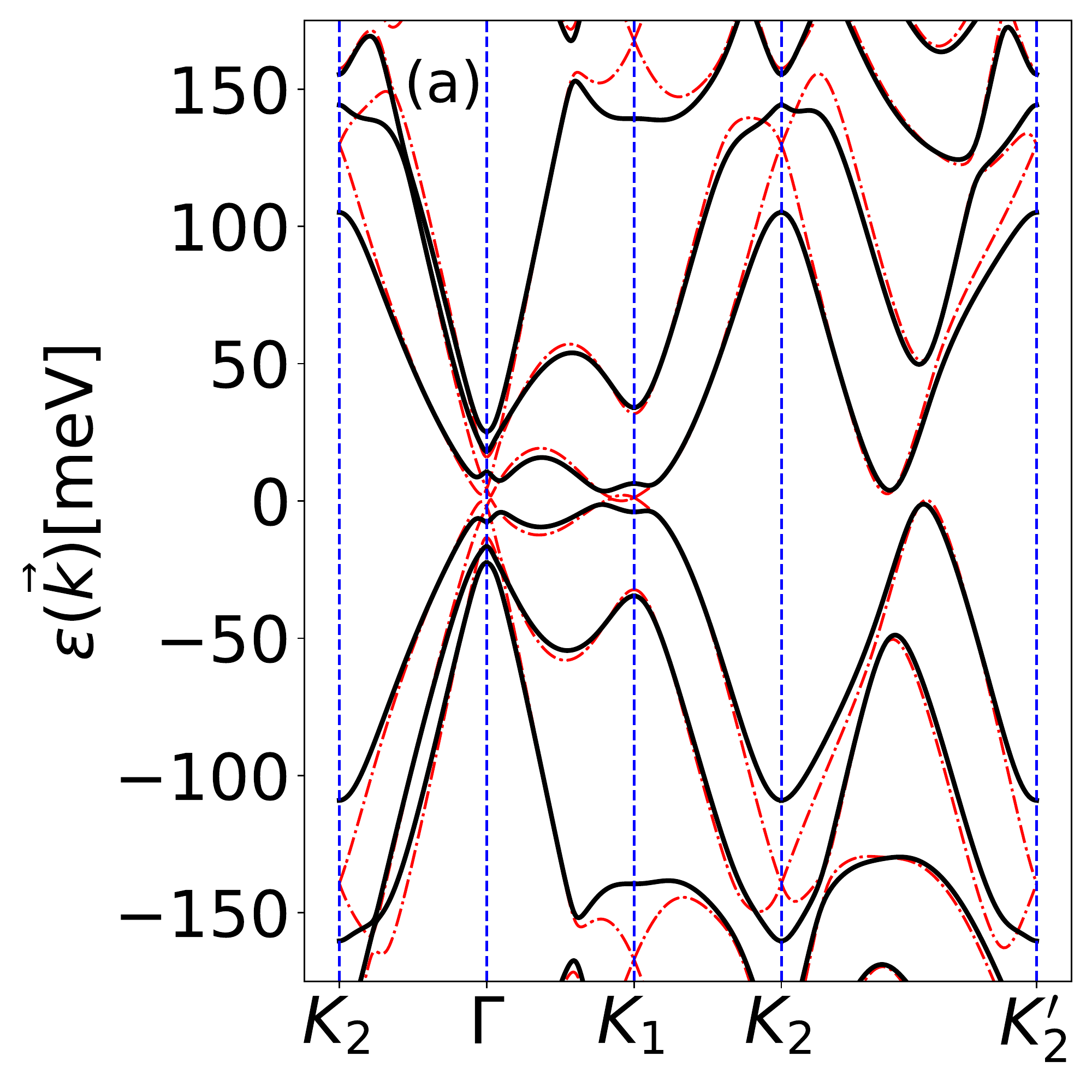}\hfill
    \includegraphics[height=0.53\linewidth,width=0.5\linewidth]{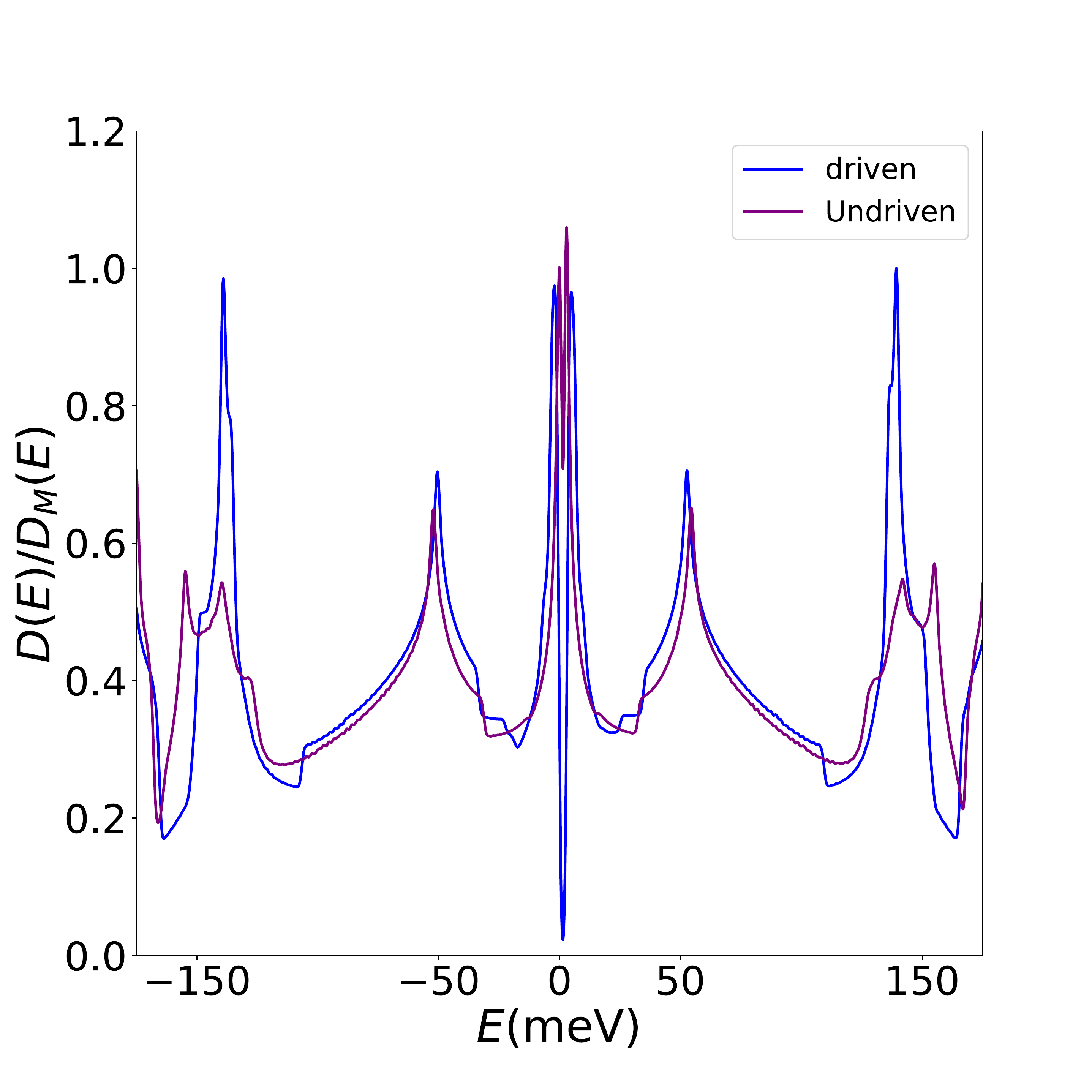}
    \hspace{0mm}
    \includegraphics[width=0.47\linewidth]{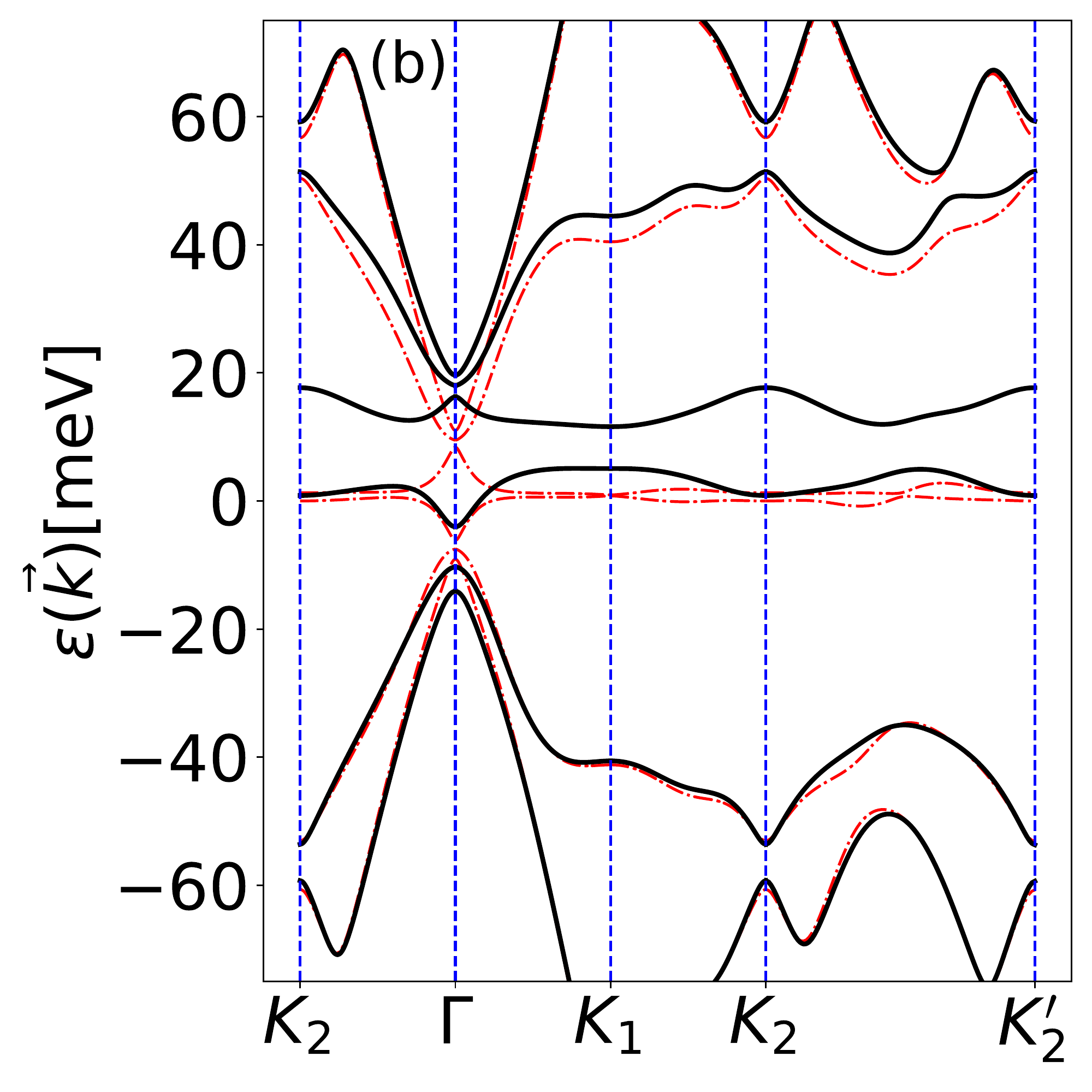}\hfill
    \includegraphics[width=0.53\linewidth]{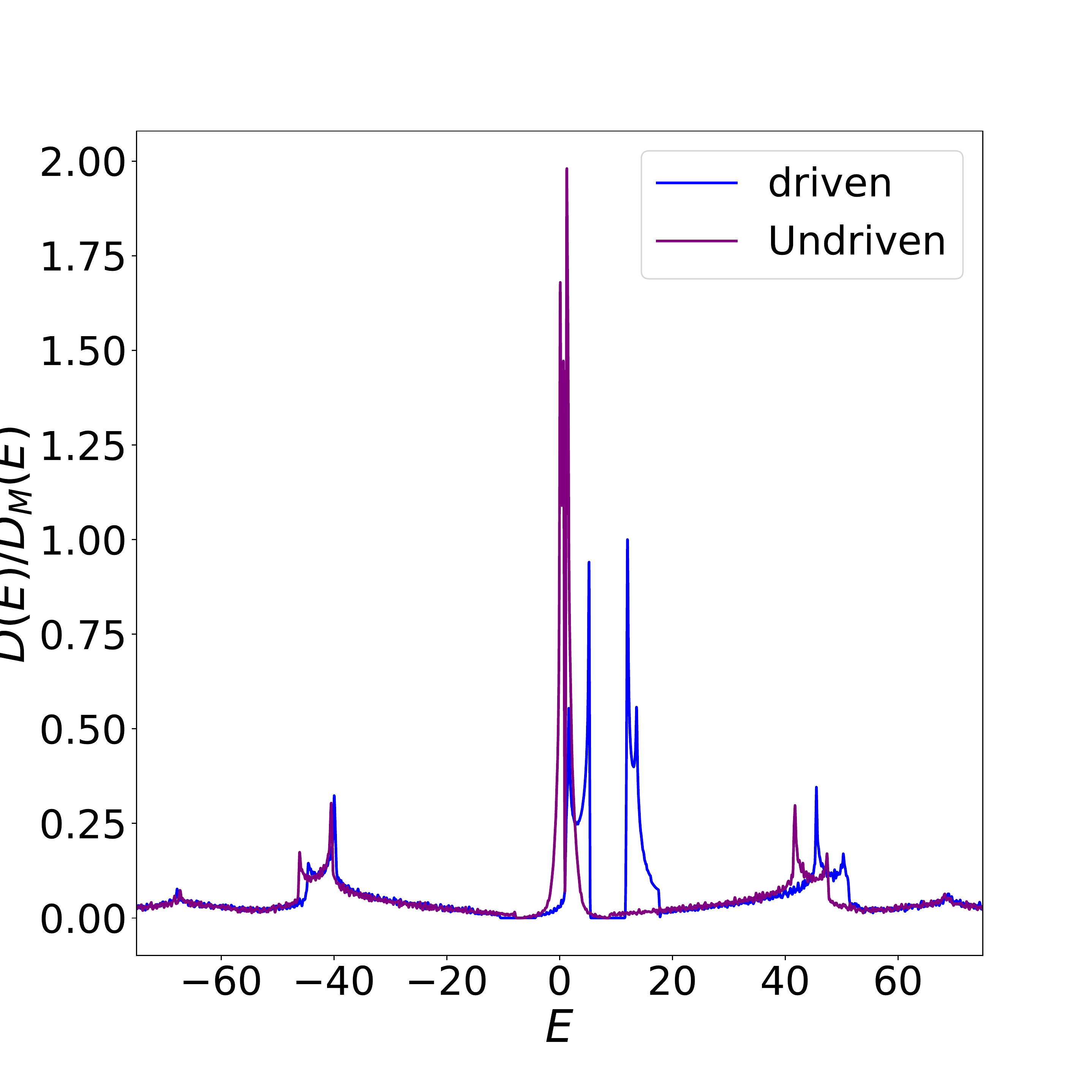}
    \hspace{0mm}
    \includegraphics[width=0.47\linewidth]{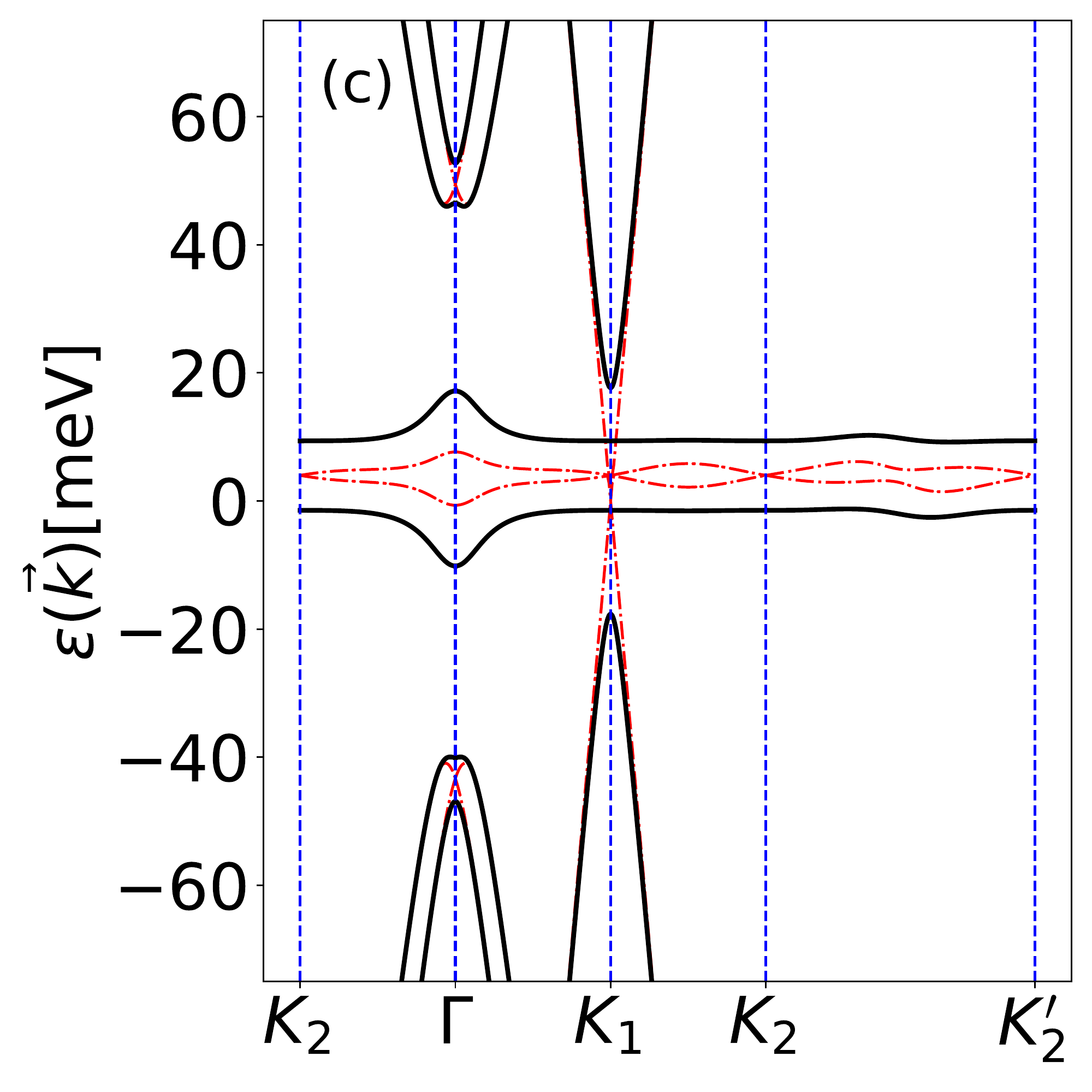}\hfill
    \includegraphics[width=0.53\linewidth]{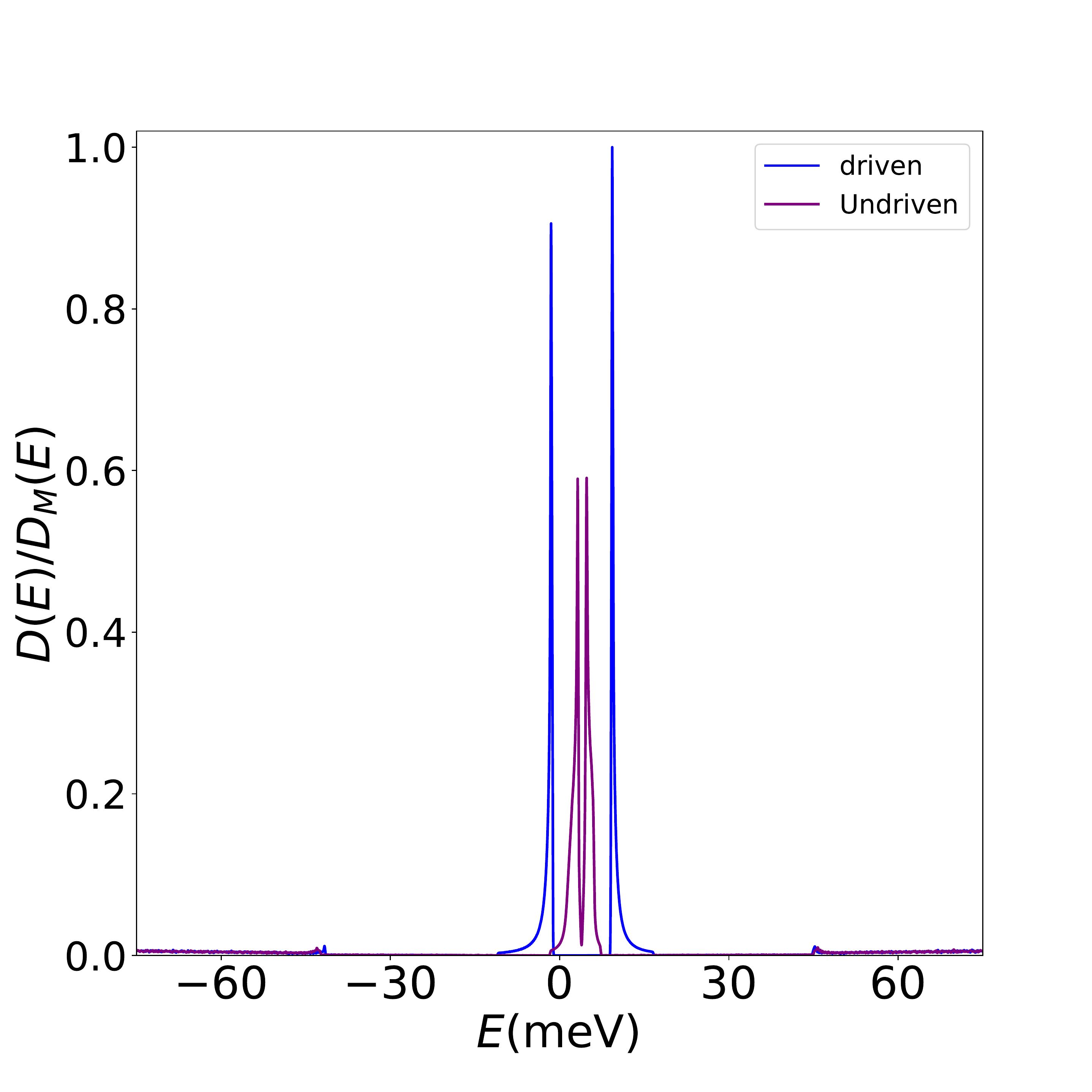}
    \hspace{0mm}
    \includegraphics[width=0.47\linewidth]{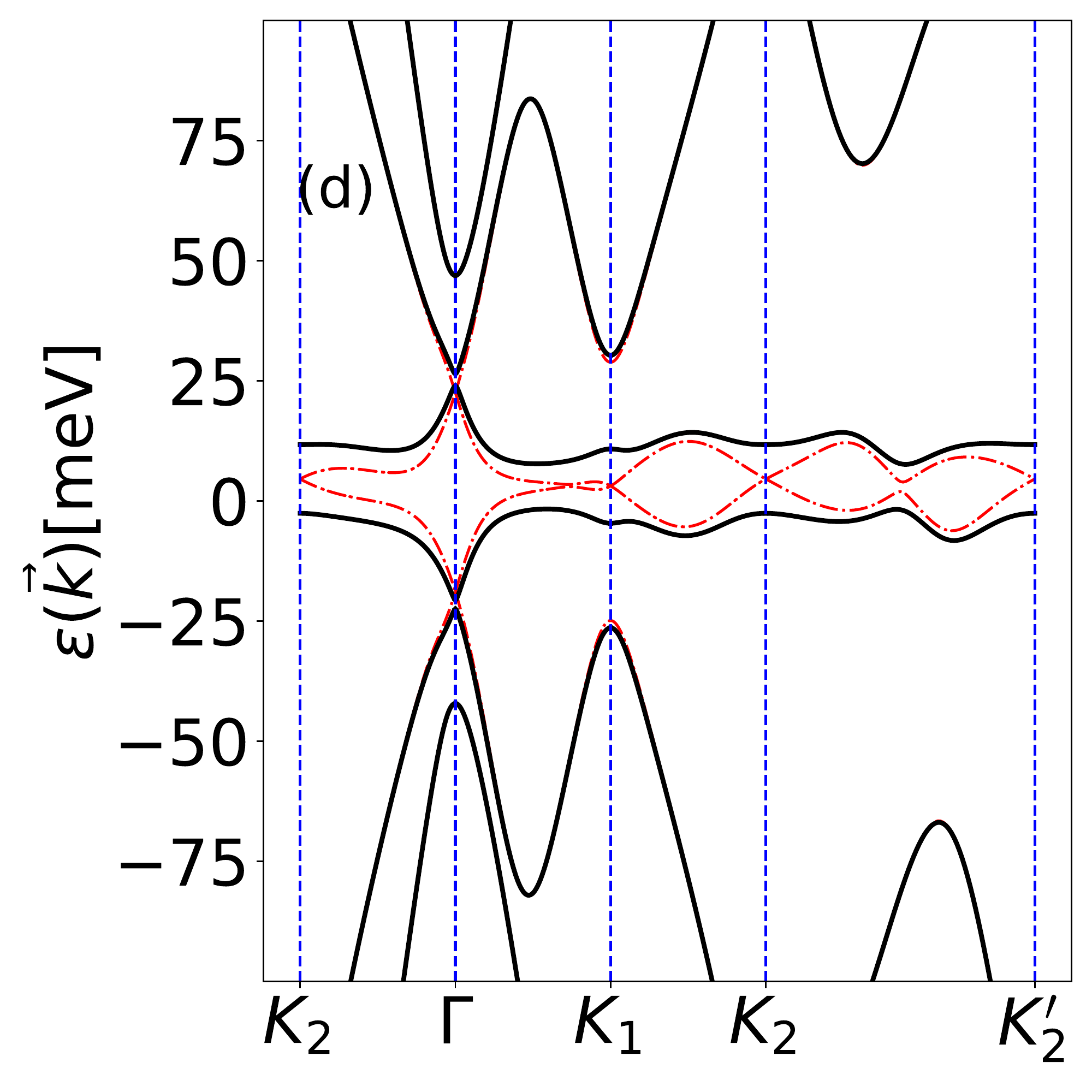}\hfill
    \includegraphics[width=0.53\linewidth]{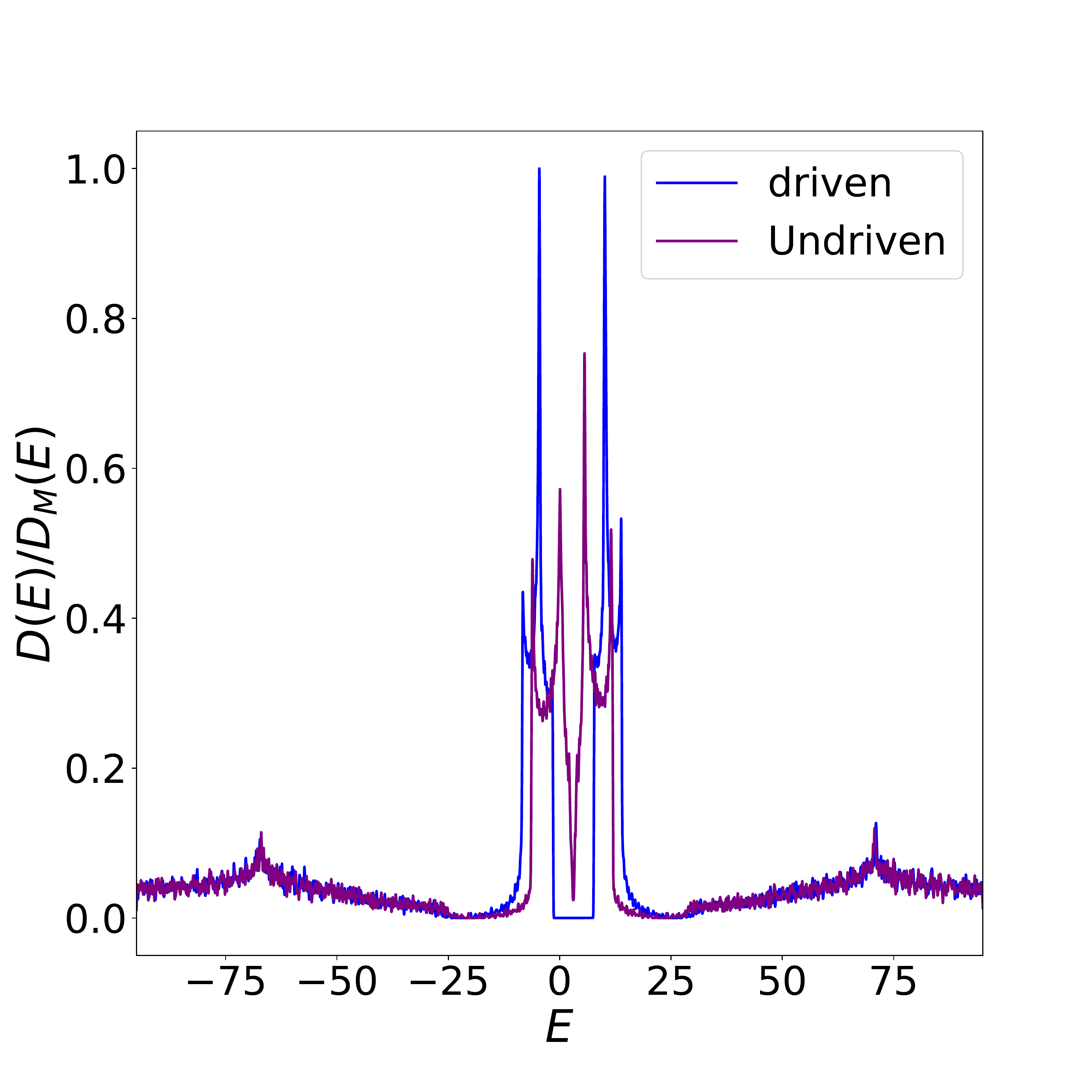}
\caption{(Color online)  Band structure (left column) for TTG driven by circularly polarized light  $w_0=0.8w_1$, $w_1=110$ meV and  $\gamma=2.364$ eV. (a) Starting with AAA stacking with TLT and parameters $(\theta,Aa_0,\omega)=(1.8^\circ,0.25,2\gamma)$, (b) the ABC with TLT and $(\theta,Aa_0,\omega)=(1.0^\circ,0.15,3\gamma)$, (c) AAA MLT and $(\theta,Aa_0,\omega)=(1.6^\circ,0.15,3\gamma)$, and (d) ABC MLT with $(\theta,Aa_0,\omega)=(1.6^\circ,0.15,3\gamma)$. In the above plots, the dash-doted lines represent the undriven case and the solid lines represent the driven case. The corresponding density of states plots (DOS) are on the right column rescaled by $D_M(E)$ which is the maximum value of the DOS of the driven case.}
\label{fig:bandstruct_driven_circ_pol}
\end{figure}

In Fig. \ref{fig:bandstruct_driven_circ_pol} we plotted the band structure for different driving strengths, driving frequencies and twist angles in the vicinity of the magic angle of the TTG. We show both the driven case as well as the undriven case to allow for a comparison. The cases we consider are twisted configurations that start from both AAA and ABC stacking and in each case we consider both top layer and middle layer twists. 

An interesting effect of circularly polarized light is that for certain choices of the twist angle, driving strength $A$ and the driving frequency $\omega$, one can flatten the central bands making it a very interesting candidate for strongly correlated phases because one can expect interactions to be dominant in this case. For example, in Fig. \ref{fig:bandstruct_driven_circ_pol} (c), we see that the two middle bands are less dispersive than their equilibrium counterparts. We observe that not all configurations in Fig. \ref{fig:bandstruct_driven_circ_pol}  yield less dispersive bands such as the case of AAA stacked top layer twist. The effect depends on the relative location of the Dirac cones, and subsequent hybridization due to interlayer couplings. In the same figure, we also plotted the density of states associated with each case. The plots reflect the flattening of the bands via the appearance of sharper peaks as can be seen in the case of AAA stacking with a middle layer twist. Moreover, Fig. \ref{fig:bandstruct_driven_circ_pol} shows the appearance of gap openings which is another rationale for using this type of light. Consequently, this opens up the possibility to discuss topological effects that are characterized by the Chern number of isolated bands. Since we work in the high frequency regime (the driving frequency is larger than the bandwidth of the model Hamiltonian) this allows access to information about topological edge states \cite{ChernNo1}. This is unlike the low frequency regime that requires the calculation of the winding numbers for a proper description of band topology - in this regime the connection between relative Chern numbers and number of edge states is not immediate anymore. Since such calculations can be computationally very expensive when working with the full quasi-energy operator, it will turn out to be convenient to rely on approximate time independent Floquet Hamiltonians that will be introduced in next section \ref{sec:eff_ham_circpol}.

\subsection{Effective Hamiltonians}
\label{sec:eff_ham_circpol}
In this section we introduce two approximations to the Floquet Hamiltonian describing the various twisted trilayer graphene systems.
Before we get to the different specific approximations let us first make a few approximations that will be convenient in both cases. It will be useful to linearize the Hamiltonian \eqref{eq:macdHam_mod}. However, since we want to capture the effect of periodic drive as accurately as possible instead of just linearizing by a Taylor expansion we first expand the Hamiltonian to first order in a Fourier series and subsequently linearize it in momenta. The result is that 
\begin{equation}
    f(\vec{k}-\vec{A})\approx a_0(k_x-ik_y)J_0(2Aa_0/3)+3J_1(2Aa_0/3)e^{i\omega t}
\label{eqn:linear}
\end{equation}
where $J_n(x)$ is the Bessel function of the first kind.
\subsubsection{Van Vleck approximation}
The first approximation we now use is the standard perturbative Van Vleck (vV) approximation \cite{Eckardt_2015,MoireFloquetRev}, where the effective Hamiltonian to first perturbative order in $1/\omega$ is given as
\begin{equation}
\begin{aligned}
H_{{\rm v}V}^{(1)}=H^{(0)}+\sum_{m\neq 0}\frac{\hat H^{(m)}\hat H^{(m)}}{m\omega}.
\end{aligned}
\end{equation}
Hereby, we made use of $H^{(n)}=\frac{1}{T}\int_0^Tdt e^{-in\omega t}H(t)$, which are Fourier modes of the Hamiltonian $H(t)$.
Within this approximation, the Hamiltonian takes almost the same form as in Eq. \eqref{eq:macdHam_mod} just with the replacement $f(\vec{k})\to a_0(k_x-ik_y)J_0(2Aa_0/3)$ and modified $2\times 2$ blocks $h_\ell\to h_\ell+{\rm diag}\left[-\Delta,+\Delta\right]$, where $\Delta=(9\gamma^2/\omega)J_1^2(2Aa_0/3)$ \cite{TBG2020}. Essentially this means that the circularly polarized light modified the Fermi velocity and introduced a Dirac gap $\Delta$ into the Hamiltonian.

\subsubsection{The Rotating Frame Approximation}
An alternative but non-perturbative scheme is to transform to a rotating frame (RF) Hamiltonian $H_R=U(t)^\dag (H-i\partial_t)U(t)$ that has a less important time dependence than the original Hamiltonian. A subsequent time average yields a Hamiltonian that is more accurate than the vV Hamiltonian. It cannot be stressed enough that special care has to be taken in that the rotating frame transformation is chosen such that the terms that are neglected in the time average do not cause breaking of  the six-fold rotational symmetry in momentum space. Here, we provide a simple generalization of the unitary transformation that was introduced in \cite{TBG2020} and that fulfills this property. We start with linearized dispersion $f(\vec{k})$ Eq. \ref{eqn:linear} in the full Hamiltonian, the time dependent Hamiltonian becomes $H(x,\vec{k},t)=H(x,\vec{k})+V(t)$. The unitary transformation can be proposed in the form \cite{TBG2020} $U_R(t)=e^{-i\int dt V_1(t)}e^{-i\int dt V_2(t)}$ with a properly chosen decomposition of the time periodic part of the Hamiltonian $V(t)=V_1(t)+V_2(t)$. Here, $V_1(t)$ is the part of the Hamiltonian that is $\propto\cos(\omega t)$, while $V_2(t)$ is $\propto\sin(\omega t)$. The transformation then is given as
\begin{equation}
\begin{aligned}
    &U_R(t)=\mathrm{diag}[u_R(\theta_1,t),u_R(\theta_2,t),u_R(\theta_2,t)]\\
    &u_R(\theta,t)=u_{R,1}(\theta,t)u_{R,2}(\theta,t)\\
    &u_{R,n}=\cos[\gamma_n(t)]-i\sin[\gamma_n(t)]\sigma_n^\theta
    \end{aligned},
\end{equation}
where $\gamma_1(t)=B_\Omega\sin(\omega t)$, $\gamma_2(t)=B_\Omega(1-\cos(\omega t))$, $B_\Omega=3J_1(2Aa_0/3)\gamma/\Omega$ and appropriately rotated Pauli matrices are given as $\sigma_n^\theta=e^{-i\theta/2\sigma_3}\sigma_ne^{i\theta/2\sigma_3}$.

After taking the time average we arrive at an effective Hamiltonian $H_{\mathrm{eff}}^{(\mathrm{bare})}$, which still has a form that is too cumbersome to display here and is hard to interpret. Therefore we apply another unitary transformation that is given by
\begin{equation}
    R=\mathrm{diag}[e^{iB_\Omega \sigma_2^{\theta_1}},e^{iB_\Omega \sigma_2^{\theta_2}},e^{iB_\Omega \sigma_2^{\theta_3}}],
\end{equation}
where we used definitions for rotated Pauli matrices from above.
Consequently, we calculate our effective Hamiltonian in the rotating frame as
\begin{equation}
\eqfitpage{
    H(\vec{x},\vec{k})=\begin{pmatrix}
    \tilde{h}(\theta_{1},\vec{k}-\vec{\kappa}_{1})&\tilde{T}_{12}(\vec{x})&0\\
    \tilde{T}_{12}^{\dagger}(\vec{x})&\tilde{h}(\theta_{2},\vec{k}-\vec{\kappa}_{2})&\tilde{T}_{23}(\vec{x})\\
    0&\tilde{T}_{23}^{\dagger}(\vec{x})&\tilde{h}(\theta_{3},\vec{k}-\vec{\kappa}_{3})
    \end{pmatrix}},
\end{equation}
where the single layer graphene blocks are modified as follows 
\begin{equation}
    \tilde{h}(\theta,\vec{k})=a_0\gamma_{\rm RF}R({\theta})\vec{k}\cdot\vec{\sigma}-\Delta_{\rm RF}\sigma_{3},
\end{equation}
the interlayer hoppings become
\begin{equation}
    \gamma_{\rm RF}=\gamma J_0\left(-\frac{6\gamma}{\Omega}J_1\left(\frac{2Aa_0}{3}\right)\right)J_0\left(\frac{2Aa_0}{3}\right)
\end{equation}
and a Dirac gap that is given as
\begin{equation}
    \Delta_{\rm RF}=-\frac{3\gamma}{\sqrt{2}}J_1\left(\frac{2Aa_0}{3}\right)J_1\left(-\frac{6\sqrt{2}\gamma}{\Omega}J_1\left(\frac{2Aa_0}{3}\right)\right)
\end{equation}
is introduced.

The effective tunneling matrices are modified as follows. We first recognize that the original hopping matrices $T_{ij}$ can be expressed in terms of Pauli matrices as $T_{ij}=T_{ij}^0\sigma_0+T_{ij}^1\sigma_1+T_{ij}^2\sigma_2+T_{ij}^3\sigma_3$, where $T_{ij}^n$ are expansion coefficients. The modified interlayer hopping matrices $\tilde{T}_{ij}$ are then found if we replace the Pauli matrices by new matrices $\sigma_i\to\tilde\sigma_i$. That is we have $\tilde{T}_{ij}=T_{ij}^0\tilde{\sigma}_0+T_{ij}^1\tilde{\sigma}_1+T_{ij}^2\tilde{\sigma}_2+T_{ij}^3\tilde{\sigma}_3$, where $\tilde{\sigma}_{1,2}=J_0(\nu)\sigma_{1,2}$ and
\begin{equation*}
    \tilde{\sigma}_0=\sigma_0+(J_0(\sqrt{2}\nu)-1)\Big[\sigma_0\sin^2\left(\frac{\theta_i-\theta_j}{2}\right)
\end{equation*}
\begin{equation}
    -\frac{i}{2}\sigma_3\sin(\theta_i-\theta_j)\Big],
\end{equation}

\begin{equation*}
    \tilde{\sigma}_3=\sigma_3+(J_0(\sqrt{2}\nu)-1)\Big[\sigma_3\cos^2\left(\frac{\theta_i-\theta_j}{2}\right)
\end{equation*}
\begin{equation}
    +\frac{i}{2}\sigma_0\sin(\theta_i-\theta_j)\Big],
\end{equation}
with $\nu=(-6\gamma/\omega)J_1(2Aa_0/3)$. 

This Hamiltonian offers a huge reduction in computational cost when compared to the exact case where the quasi-energy operator - if we include a large number of Fourier modes - is very large.  The approximation offers highly reliable results for the experimentally accessible range of driving strengths in the high frequency regime. 

\subsubsection{Comparison of the effective Hamiltonian spectrum and the exact quasi-energies} 
\begin{figure}[!htbp]
\centering
    \includegraphics[width=0.47\linewidth]{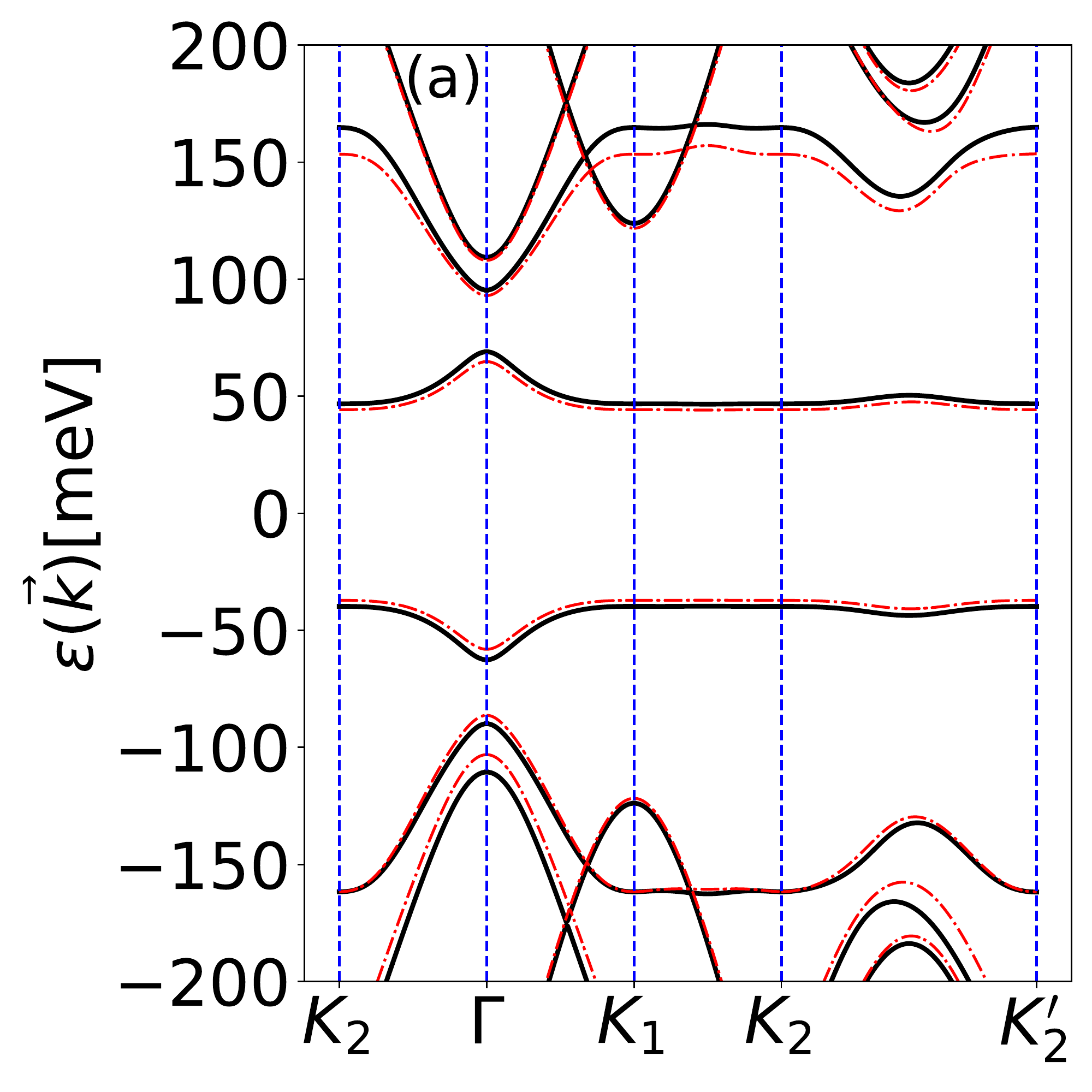}\hfil
    \includegraphics[width=0.47\linewidth]{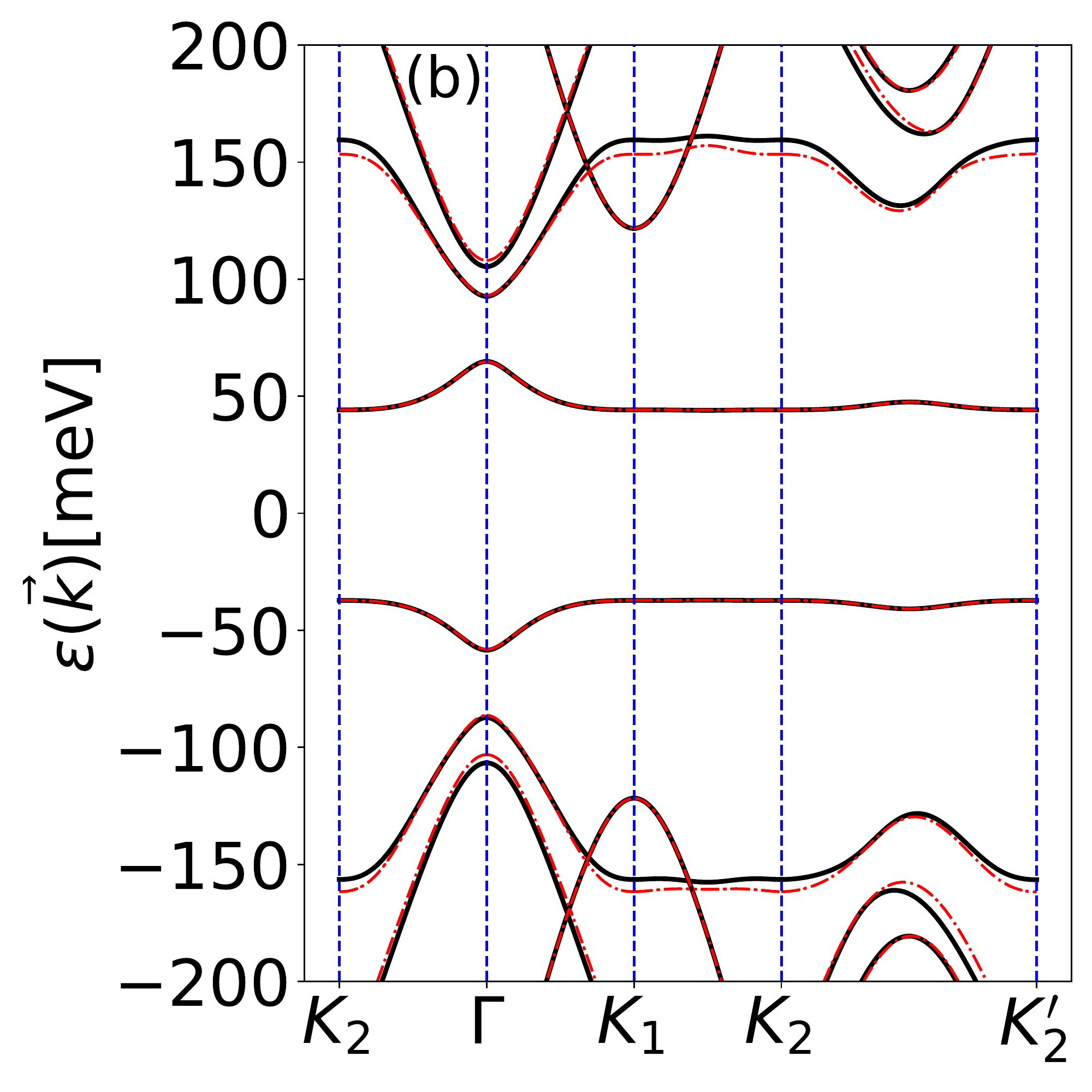}\par\medskip
\caption{(Color online) Comparison between the exact quasi-energies and the energies from the effective Hamiltonians for the band structure of the TTG with AAA stacking and middle layer twisted, driven by circularly polarized light with $w_0=0.8w_1$, $w_1=110$ meV and $\theta=1.6^\circ$, $Aa_0=0.4$, $\Omega=3\gamma$ and  $\gamma=2.364$ eV. (a) Van Vleck (solid lines) and exact quasi-energies compared (dashed lines) (b) Rotating frame (solid lines) versus the quasi-energies (dashed lines).}
\label{fig:RFvsVVvsExact}
\end{figure}

We compare each approximation to the exact result determined by the full quasi-energy operator Eq.~(\ref{eq:Floquet_schroedinger}) in order to probe the efficiency of each approximation scheme. To illustrate their accuracy we compare the full band structure generated plots of which we show one example in which we show both the result from the exact quasi energy operator versus the data found using the above approximate Hamiltonians. The results for an AAA stacked middle twist can be seen in Fig. \ref{fig:RFvsVVvsExact}.

\begin{figure}[!htbp]
\centering
    \includegraphics[width=0.47\linewidth]{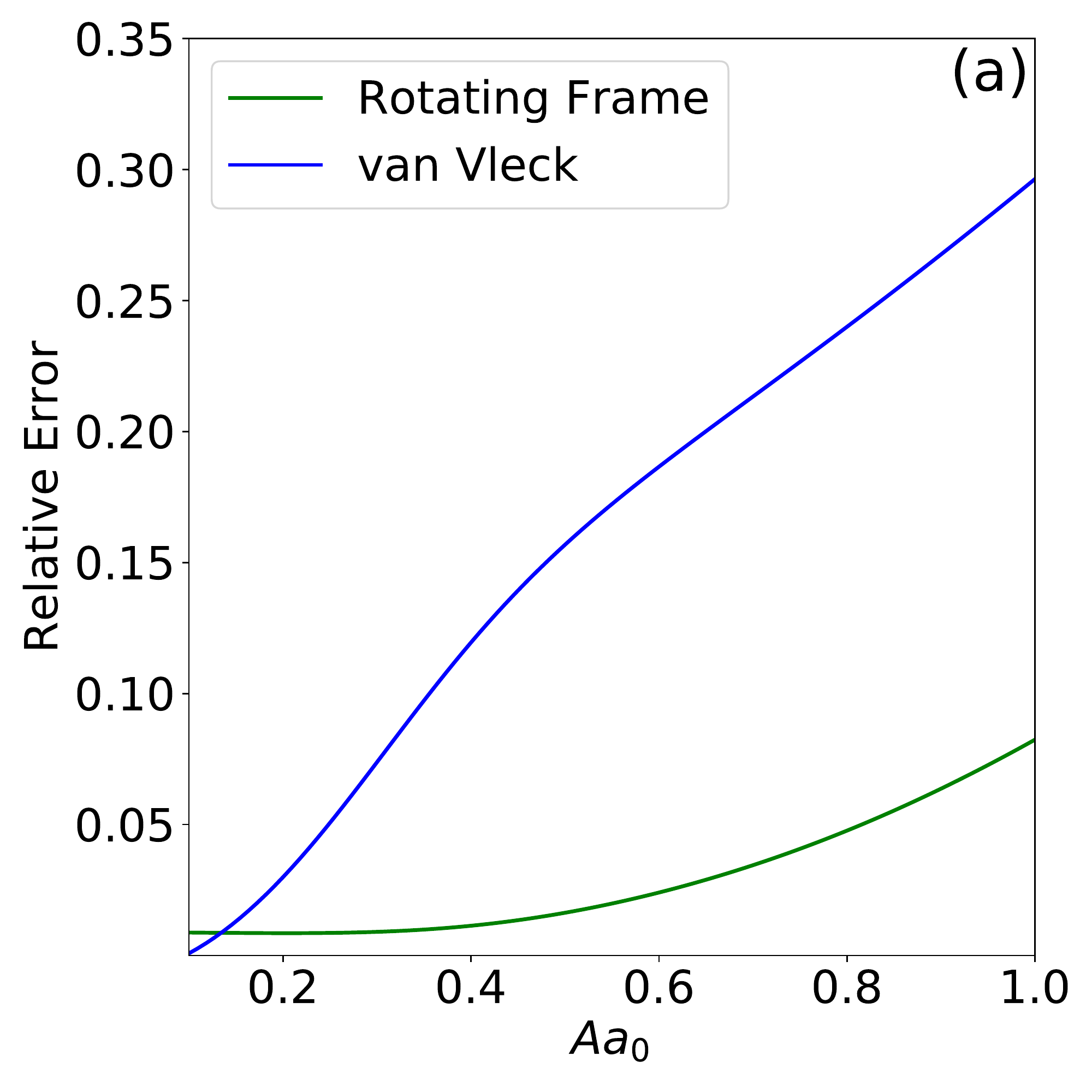}\hfil
    \includegraphics[width=0.47\linewidth]{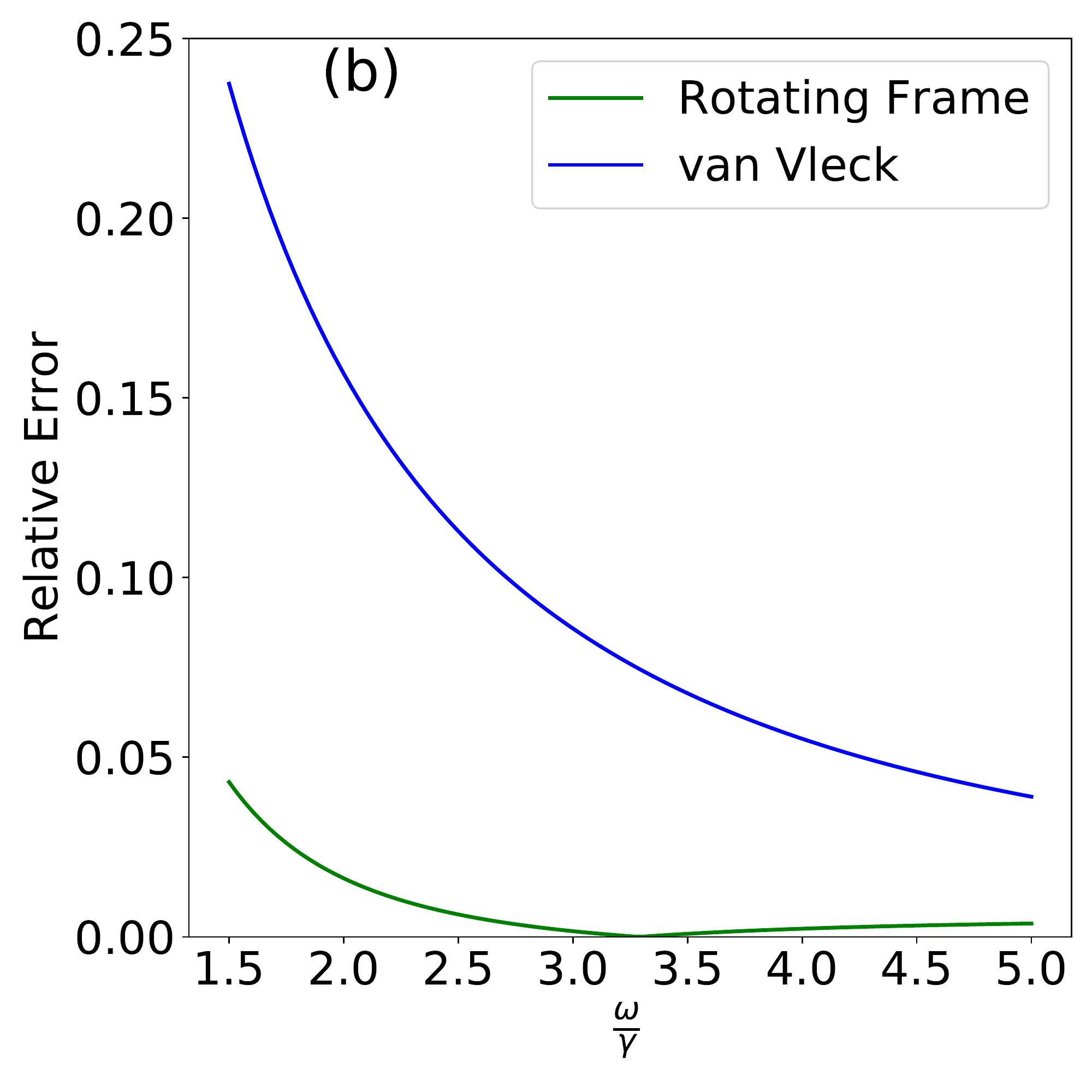}\par\medskip
\caption{(Color online) Plot of relative error for the gap at the $K_1$ symmetry point for the TTG system with AAA stacking and middle layer twisted driven by circularly polarized light (a) as function of $Aa_0$, $w_0=0.8w_1$, $w_1=110$ meV and $\theta=1.6^\circ$,  $\Omega=2\gamma$ and  $\gamma=2.36$, (b) and as a function of $\omega/\gamma$ with $Aa_0=0.5$.}
\label{fig:relativeE}
\end{figure}

We find that the rotating frame effective Hamiltonian is systematically closer to the exact quasi-energy than the truncated Van Vleck Hamiltonian. To further quantify  the efficiency of both approximations and get a feel for their range of validity, we computed the relative error in the center gap at the $K_1$ point for both methods and plotted the result as a function of driving strengths and frequencies as shown in Fig. \ref{fig:relativeE}. 

Here, we find that the vV approximation works well for very weak drives and large frequencies and quickly deteriorates otherwise, while the RF approximation works well for a much larger range of driving strengths $Aa_0$ and frequencies $\omega$. It therefore seems reasonable to use the rotating frame approximation in all that follows. This is further substantiated by the fact that the vV approximation can be obtained from the RF approximation by means of a Taylor expansion in small $J_1(2Aa_0/3)$. In this sense the RF approximation can be seen as a partial re-summation of the the vV expansion.

\subsection{Topological Phase Diagrams}

Next, we make use of the rotating frame Hamiltonian to compute maps of Chern numbers. First, however, let us give a brief summary of the algorithm due to Fukui \cite{ChernNo1} that we used in our computations.

To compute the Chern number we divide the moiré Brillouin zone into uniform small rectangles of size $dk_x\times dk_y$. Then we compute the so-called link variables $U_{\vec{k}_j\hat{u}}^{(\ell)}$\cite{ChernNo1},
\begin{equation}
    U_{\vec{k}_j\hat{u}}^{(\ell)}=\frac{\langle\psi^{(\ell)}(\vec{k}_j)|\psi^{(\ell)}(\vec{k}_j+\hat{u})\rangle}{|\langle\psi^{(\ell)}(\vec{k}_j)|\psi^{(\ell)}(\vec{k}_j+\hat{u})\rangle|}
\end{equation}
where $\psi^{(\ell)}(\vec{k}_j)$ is the eigenvector of the Hamiltonian corresponding to the band with index $\ell$, and $\hat{u}:=\hat{u}_{x}=\left(dk_x,0\right)$ or $\hat{u}:=\hat{u}_{y}=\left(0,dk_y\right)$. Next, we calculate the field strength \cite{ChernNo1}
\begin{equation}
    F_{\vec{k}_j}^{(\ell)}=\ln\left[U_{\vec{k}_j\hat{u}_{x}}^{(\ell)}U_{\vec{k}_j+\hat{u}_{x},\hat{u}_{y}}^{(\ell)}U_{\vec{k}_j+\hat{u}_{x}+\hat{u}_{y},\vec{k}_j+\hat{u}_{y}}^{(\ell)}U_{\vec{k}_j+\hat{u}_{y},\vec{k}_j}^{(\ell)}\right].
\end{equation}
Finally, the Chern number for the $\ell$th band is given as
\begin{equation}
    c_{\ell}=\frac{1}{2\pi i}\sum_{\vec{k}_j}F_{\vec{k}_j}^{(\ell)},
\end{equation}

where the sum is taken over all plaquettes in the Brillouin zone.

\begin{figure}[!htbp]
\centering
    \includegraphics[width=0.47\linewidth]{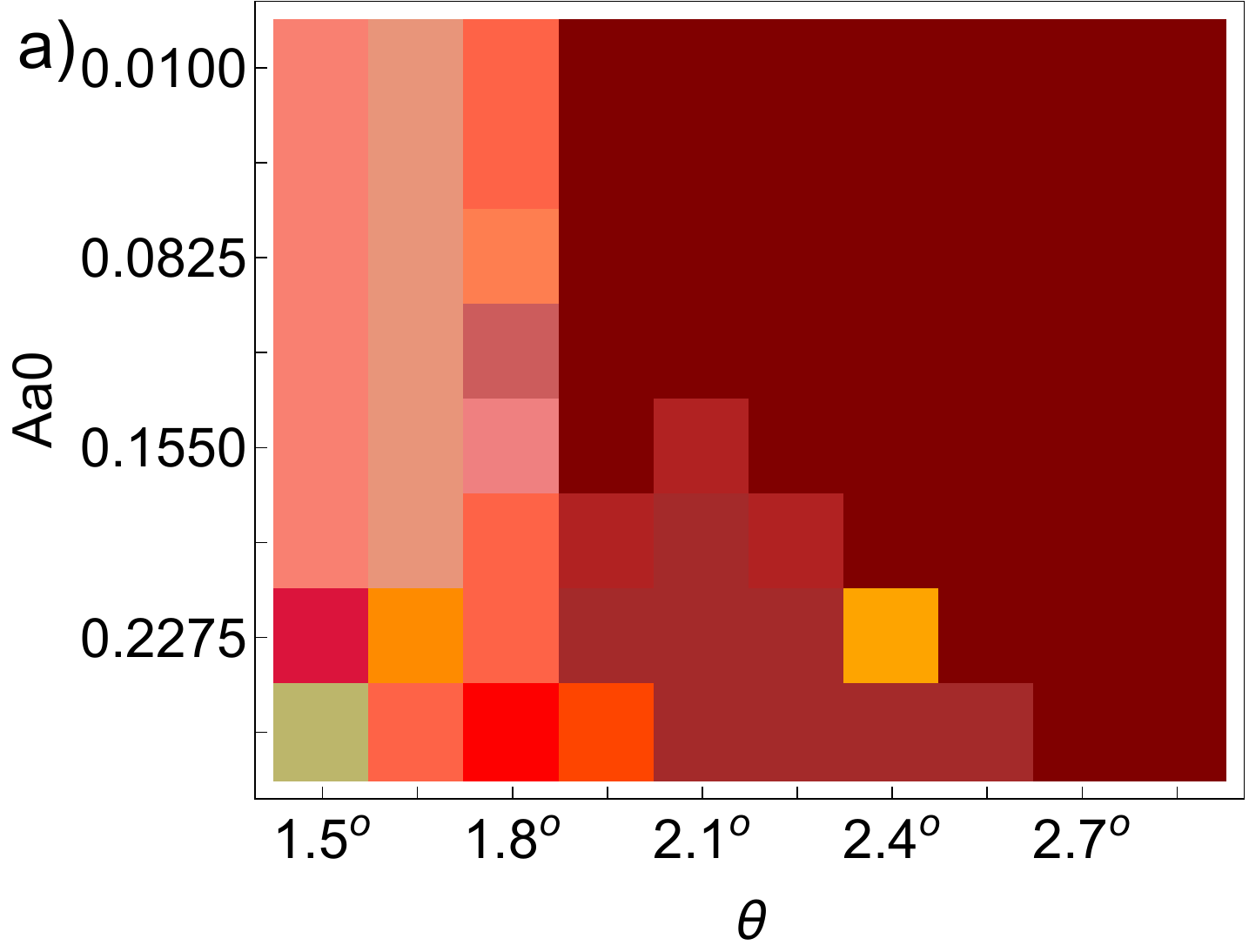}\hfill
    \includegraphics[width=0.47\linewidth]{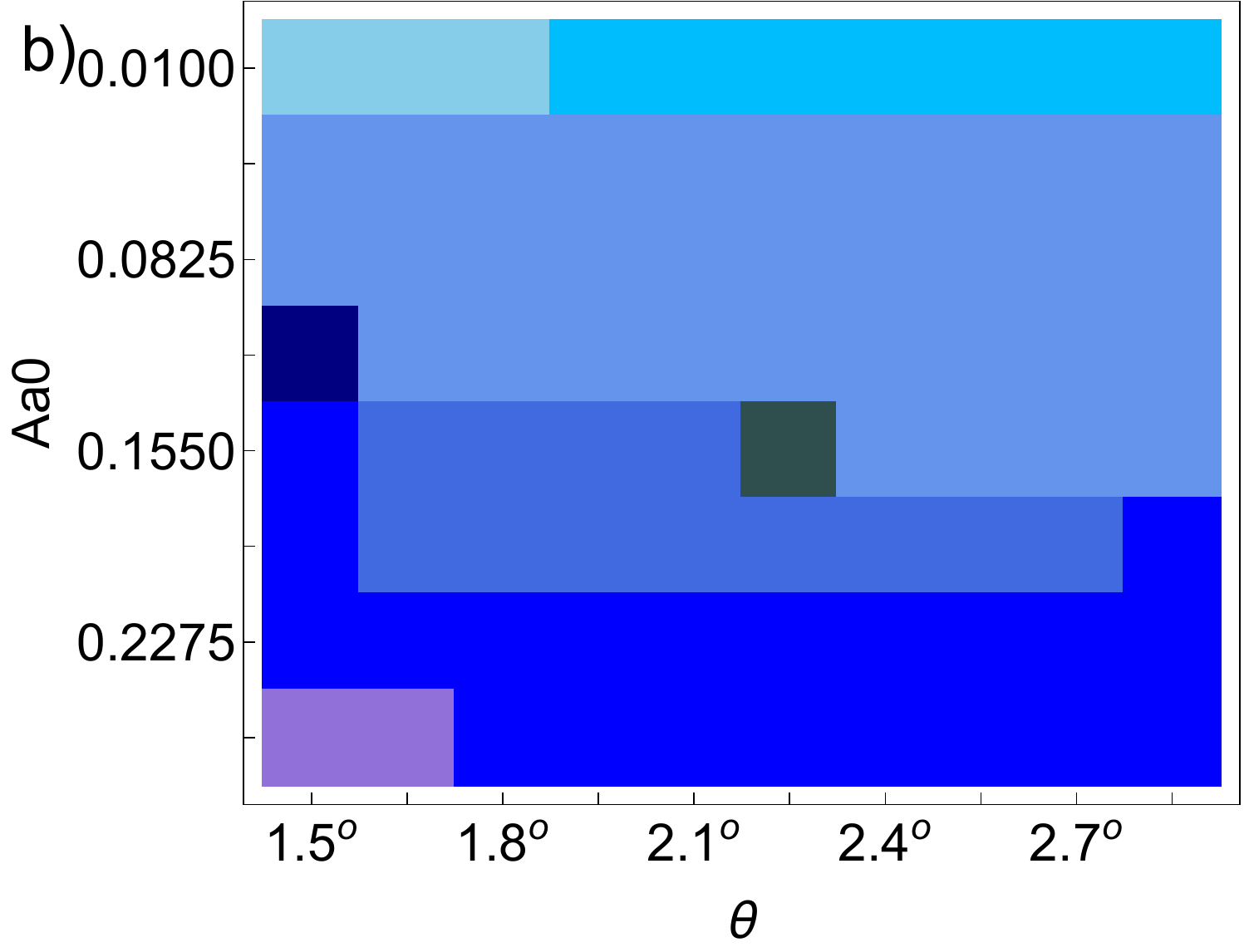}
    \hspace{0mm}
    \includegraphics[width=0.47\linewidth]{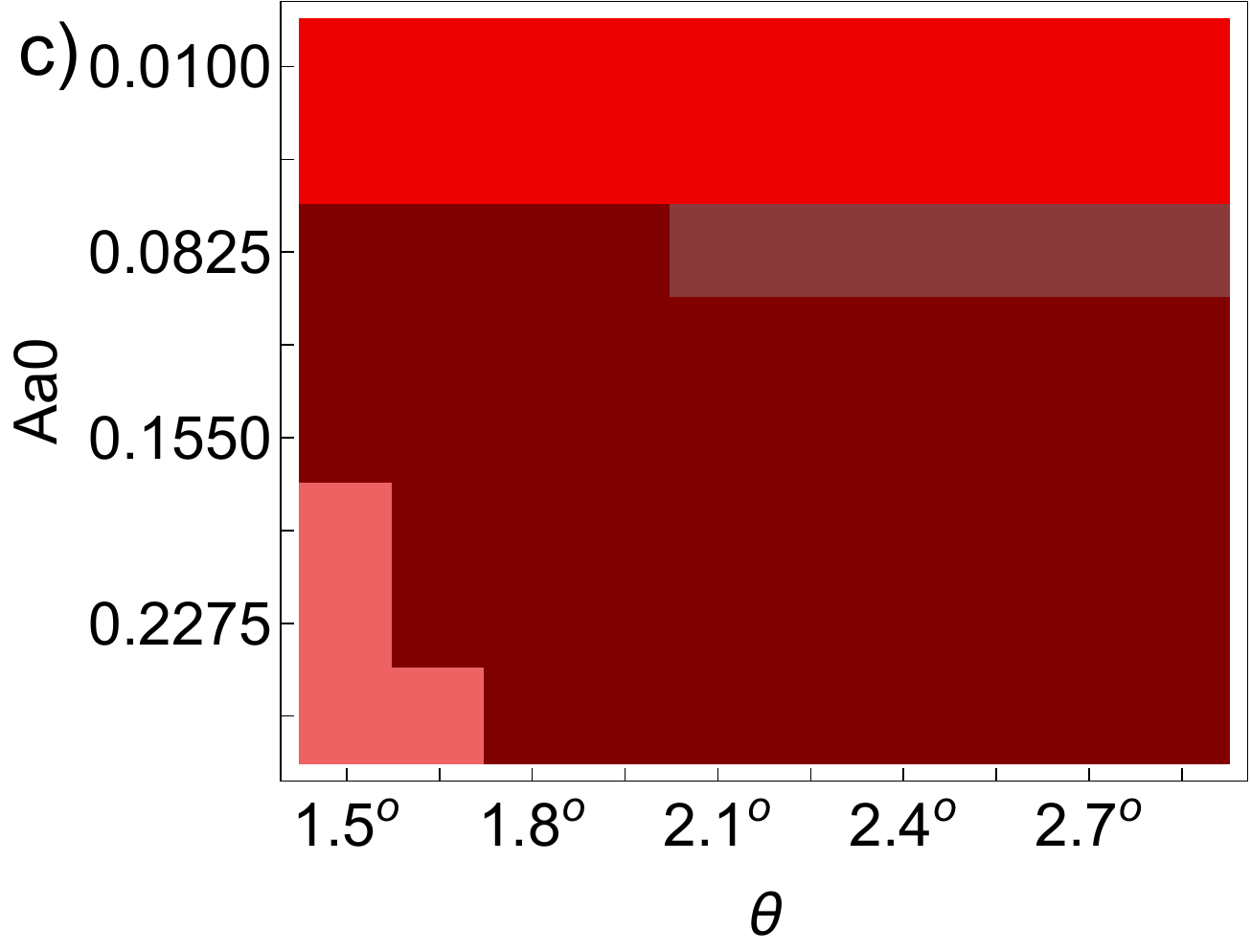}\hfill
    \includegraphics[width=0.47\linewidth]{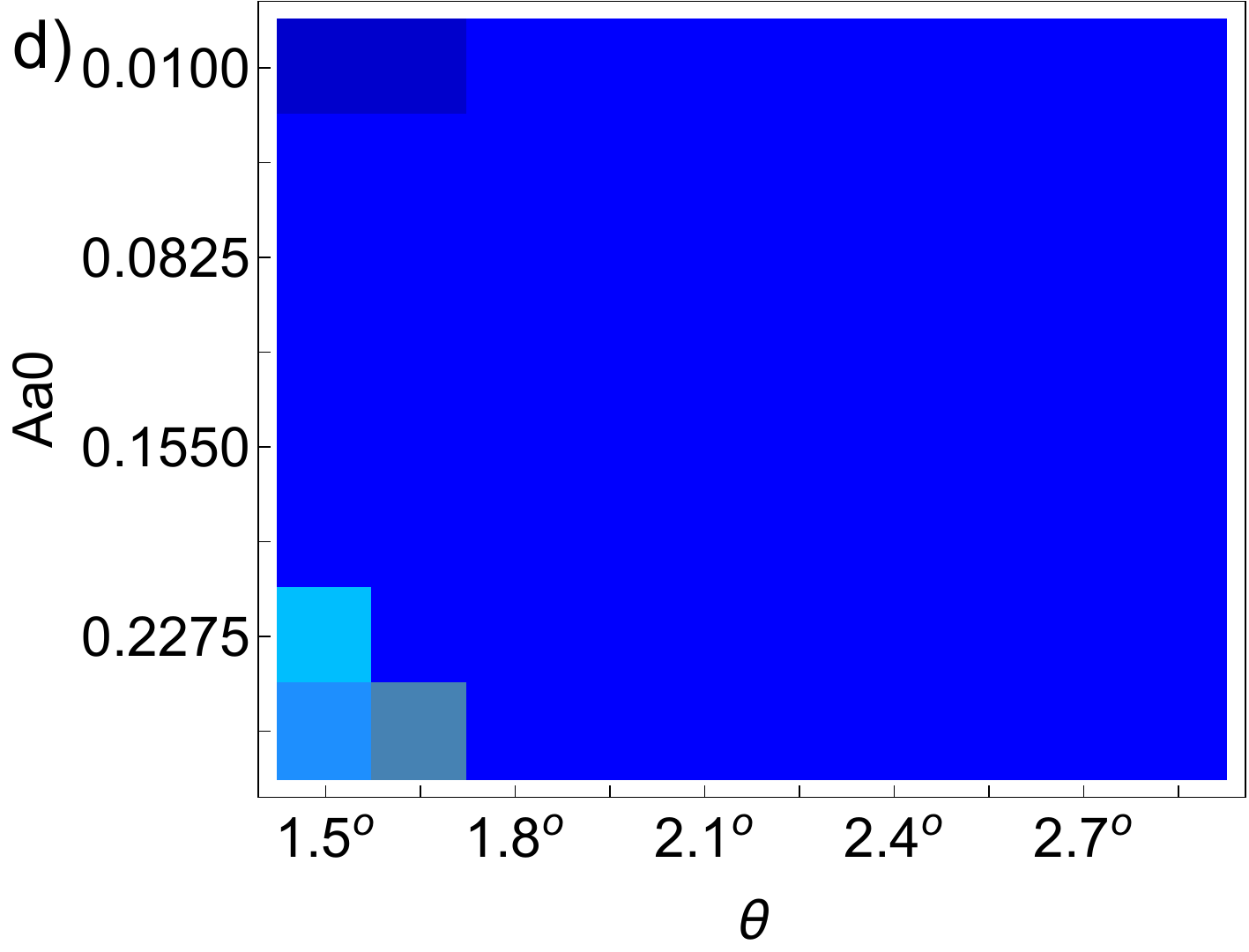}
\caption{(Color online) The topological phase diagrams for the TTG system for a range  of values $Aa_0$ and the twist angle $\theta$ with $\omega=2\gamma$. (a) AAA stacking and top layer twist, (b) ABC stacking with top layer twisted, (c) AAA stacking with middle layer twisted, and (d) ABC stacking with middle layer twisted. Each color represents a set of values for the Chern numbers of the central six bands as indicated in Table.~\ref{tab:colors}. }
\label{fig:phaseI}
\end{figure}

In this study, we restrict ourselves to the topology of the six central bands and we use the rotating frame Hamiltonian to be able to compute Chern numbers sufficiently quickly. We have spot-checked our results against results that we obtained when we were working with the full quasi-energy operator. Since we work in the high frequency regime, where the frequency is larger than the bandwidth of the included bands, it is sufficient to consider the Chern numbers to learn more about topological properties.  We computed Chern numbers for the six central bands and for various values of the driving strength $Aa_0$ and the twist angle $\theta$ at a fixed driving frequency $\omega=2\gamma$. The resulting topological phase diagrams are shown in Fig.\ref{fig:phaseI}. Due to the high computational cost, we limited ourselves to driving strengths between $Aa_0=0$ and $Aa_0=0.3$ (this for frequencies larger than the bandwidth which is the experimentally favorable regime), and twist angle ranging from $1.5^\circ$ to $2.85^\circ$.

\begin{figure}
	\begin{center}
		\includegraphics[width=1\linewidth]{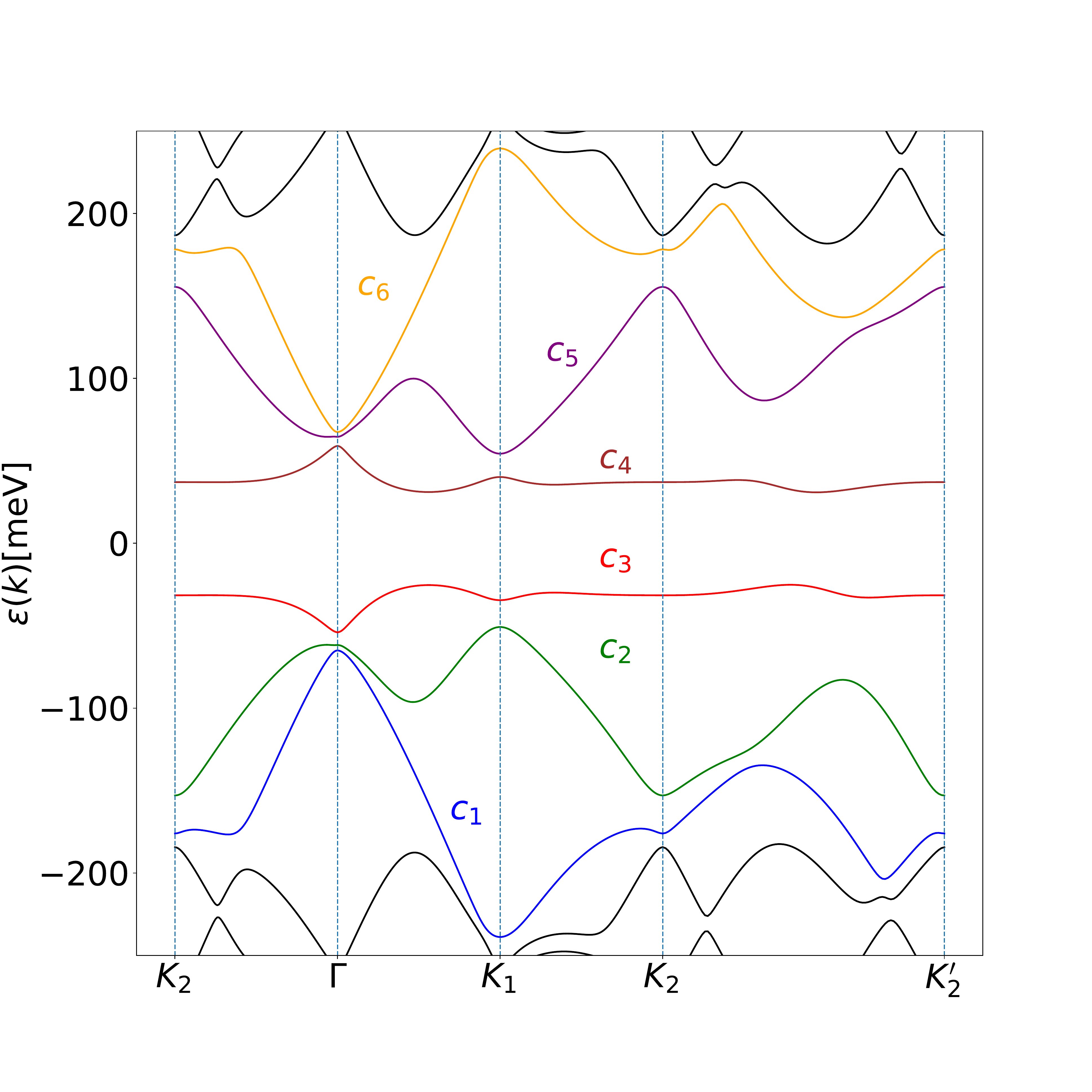}
		\caption{Band structure for the ABC MLT showing the six bands labeled $c_1\to c_6$. Parameters $Aa_0=0.26375$, $\theta=1.65^\circ$ and $\omega=2\gamma$ were chosen for the plot.}
		\label{fig:bandlabels}
	\end{center}
\end{figure}

Each color in the diagrams represents a list of Chern numbers for the six central bands $C=\{c_1,c_2,c_3,c_4,c_5,c_6\}$. Mathematically, if we have $N$ bands $\{\varepsilon_i(\vec{k})\}_{i=0}^{i=N}$, then the six bands are $\varepsilon_{N/2-3}(\vec{k})$, $\varepsilon_{N/2-2}(\vec{k})$, $\varepsilon_{N/2-1}(\vec{k})$, $\varepsilon_{N/2}(\vec{k})$, $\varepsilon_{N/2+1}(\vec{k})$, and $\varepsilon_{N/2+2}(\vec{k})$. As an illustration, we plotted the band structure for the ABC stacking case with middle layer twist in Fig. \ref{fig:bandlabels}, were we labeled the six central bands. We have tabulated the corresponding values for each color in Table \ref{tab:colors}.  
\begin{table}[htbp]
\centering
\begin{tabular*}{0.5\textwidth}{c@{\extracolsep{\fill}}ccccccc}
\hline
color&$c_{1}$&$c_{2}$&$c_3$&$c_4$&$c_{5}$&$c_{6}$\\
\hline
\multicolumn{7}{c}{${\rm AAA\,top\,layer\,twisted}$} \\
\hline
\textcolor{AAATTc1}{$\blacksquare$} &-5& 3& -1& 1& -3& 5\\
\textcolor{AAATTc2}{$\blacksquare$} &-5&3&-1&1&0&2\\
\textcolor{AAATTc3}{$\blacksquare$}&-4&1& 0& 0&-1&4\\
\textcolor{AAATTc4}{$\blacksquare$}&-4& 1& 0&0&2&1\\
\textcolor{AAATTc5}{$\blacksquare$}&-4&2&-1&1&0&2\\
\textcolor{AAATTc6}{$\blacksquare$}&-3&0&0&0&2&1\\
\textcolor{AAATTc7}{$\blacksquare$}&-2&0&-1&1&0&2\\
\textcolor{AAATTc8}{$\blacksquare$}&-2& 0&2&-2&0&2\\
\textcolor{AAATTc9}{$\blacksquare$}&-2&3&-1&1&-3&2\\
\textcolor{AAATTc10}{$\blacksquare$}&-2&3&-1&1&0&-1\\
\textcolor{AAATTc11}{$\blacksquare$}&-1&-2&0&0&2&1\\
\textcolor{AAATTc12}{$\blacksquare$}&-1&-1&-1&1&1&1\\
\textcolor{AAATTc13}{$\blacksquare$}&1&-3&-1&1&3&-1\\
\textcolor{AAATTc14}{$\blacksquare$}&1&3&-1&1&-3&-1\\
\textcolor{AAATTc15}{$\blacksquare$}&$\nu$&$\nu$&-1&1&0&2\\
\hline
\multicolumn{7}{c}{${\rm AAA\,middle\,layer\,twisted}$} \\
\hline
\textcolor{AAAMTc1}{$\blacksquare$}&v&v&0&0&v&v\\
\textcolor{AAAMTc2}{$\blacksquare$}&v&v&1&-1&v&v\\
\textcolor{AAAMTc3}{$\blacksquare$}&v&v&1&v&v&v\\
\textcolor{AAAMTc4}{$\blacksquare$}&v&v&v&v&v&v\\
\hline
\multicolumn{7}{c}{${\rm ABC\,top\,layer\,twisted (RH)}$} \\
\hline
\textcolor{ABCTTc1}{$\blacksquare$}&-3&0&-1&2&0&-3\\
\textcolor{ABCTTc2}{$\blacksquare$}&-3&0&1&0&0&-3\\
\textcolor{ABCTTc3}{$\blacksquare$}&-3&0&2&-1&0&-3\\
\textcolor{ABCTTc4}{$\blacksquare$}&-2&0&-1&2&0&-3\\
\textcolor{ABCTTc5}{$\blacksquare$}&-2&0&-1&2&1&-4\\
\textcolor{ABCTTc6}{$\blacksquare$}&0&0&-1&2&0&-3\\
\textcolor{ABCTTc7}{$\blacksquare$}&1&-1&0&2&-2&-1\\
\textcolor{ABCTTc8}{$\blacksquare$}&1&0&-1&2&1&-4\\
\hline
\multicolumn{7}{c}{${\rm ABC\,top\,layer\,twisted (LH)}$} \\
\hline
\textcolor{ABCTTflipc1}{$\blacksquare$}&-4&1&2&-1&0&-3\\
\textcolor{ABCTTflipc2}{$\blacksquare$}&-4&1&2&-1&0&-2\\
\textcolor{ABCTTflipc3}{$\blacksquare$}&-4&1&2&-1&0&1\\
\textcolor{ABCTTflipc4}{$\blacksquare$}&-3&0&2&-1&0&-3\\
\textcolor{ABCTTflipc5}{$\blacksquare$}&-3&0&2&-1&0&0\\
\textcolor{ABCTTflipc6}{$\blacksquare$}&-1&-2&2&-1&0&1\\
\textcolor{ABCTTflipc7}{$\blacksquare$}&-1&-2&2&0&-1&1\\
\hline
\multicolumn{7}{c}{${\rm ABC\,middle\,layer\,twisted}$} \\
\hline
\textcolor{ABCMTc1}{$\blacksquare$}&-1&0&1&-1&0&1\\
\textcolor{ABCMTc2}{$\blacksquare$}&-1&1&0&0&2&-2\\
\textcolor{ABCMTc3}{$\blacksquare$}&0&-1&1&-1&1&0\\
\textcolor{ABCMTc4}{$\blacksquare$}&0&-1&1&-1&1&$\nu$\\
\textcolor{ABCMTc5}{$\blacksquare$}&2&-3&1&-1&3&-2\\
\hline
\end{tabular*}
\caption{Color codes for the topological phase diagrams Figs. \ref{fig:phaseI}\&\ref{fig:phaseII}. Here, the term v represents a band closing that was confirmed up to numerical accuracy. The term $\nu$ corresponds to a Chern number that did not converge even when more than $10^4$ k points were used in the Chern number computation. RH: right-handed polarized light, and LH: left handed polarized light. }
\label{tab:colors}
\end{table}

We find that each of the different TTG realizations has its own unique topological structure. The bands in most cases - except for the AAA middle twist - are also found to be gaped. Quite generally, we find a rich structure of Chern numbers and for certain parameter pairs $(Aa_0,\theta)$ we find very large Chern number of 4 or 5 for some of the bands.

\begin{figure}[!htbp]
\centering
    \includegraphics[width=0.47\linewidth]{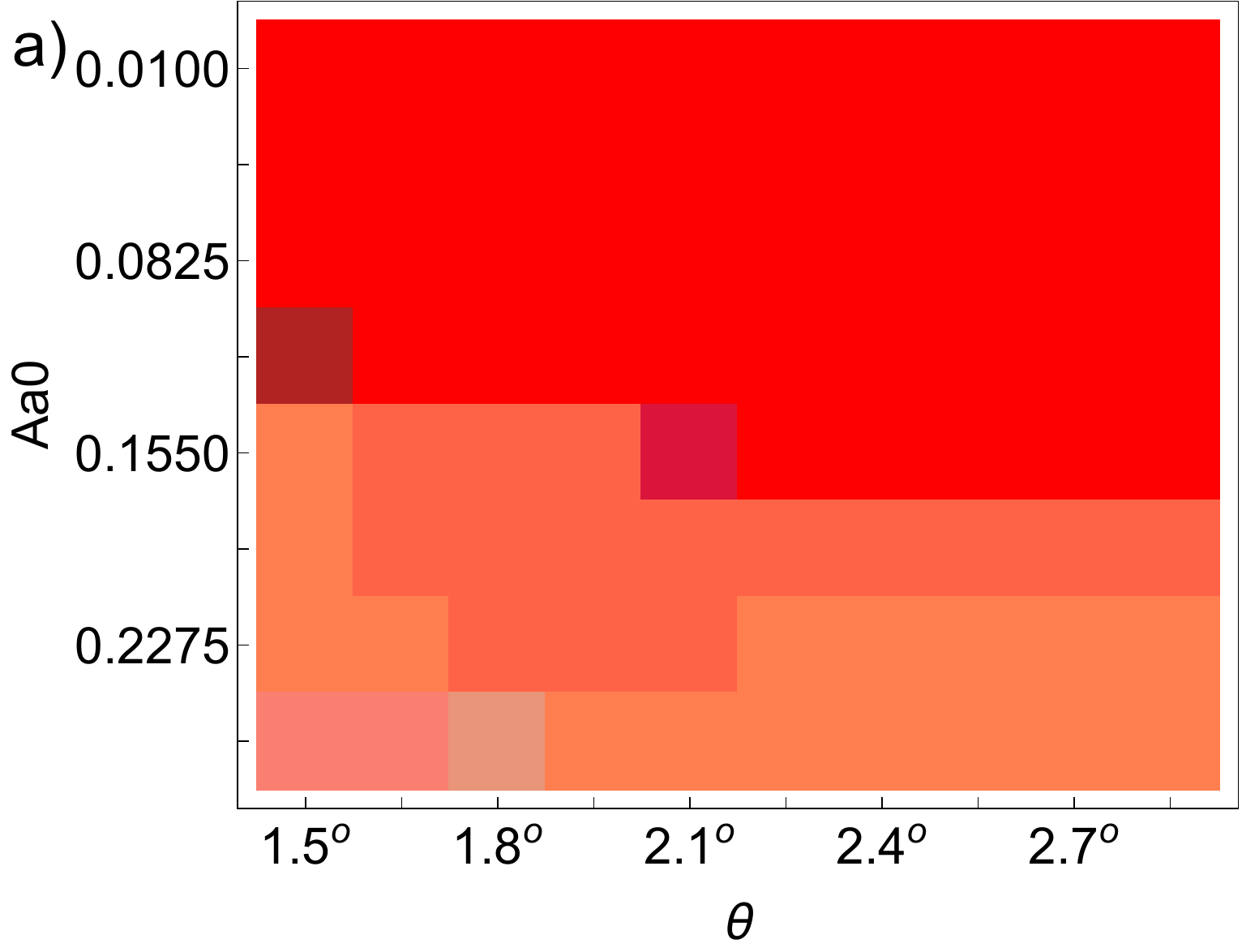}\hfill
    \includegraphics[width=0.47\linewidth]{ABCTop-Fin.pdf}
\caption{(Color online) The topological phase diagrams for the TTG system with ABC stacking and top layer twisted where (a) for the left-handed circularly polarized light and (b) the right-handed circularly polarized light, where we took $\omega=2\gamma$. For color codes, please refer to Table.~\ref{tab:colors}.}
\label{fig:phaseII}
\end{figure}

For all cases except the ABC top twist we find that the handedness of the incident circularly polarized light has no influence on the topological structure. In Fig.\ref{fig:phaseII} we see that for this case, however, there are large changes in the topological structure if we change from left handed (LH) to right handed (RH) circularly polarized light.

\subsection{Experimental proposal}

\begin{figure}[!htbp]
	\begin{center}
		\includegraphics[width=1\linewidth]{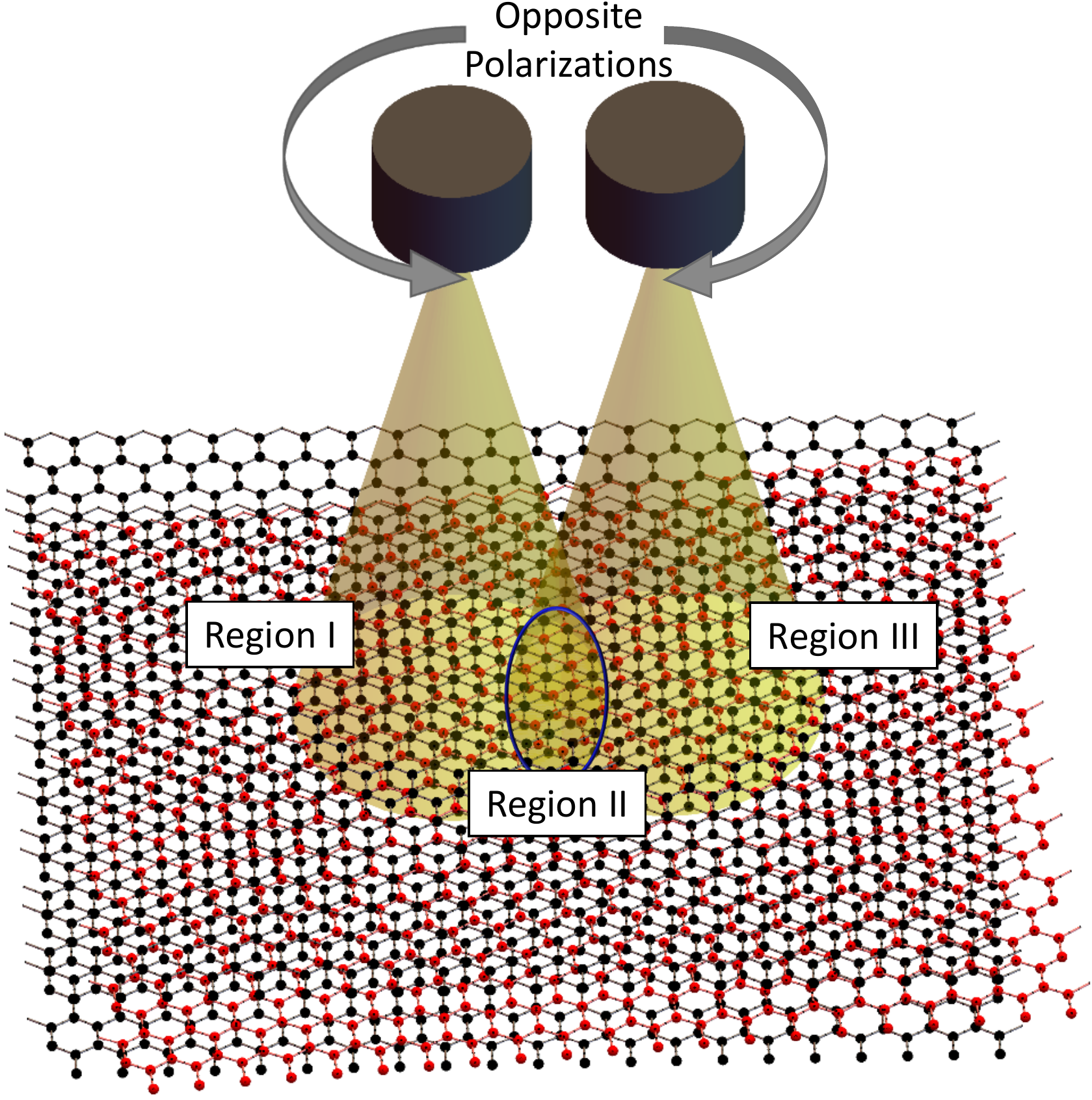}
		\caption{Sketch of a two laser procedure to create light-induced topological boundaries in ABC TTG samples.}
		\label{fig:exp_fig}
	\end{center}
\end{figure}

Motivated by the rich topological structure of the different types of TTGs (see Fig. \ref{fig:phaseI}), and the different responses of ABC to left and right-handed light we also propose the following experiment. For a large-enough ABC TTG sample, we shine light with opposite handedness next to each other, as depicted in Fig. \ref{fig:exp_fig}. By bringing the light beam edges very close to each other, we can expect to create three distinct topological regions, one for each laser pulse and one more at the intersection between the laser pulses. From the bulk-edge correspondence, we expect edge states at the boundary between the driven and undriven regions and at boundaries between the topological regions. These boundary states, indicated in blue in Fig. \ref{fig:exp_fig}, could be manifest in optical conductivity measurements \cite{Dehghani_2015,Kumar_2020}. This measurement would require employing a pump-probe experimental setup, where the probe amplitude $a_0 A_{\rm probe}$ is weak compared with the pump pulse $a_0 A$ employed to create the Floquet states.   

\section{Waveguide light}
\label{sec:waveguide}
\subsection{Numeric Results}
Next, we discuss the effect that light coming from a waveguide has on the bandstructure. This effect is included in the Hamiltonian via the time dependent maps of $w_{0,1}$ that were mentioned in Sec. \ref{sec:III}. We treat it numerically by solving equation \eqref{eq:Floquet_schroedinger} that was truncated to finite order. For all cases the resultant band structure reaches convergence when we include the first 3 Floquet copies ($n=-1,0,1)$.

\begin{figure}[!htbp]
\centering
    \includegraphics[width=0.47\linewidth]{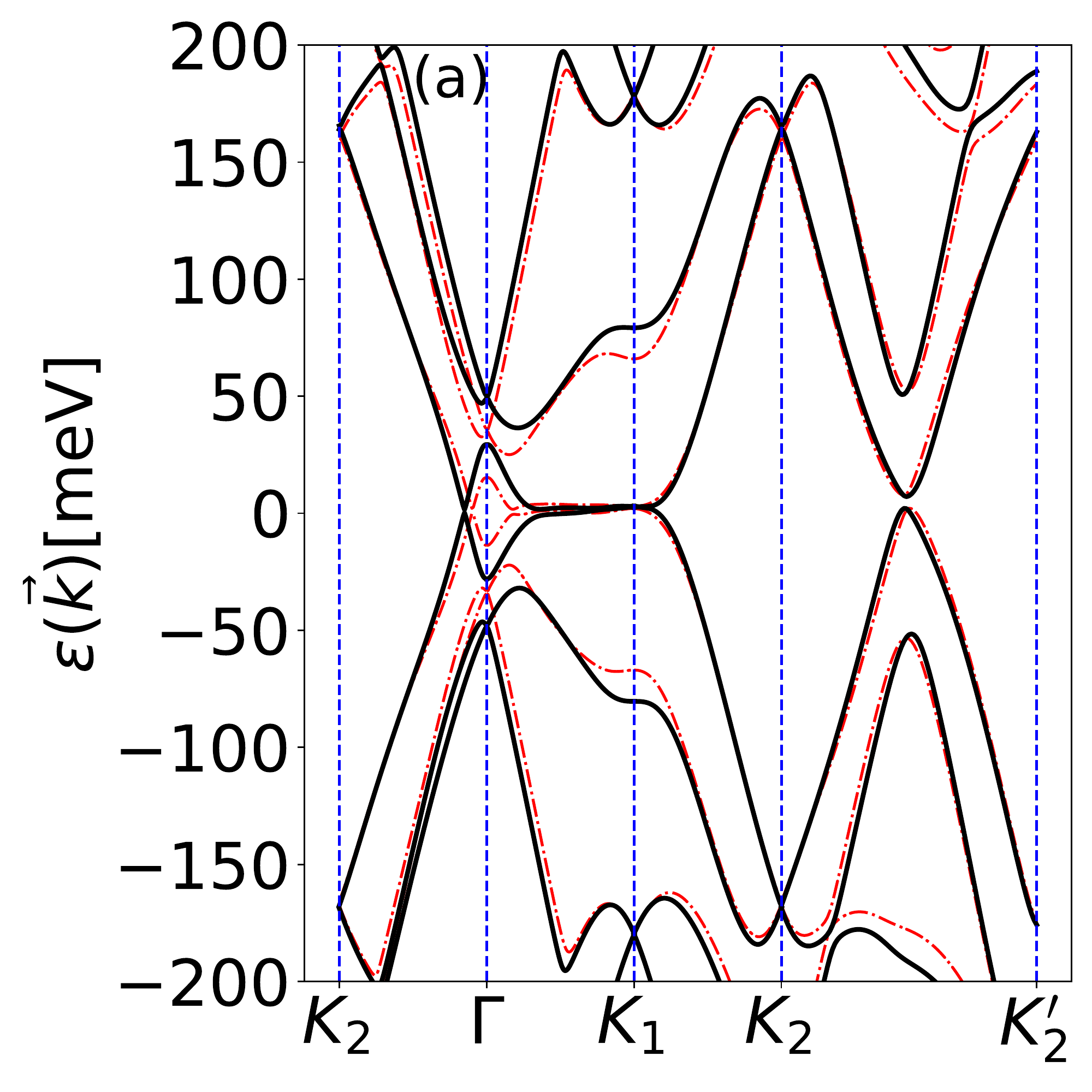}\hfil
    \includegraphics[width=0.53\linewidth]{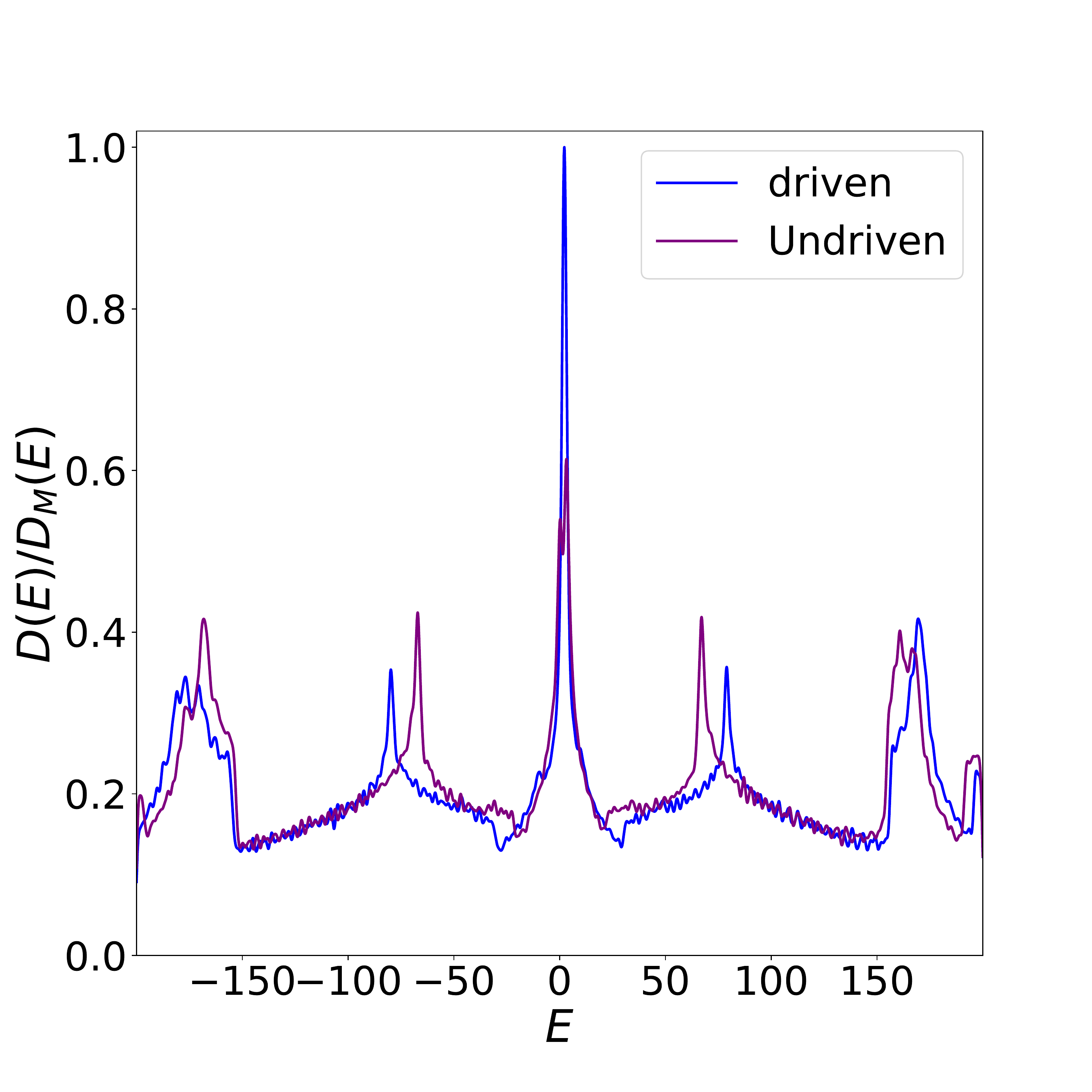}\par\medskip
    
    \includegraphics[width=0.47\linewidth]{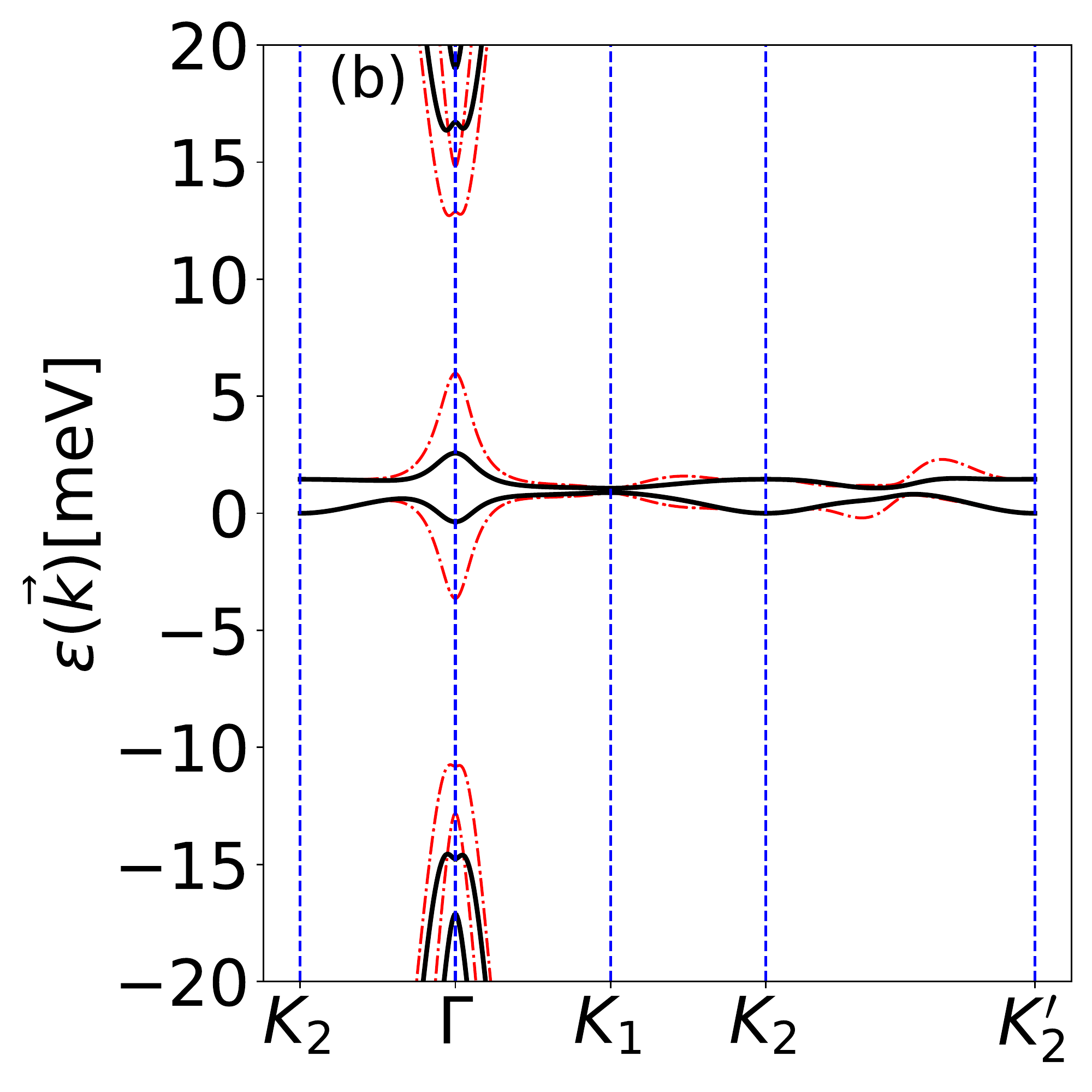}\hfil
    \includegraphics[width=0.53\linewidth]{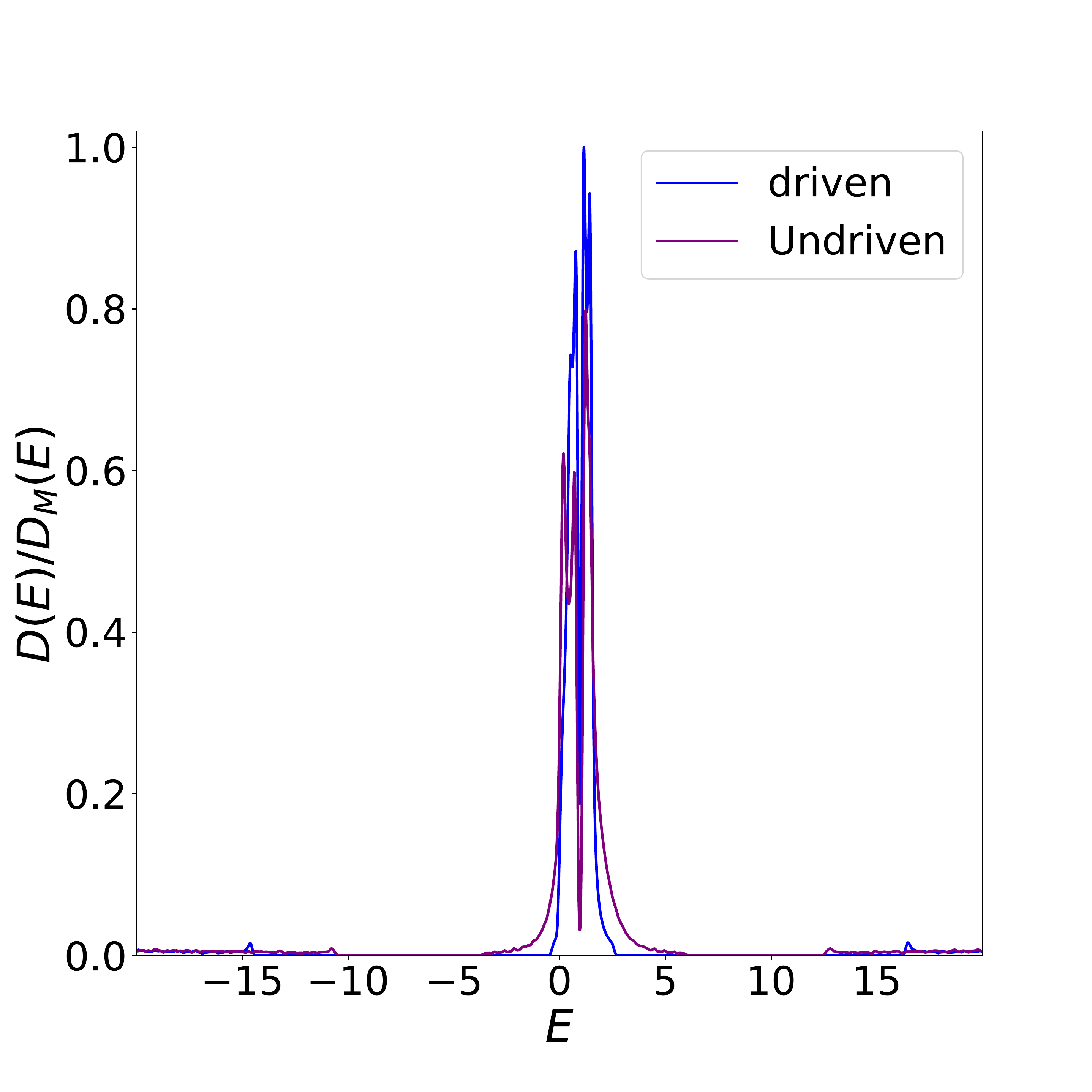}\par\medskip
    
    \includegraphics[width=0.47\linewidth]{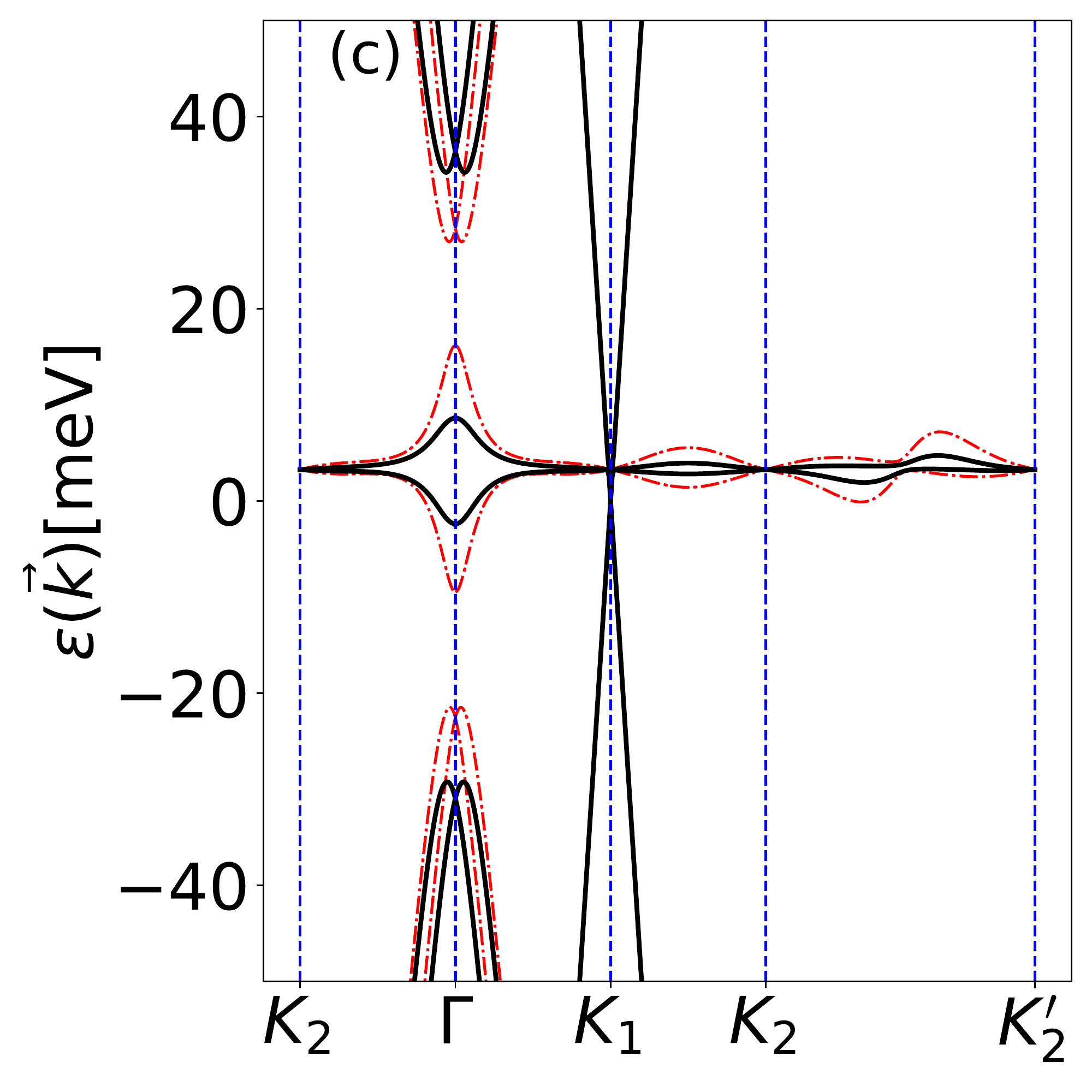}\hfil
    \includegraphics[width=0.53\linewidth]{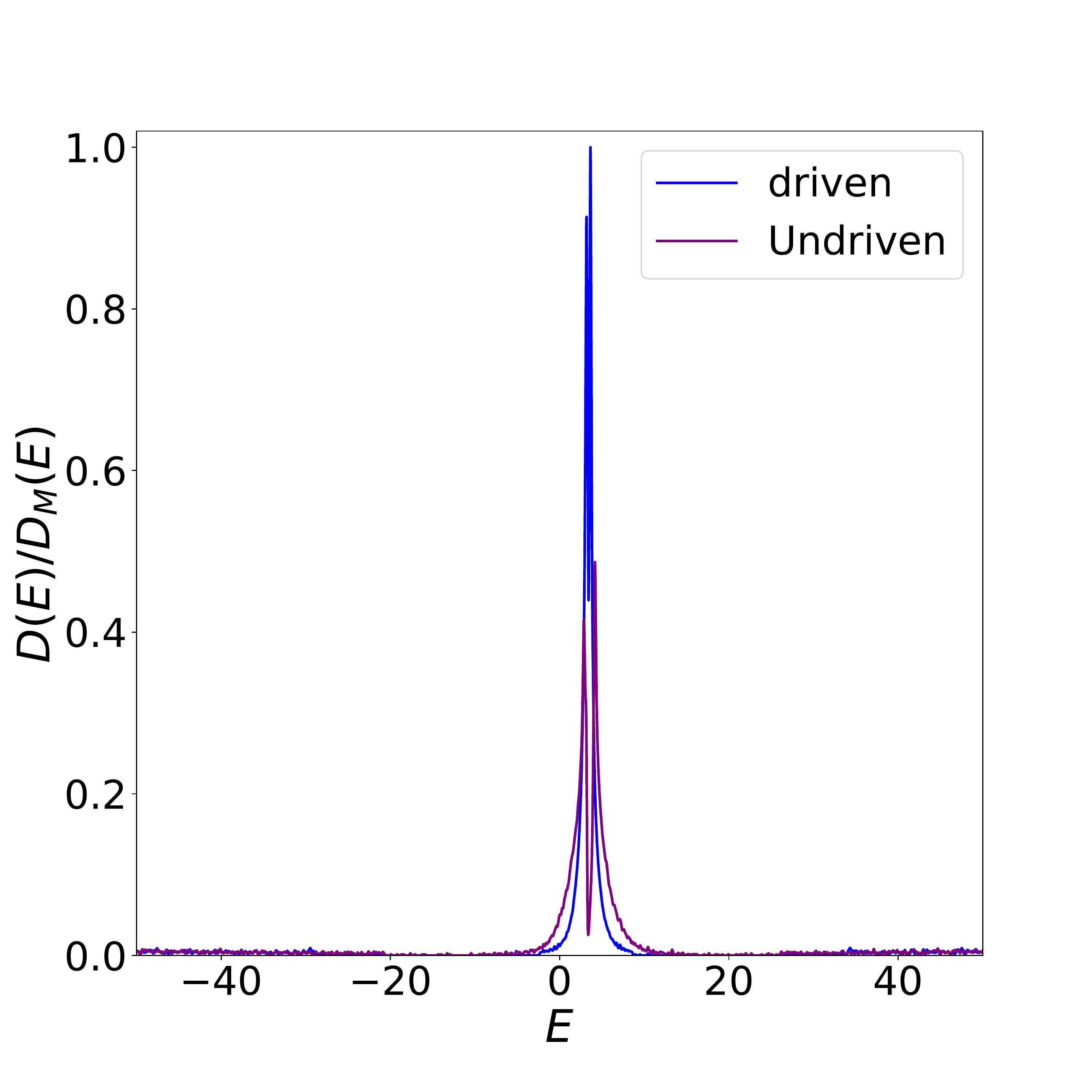}\par\medskip
    
    \includegraphics[width=0.47\linewidth]{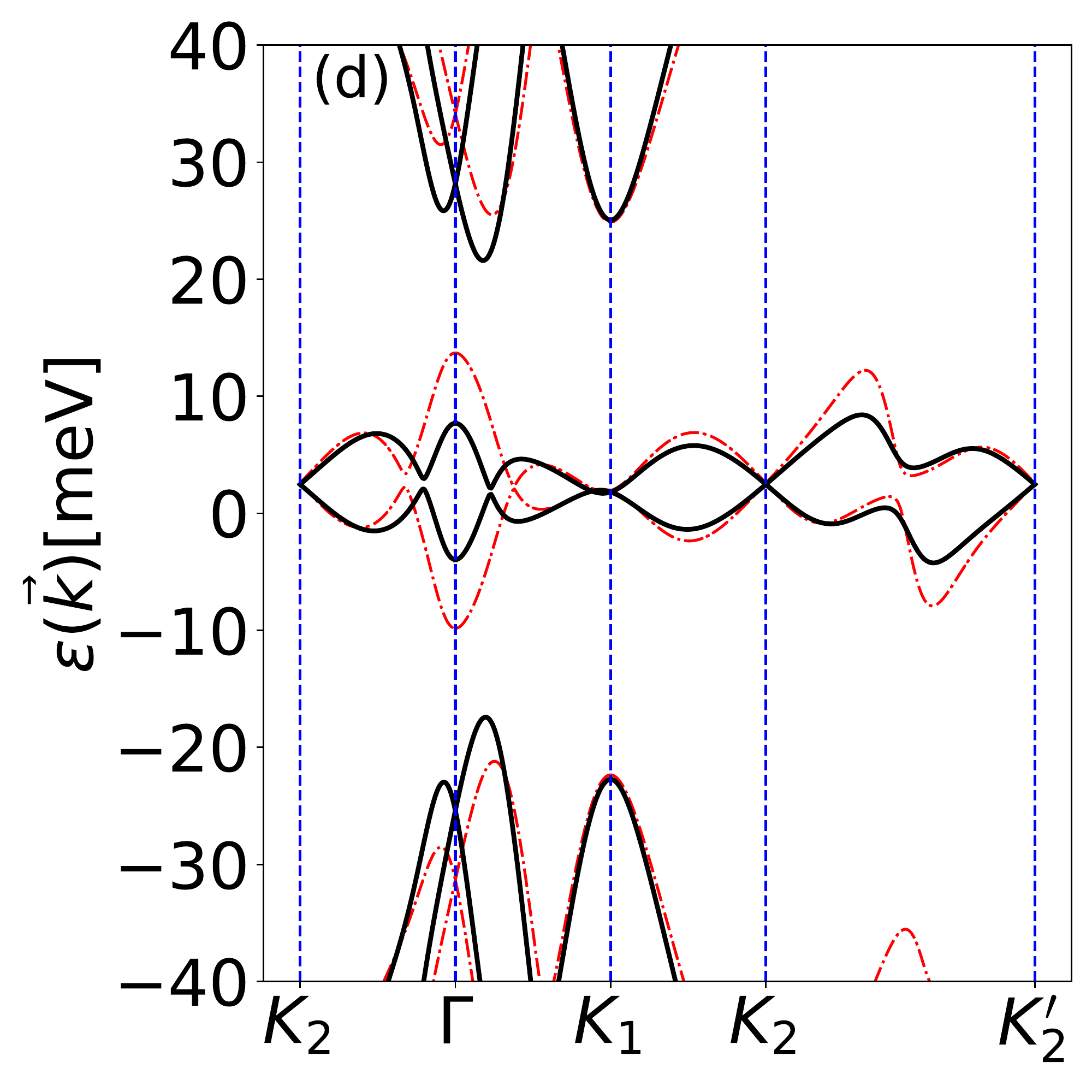}\hfil
    \includegraphics[width=0.53\linewidth]{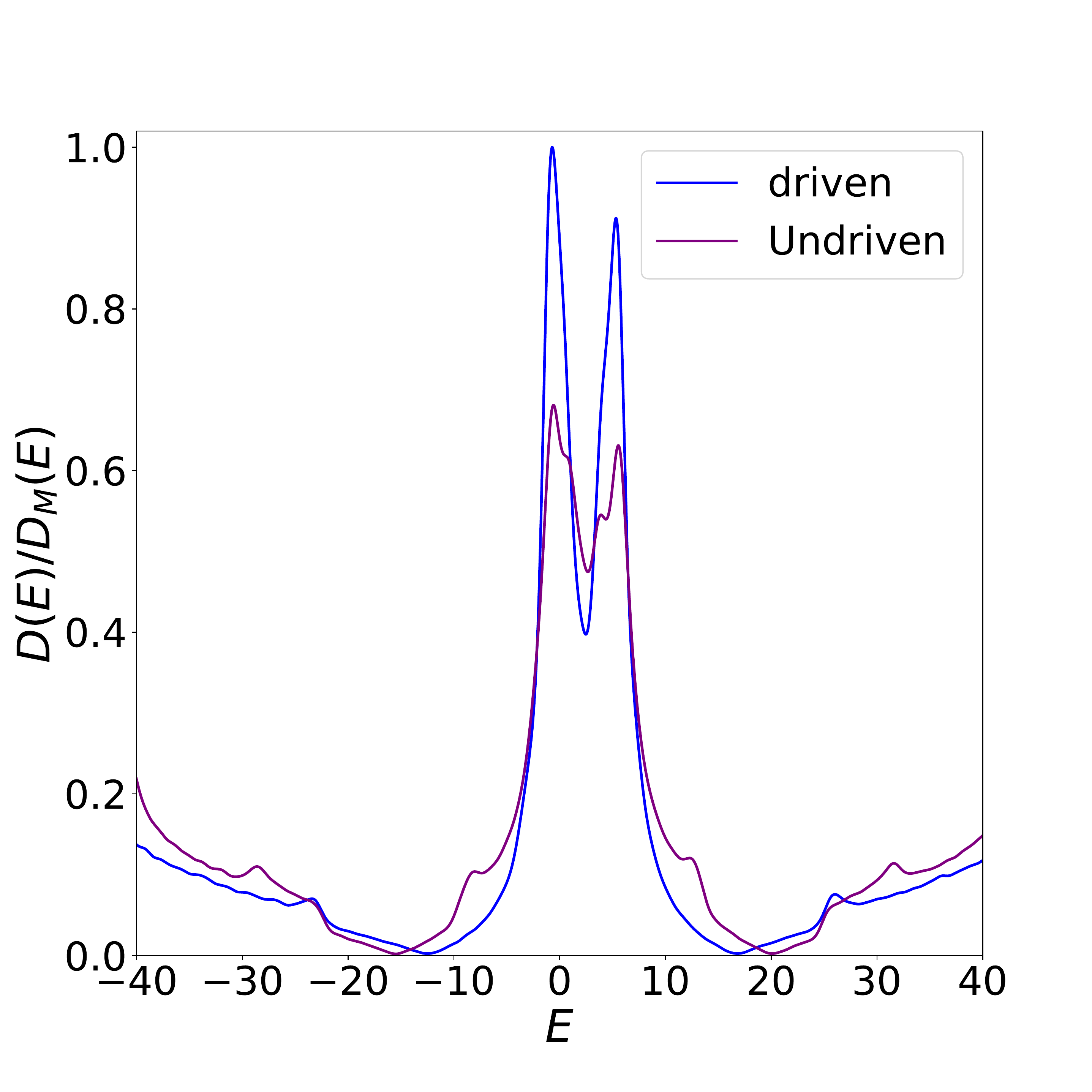}
\caption{(Color online) Left Column: Band structure of the TTG driven by waveguide light with  $w_0=0.8w_1$, $w_1=110$ meV and $Aa_0=0.3$. (a) The AAA stacking with TLT and $(\theta,\omega)=(2.000^\circ,2\gamma)$, (b) the ABC configuration with TLT and $(\theta,\omega)=(1.040^\circ,2\gamma)$, (c) AAA with MLT and parameters $(\theta,\omega)=(1.450^\circ,3\gamma)$, and (d) the ABC stacking with MLT and $(\theta,\omega)=(1.157^\circ,2\gamma)$. The solid lines are for the driven and the dash-doted lines for the undriven case. Right Column: The density of states plots associated with the configurations on the left column rescaled by the maximum value $D_M(E)$ of the driven case.}
\label{fig:WGBands}
\end{figure}

In Fig.\ref{fig:WGBands} we plotted the band structure for four different configurations (AAA/ABC stacking, top and middle layer twists). For convenience we also included the undriven case for comparison.

When comparing the ABC case (top and middle layers twists) and the AAA stacking (middle layer twisted) for the undriven (dashed lines) against the driven case (solid lines), we find that we can flatten the central bands without introducing a band opening as in the case of circularly polarized light. This means that we can tune band flatness, which can be convenient when trying to realize strongly correlated phases such as superconductivity.

\begin{figure}[h]
	\begin{center}
	\subfloat[AAA (MLT)]{\includegraphics[width = 0.5\linewidth]{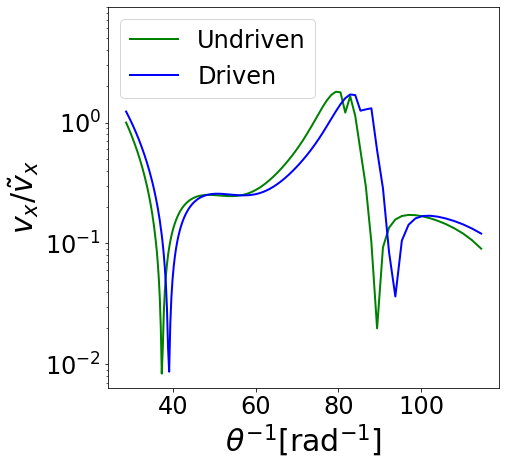}}
	\subfloat[ABC (MLT)]{\includegraphics[width = 0.5\linewidth]{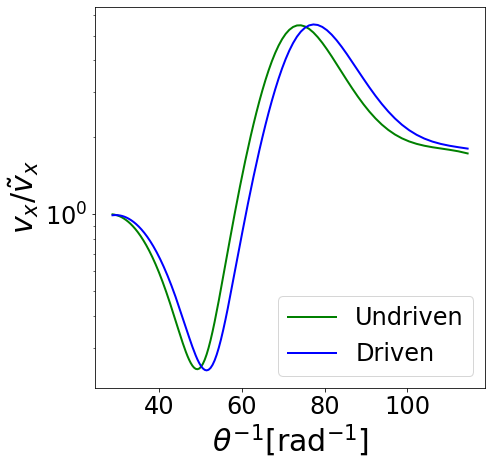}}
	\caption{Plot of velocity as function of $1/\theta$ for the center bands near K1 in the twisted TTG system driven by waveguide light compared with the undriven case  $w_0=0.8w_1$, $w_1=110$ meV, $Aa_0=0.3$, $\Omega=3\gamma$ and  $\gamma=2.36$ eV. (a) Starting from AAA stacking middle layer twist (b) starting from ABC stacking middle layer twist. Here, $\rm \tilde{v}_{\rm x}$ is the velocity at $\theta=2.0^\circ$.}
	\label{fig:velocity}
	\end{center}
\end{figure}

To understand this effect even better, we have plotted the velocity of electrons near the $K_1$ symmetry point for the AAA and ABC stacking with middle layer twisted versus $\theta^{-1}$ in Fig. \ref{fig:velocity}.

We find that the introduction of light from a waveguide can shift the magic angles. This gives us the opportunity to speculate a bit about possible applications. For instance, this observation could be useful in an experiment where one wants to realize strongly correlated phases. This is because when one produces a twisted trilayer graphene sample for use in experiments with strongly correlated phases one has to try to match the magic angle as precisely as possible. If there is a small deviation from the angle with flat bands the setup with waveguide light could be used to correct for these deviations. Alternatively it could even be possible to use light of this sort to switch between strongly correlated phases and other phases.

\subsection{Effective Hamiltonian}
It is beneficial to better understand the effects that light from a waveguide has on the band structure from an analytical perspective. For this purpose, we consider an appropriate effective time independent Hamiltonian. The vV Hamiltonian to first order is given by \cite{MoireFloquetRev}
\begin{equation}
    H_{\rm eff}=H_0+\sum_{m\neq 0}\frac{H_mH_{-m}}{m\omega},
\end{equation}
where $H_m$ is defined by $H_n=1/T\int_0^T e^{-in\omega t} H(t)$. We find that in our case the term $\sum_{m\neq 0}\frac{H_mH_{-m}}{m\omega}=0$ vanishes and therefore the Floquet Hamiltonian is given by $H_F=H_0+\mathcal{O}(\omega^{-2})$.

That is in the Hamiltonian one simply has to replace interlayer couplings by
\begin{equation}
    w_0\to w_0J_0(Aa_{AA});\quad w_1\to w_1J_0(Aa_{AB}),
\end{equation}
The interlayer hoppings are weakened by Bessel functions as it was also found in \cite{Vogl_2020_interlayer}. In twisted bilayer graphene it was found that the value of the magic angles was interlayer hopping dependent \cite{Tarnopolsky_2019}. If we assume the same is true for twisted trilayer graphene then this explains the shift of magic angles that we observed in Fig.\ref{fig:velocity}.

\begin{figure}[h]
	\begin{center}
		\includegraphics[width=1\linewidth]{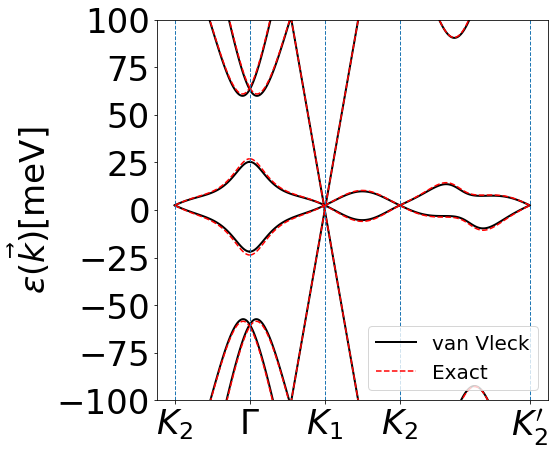}
		\caption{Comparison of the band structure for TTG with ABA stacking and middle layer twisted driven by waveguide light  $w_0=0.8w_1$, $w_1=110$ meV and $\theta=1.3^\circ$, $Aa_0=0.8$, $\Omega=1.5\gamma$ and  $\gamma=2.364$ eV. obtained via the quasi-energy operator Eq.~\ref{eq:Floquet_schroedinger} (dashed lines) versus the zeroth order van Vleck Hamiltonian $H_0$ (solid lines).}
		\label{fig:gapabab}
	\end{center}
\end{figure}

Lastly, to see quantitatively how good this approximation is we have plotted both the  exact quasi-energy spectrum as well as the one from this simple approximation. The result is shown in Fig. \ref{fig:gapabab}.

We find that the result even for a very large driving strength $Aa_0=0.8$ and relatively low frequencies $\Omega=1.5\gamma$ is almost perfect. For weaker driving strengths and higher frequencies it was even hard to see any discernible difference between the approximate  and exact quasi-energies.

\section{Conclusions}
\label{sec:conclusion}
In this work, we have studied four different stacking configurations of twisted trilayer graphene (TTG). First we reviewed the equilibrium properties and then went on to study various non-equilibrium scenarios. In the presence of a circularly polarized light, we found that we were able to flatten the two middle bands compared to the equilibrium case which could open the door to experiments with strongly correlated non-equlibrium phases. Even more exciting, is the fact that this type of light causes gap openings, which then allowed us to study the topological properties of the TTG configurations that are characterized by single-band Chern numbers. Here, we focused on the six middle bands and generated topological phase diagrams for a range of different values of the driving strength and the twist angle. These diagrams revealed that TTG has a rich topological structure. Moreover, we found that the topological structure of ABC stacked TTG with a top-layer twist is sensitive to the handedness of circularly polarized light. This is in stark contrast with the other three configurations that we have studied. 

The fact that top twisted ABC stacked bilayer graphene has different topological phase diagrams depending on the the polarization of circularly polarized light led us to propose an experiment where this difference in typologies could be captured via optical conductivity measurements. Here, one takes two sources of circularly polarized light with different handedness and let them shine on TTG such that the two illuminated regions intersect. This creates three distinct regions of different topology, and thus edge states are expected on the boundaries between these regions and their boundary with the undriven regions. The different topological properties are expected to be measurable in optical conductivity measurements. 

In addition, to circularly polarized light we also studied the effects that longitudinally polarized light, coming from a  waveguide, has on the band structure. We found that the presence of this light source can make the central bands less dispersive without the introduction of band gap openings like in the case of circularly polarized light.  This means that effectively we were able to shift the magic angle where flat-bands appear without introducing additional side-effects that complicate the Hamiltonian description. This  observation might be useful for the realization of strongly correlated phases in TTG that closely resemble the equilibrium case at a different twist angle.

\section{Acknowledgments} M. R-V. was supported by LANL LDRD Program and
by the U.S. Department of Energy, Office of Science, Basic Energy Sciences, Materials Sciences and Engineering Division, Condensed Matter Theory Program. M.V. and H.B. acknowledge the support of King Fahd University of Petroleum and Minerals.

\bibliography{literature}

\begin{thebibliography}{90}%
\makeatletter
\providecommand \@ifxundefined [1]{%
 \@ifx{#1\undefined}
}%
\providecommand \@ifnum [1]{%
 \ifnum #1\expandafter \@firstoftwo
 \else \expandafter \@secondoftwo
 \fi
}%
\providecommand \@ifx [1]{%
 \ifx #1\expandafter \@firstoftwo
 \else \expandafter \@secondoftwo
 \fi
}%
\providecommand \natexlab [1]{#1}%
\providecommand \enquote  [1]{``#1''}%
\providecommand \bibnamefont  [1]{#1}%
\providecommand \bibfnamefont [1]{#1}%
\providecommand \citenamefont [1]{#1}%
\providecommand \href@noop [0]{\@secondoftwo}%
\providecommand \href [0]{\begingroup \@sanitize@url \@href}%
\providecommand \@href[1]{\@@startlink{#1}\@@href}%
\providecommand \@@href[1]{\endgroup#1\@@endlink}%
\providecommand \@sanitize@url [0]{\catcode `\\12\catcode `\$12\catcode
  `\&12\catcode `\#12\catcode `\^12\catcode `\_12\catcode `\%12\relax}%
\providecommand \@@startlink[1]{}%
\providecommand \@@endlink[0]{}%
\providecommand \url  [0]{\begingroup\@sanitize@url \@url }%
\providecommand \@url [1]{\endgroup\@href {#1}{\urlprefix }}%
\providecommand \urlprefix  [0]{URL }%
\providecommand \Eprint [0]{\href }%
\providecommand \doibase [0]{https://doi.org/}%
\providecommand \selectlanguage [0]{\@gobble}%
\providecommand \bibinfo  [0]{\@secondoftwo}%
\providecommand \bibfield  [0]{\@secondoftwo}%
\providecommand \translation [1]{[#1]}%
\providecommand \BibitemOpen [0]{}%
\providecommand \bibitemStop [0]{}%
\providecommand \bibitemNoStop [0]{.\EOS\space}%
\providecommand \EOS [0]{\spacefactor3000\relax}%
\providecommand \BibitemShut  [1]{\csname bibitem#1\endcsname}%
\let\auto@bib@innerbib\@empty
\bibitem [{\citenamefont {Novoselov}\ \emph {et~al.}(2004)\citenamefont
  {Novoselov}, \citenamefont {Geim}, \citenamefont {Morozov}, \citenamefont
  {Jiang}, \citenamefont {Zhang}, \citenamefont {Dubonos}, \citenamefont
  {Grigorieva},\ and\ \citenamefont {Firsov}}]{Novoselov666}%
  \BibitemOpen
  \bibfield  {author} {\bibinfo {author} {\bibfnamefont {K.~S.}\ \bibnamefont
  {Novoselov}}, \bibinfo {author} {\bibfnamefont {A.~K.}\ \bibnamefont {Geim}},
  \bibinfo {author} {\bibfnamefont {S.~V.}\ \bibnamefont {Morozov}}, \bibinfo
  {author} {\bibfnamefont {D.}~\bibnamefont {Jiang}}, \bibinfo {author}
  {\bibfnamefont {Y.}~\bibnamefont {Zhang}}, \bibinfo {author} {\bibfnamefont
  {S.~V.}\ \bibnamefont {Dubonos}}, \bibinfo {author} {\bibfnamefont {I.~V.}\
  \bibnamefont {Grigorieva}},\ and\ \bibinfo {author} {\bibfnamefont {A.~A.}\
  \bibnamefont {Firsov}},\ }\bibfield  {title} {\bibinfo {title} {Electric
  field effect in atomically thin carbon films},\ }\href
  {https://doi.org/10.1126/science.1102896} {\bibfield  {journal} {\bibinfo
  {journal} {Science}\ }\textbf {\bibinfo {volume} {306}},\ \bibinfo {pages}
  {666} (\bibinfo {year} {2004})},\ \Eprint
  {https://arxiv.org/abs/https://science.sciencemag.org/content/306/5696/666.full.pdf}
  {https://science.sciencemag.org/content/306/5696/666.full.pdf} \BibitemShut
  {NoStop}%
\bibitem [{\citenamefont {Castro~Neto}\ \emph {et~al.}(2009)\citenamefont
  {Castro~Neto}, \citenamefont {Guinea}, \citenamefont {Peres}, \citenamefont
  {Novoselov},\ and\ \citenamefont {Geim}}]{RevModPhys.81.109}%
  \BibitemOpen
  \bibfield  {author} {\bibinfo {author} {\bibfnamefont {A.~H.}\ \bibnamefont
  {Castro~Neto}}, \bibinfo {author} {\bibfnamefont {F.}~\bibnamefont {Guinea}},
  \bibinfo {author} {\bibfnamefont {N.~M.~R.}\ \bibnamefont {Peres}}, \bibinfo
  {author} {\bibfnamefont {K.~S.}\ \bibnamefont {Novoselov}},\ and\ \bibinfo
  {author} {\bibfnamefont {A.~K.}\ \bibnamefont {Geim}},\ }\bibfield  {title}
  {\bibinfo {title} {The electronic properties of graphene},\ }\href
  {https://doi.org/10.1103/RevModPhys.81.109} {\bibfield  {journal} {\bibinfo
  {journal} {Rev. Mod. Phys.}\ }\textbf {\bibinfo {volume} {81}},\ \bibinfo
  {pages} {109} (\bibinfo {year} {2009})}\BibitemShut {NoStop}%
\bibitem [{\citenamefont {Klein}(1929)}]{Klein1929}%
  \BibitemOpen
  \bibfield  {author} {\bibinfo {author} {\bibfnamefont {O.}~\bibnamefont
  {Klein}},\ }\bibfield  {title} {\bibinfo {title} {Die reflexion von
  elektronen an einem potentialsprung nach der relativistischen dynamik von
  dirac},\ }\href {https://doi.org/10.1007/BF01339716} {\bibfield  {journal}
  {\bibinfo  {journal} {Zeitschrift f{\"u}r Physik}\ }\textbf {\bibinfo
  {volume} {53}},\ \bibinfo {pages} {157} (\bibinfo {year} {1929})}\BibitemShut
  {NoStop}%
\bibitem [{\citenamefont {Katsnelson}\ \emph {et~al.}(2006)\citenamefont
  {Katsnelson}, \citenamefont {Novoselov},\ and\ \citenamefont
  {Geim}}]{Katsnelson_2006}%
  \BibitemOpen
  \bibfield  {author} {\bibinfo {author} {\bibfnamefont {M.~I.}\ \bibnamefont
  {Katsnelson}}, \bibinfo {author} {\bibfnamefont {K.~S.}\ \bibnamefont
  {Novoselov}},\ and\ \bibinfo {author} {\bibfnamefont {A.~K.}\ \bibnamefont
  {Geim}},\ }\bibfield  {title} {\bibinfo {title} {Chiral tunnelling and the
  klein paradox in graphene},\ }\href {https://doi.org/10.1038/nphys384}
  {\bibfield  {journal} {\bibinfo  {journal} {Nature Physics}\ }\textbf
  {\bibinfo {volume} {2}},\ \bibinfo {pages} {620–625} (\bibinfo {year}
  {2006})}\BibitemShut {NoStop}%
\bibitem [{\citenamefont {Zhang}\ \emph {et~al.}(2009)\citenamefont {Zhang},
  \citenamefont {Tang}, \citenamefont {Girit}, \citenamefont {Hao},
  \citenamefont {Martin}, \citenamefont {Zettl}, \citenamefont {Crommie},
  \citenamefont {Shen},\ and\ \citenamefont {Wang}}]{Zhang2009}%
  \BibitemOpen
  \bibfield  {author} {\bibinfo {author} {\bibfnamefont {Y.}~\bibnamefont
  {Zhang}}, \bibinfo {author} {\bibfnamefont {T.-T.}\ \bibnamefont {Tang}},
  \bibinfo {author} {\bibfnamefont {C.}~\bibnamefont {Girit}}, \bibinfo
  {author} {\bibfnamefont {Z.}~\bibnamefont {Hao}}, \bibinfo {author}
  {\bibfnamefont {M.~C.}\ \bibnamefont {Martin}}, \bibinfo {author}
  {\bibfnamefont {A.}~\bibnamefont {Zettl}}, \bibinfo {author} {\bibfnamefont
  {M.~F.}\ \bibnamefont {Crommie}}, \bibinfo {author} {\bibfnamefont {Y.~R.}\
  \bibnamefont {Shen}},\ and\ \bibinfo {author} {\bibfnamefont
  {F.}~\bibnamefont {Wang}},\ }\bibfield  {title} {\bibinfo {title} {Direct
  observation of a widely tunable bandgap in bilayer graphene},\ }\href
  {https://doi.org/10.1038/nature08105} {\bibfield  {journal} {\bibinfo
  {journal} {Nature}\ }\textbf {\bibinfo {volume} {459}},\ \bibinfo {pages}
  {820} (\bibinfo {year} {2009})}\BibitemShut {NoStop}%
\bibitem [{\citenamefont {Zuo}\ \emph {et~al.}(2018)\citenamefont {Zuo},
  \citenamefont {Qiao}, \citenamefont {Ma}, \citenamefont {Yin}, \citenamefont
  {Sun}, \citenamefont {Zhang}, \citenamefont {Guan},\ and\ \citenamefont
  {He}}]{STMTTG}%
  \BibitemOpen
  \bibfield  {author} {\bibinfo {author} {\bibfnamefont {W.-J.}\ \bibnamefont
  {Zuo}}, \bibinfo {author} {\bibfnamefont {J.-B.}\ \bibnamefont {Qiao}},
  \bibinfo {author} {\bibfnamefont {D.-L.}\ \bibnamefont {Ma}}, \bibinfo
  {author} {\bibfnamefont {L.-J.}\ \bibnamefont {Yin}}, \bibinfo {author}
  {\bibfnamefont {G.}~\bibnamefont {Sun}}, \bibinfo {author} {\bibfnamefont
  {J.-Y.}\ \bibnamefont {Zhang}}, \bibinfo {author} {\bibfnamefont {L.-Y.}\
  \bibnamefont {Guan}},\ and\ \bibinfo {author} {\bibfnamefont
  {L.}~\bibnamefont {He}},\ }\bibfield  {title} {\bibinfo {title} {Scanning
  tunneling microscopy and spectroscopy of twisted trilayer graphene},\
  }\href@noop {} {\bibfield  {journal} {\bibinfo  {journal} {Phys. Rev. B}\
  }\textbf {\bibinfo {volume} {97}},\ \bibinfo {pages} {035440} (\bibinfo
  {year} {2018})}\BibitemShut {NoStop}%
\bibitem [{\citenamefont {Pong}\ and\ \citenamefont
  {Durkan}(2005)}]{Pong_2005}%
  \BibitemOpen
  \bibfield  {author} {\bibinfo {author} {\bibfnamefont {W.-T.}\ \bibnamefont
  {Pong}}\ and\ \bibinfo {author} {\bibfnamefont {C.}~\bibnamefont {Durkan}},\
  }\bibfield  {title} {\bibinfo {title} {A review and outlook for an anomaly of
  scanning tunnelling microscopy ({STM}): superlattices on graphite},\ }\href
  {https://doi.org/10.1088/0022-3727/38/21/r01} {\bibfield  {journal} {\bibinfo
   {journal} {Journal of Physics D: Applied Physics}\ }\textbf {\bibinfo
  {volume} {38}},\ \bibinfo {pages} {R329} (\bibinfo {year}
  {2005})}\BibitemShut {NoStop}%
\bibitem [{\citenamefont {Bistritzer}\ and\ \citenamefont
  {MacDonald}(2011)}]{Bistritzer12233}%
  \BibitemOpen
  \bibfield  {author} {\bibinfo {author} {\bibfnamefont {R.}~\bibnamefont
  {Bistritzer}}\ and\ \bibinfo {author} {\bibfnamefont {A.~H.}\ \bibnamefont
  {MacDonald}},\ }\bibfield  {title} {\bibinfo {title} {Moir{\'e} bands in
  twisted double-layer graphene},\ }\href
  {https://doi.org/10.1073/pnas.1108174108} {\bibfield  {journal} {\bibinfo
  {journal} {Proceedings of the National Academy of Sciences}\ }\textbf
  {\bibinfo {volume} {108}},\ \bibinfo {pages} {12233} (\bibinfo {year}
  {2011})},\ \Eprint
  {https://arxiv.org/abs/https://www.pnas.org/content/108/30/12233.full.pdf}
  {https://www.pnas.org/content/108/30/12233.full.pdf} \BibitemShut {NoStop}%
\bibitem [{\citenamefont {Eisenstein}\ and\ \citenamefont
  {MacDonald}(2004)}]{Eisenstein2004}%
  \BibitemOpen
  \bibfield  {author} {\bibinfo {author} {\bibfnamefont {J.~P.}\ \bibnamefont
  {Eisenstein}}\ and\ \bibinfo {author} {\bibfnamefont {A.~H.}\ \bibnamefont
  {MacDonald}},\ }\bibfield  {title} {\bibinfo {title} {Bose--einstein
  condensation of excitons in bilayer electron systems},\ }\href
  {https://doi.org/10.1038/nature03081} {\bibfield  {journal} {\bibinfo
  {journal} {Nature}\ }\textbf {\bibinfo {volume} {432}},\ \bibinfo {pages}
  {691} (\bibinfo {year} {2004})}\BibitemShut {NoStop}%
\bibitem [{\citenamefont {Min}\ \emph {et~al.}(2008)\citenamefont {Min},
  \citenamefont {Bistritzer}, \citenamefont {Su},\ and\ \citenamefont
  {MacDonald}}]{Min_2008}%
  \BibitemOpen
  \bibfield  {author} {\bibinfo {author} {\bibfnamefont {H.}~\bibnamefont
  {Min}}, \bibinfo {author} {\bibfnamefont {R.}~\bibnamefont {Bistritzer}},
  \bibinfo {author} {\bibfnamefont {J.-J.}\ \bibnamefont {Su}},\ and\ \bibinfo
  {author} {\bibfnamefont {A.~H.}\ \bibnamefont {MacDonald}},\ }\bibfield
  {title} {\bibinfo {title} {Room-temperature superfluidity in graphene
  bilayers},\ }\bibfield  {journal} {\bibinfo  {journal} {Physical Review B}\
  }\textbf {\bibinfo {volume} {78}},\ \href
  {https://doi.org/10.1103/physrevb.78.121401} {10.1103/physrevb.78.121401}
  (\bibinfo {year} {2008})\BibitemShut {NoStop}%
\bibitem [{\citenamefont {Arita}\ \emph {et~al.}(2002)\citenamefont {Arita},
  \citenamefont {Suwa}, \citenamefont {Kuroki},\ and\ \citenamefont
  {Aoki}}]{PhysRevLett.88.127202}%
  \BibitemOpen
  \bibfield  {author} {\bibinfo {author} {\bibfnamefont {R.}~\bibnamefont
  {Arita}}, \bibinfo {author} {\bibfnamefont {Y.}~\bibnamefont {Suwa}},
  \bibinfo {author} {\bibfnamefont {K.}~\bibnamefont {Kuroki}},\ and\ \bibinfo
  {author} {\bibfnamefont {H.}~\bibnamefont {Aoki}},\ }\bibfield  {title}
  {\bibinfo {title} {Gate-induced band ferromagnetism in an organic polymer},\
  }\href {https://doi.org/10.1103/PhysRevLett.88.127202} {\bibfield  {journal}
  {\bibinfo  {journal} {Phys. Rev. Lett.}\ }\textbf {\bibinfo {volume} {88}},\
  \bibinfo {pages} {127202} (\bibinfo {year} {2002})}\BibitemShut {NoStop}%
\bibitem [{\citenamefont {Cao}\ \emph {et~al.}(2018{\natexlab{a}})\citenamefont
  {Cao}, \citenamefont {Fatemi}, \citenamefont {Fang}, \citenamefont
  {Watanabe}, \citenamefont {Taniguchi}, \citenamefont {Kaxiras},\ and\
  \citenamefont {Jarillo-Herrero}}]{Cao2018}%
  \BibitemOpen
  \bibfield  {author} {\bibinfo {author} {\bibfnamefont {Y.}~\bibnamefont
  {Cao}}, \bibinfo {author} {\bibfnamefont {V.}~\bibnamefont {Fatemi}},
  \bibinfo {author} {\bibfnamefont {S.}~\bibnamefont {Fang}}, \bibinfo {author}
  {\bibfnamefont {K.}~\bibnamefont {Watanabe}}, \bibinfo {author}
  {\bibfnamefont {T.}~\bibnamefont {Taniguchi}}, \bibinfo {author}
  {\bibfnamefont {E.}~\bibnamefont {Kaxiras}},\ and\ \bibinfo {author}
  {\bibfnamefont {P.}~\bibnamefont {Jarillo-Herrero}},\ }\bibfield  {title}
  {\bibinfo {title} {Unconventional superconductivity in magic-angle graphene
  superlattices},\ }\href {https://doi.org/10.1038/nature26160} {\bibfield
  {journal} {\bibinfo  {journal} {Nature}\ }\textbf {\bibinfo {volume} {556}},\
  \bibinfo {pages} {43} (\bibinfo {year} {2018}{\natexlab{a}})}\BibitemShut
  {NoStop}%
\bibitem [{\citenamefont {Cao}\ \emph {et~al.}(2020)\citenamefont {Cao},
  \citenamefont {Rodan-Legrain}, \citenamefont {Rubies-Bigorda}, \citenamefont
  {Park}, \citenamefont {Watanabe}, \citenamefont {Taniguchi},\ and\
  \citenamefont {Jarillo-Herrero}}]{Cao2020}%
  \BibitemOpen
  \bibfield  {author} {\bibinfo {author} {\bibfnamefont {Y.}~\bibnamefont
  {Cao}}, \bibinfo {author} {\bibfnamefont {D.}~\bibnamefont {Rodan-Legrain}},
  \bibinfo {author} {\bibfnamefont {O.}~\bibnamefont {Rubies-Bigorda}},
  \bibinfo {author} {\bibfnamefont {J.~M.}\ \bibnamefont {Park}}, \bibinfo
  {author} {\bibfnamefont {K.}~\bibnamefont {Watanabe}}, \bibinfo {author}
  {\bibfnamefont {T.}~\bibnamefont {Taniguchi}},\ and\ \bibinfo {author}
  {\bibfnamefont {P.}~\bibnamefont {Jarillo-Herrero}},\ }\bibfield  {title}
  {\bibinfo {title} {Tunable correlated states and spin-polarized phases in
  twisted bilayer--bilayer graphene},\ }\bibfield  {journal} {\bibinfo
  {journal} {Nature}\ }\href {https://doi.org/10.1038/s41586-020-2260-6}
  {10.1038/s41586-020-2260-6} (\bibinfo {year} {2020})\BibitemShut {NoStop}%
\bibitem [{\citenamefont {Shen}\ \emph {et~al.}(2020)\citenamefont {Shen},
  \citenamefont {Chu}, \citenamefont {Wu}, \citenamefont {Li}, \citenamefont
  {Wang}, \citenamefont {Zhao}, \citenamefont {Tang}, \citenamefont {Liu},
  \citenamefont {Tian}, \citenamefont {Watanabe}, \citenamefont {Taniguchi},
  \citenamefont {Yang}, \citenamefont {Meng}, \citenamefont {Shi},
  \citenamefont {Yazyev},\ and\ \citenamefont {Zhang}}]{Shen2020}%
  \BibitemOpen
  \bibfield  {author} {\bibinfo {author} {\bibfnamefont {C.}~\bibnamefont
  {Shen}}, \bibinfo {author} {\bibfnamefont {Y.}~\bibnamefont {Chu}}, \bibinfo
  {author} {\bibfnamefont {Q.}~\bibnamefont {Wu}}, \bibinfo {author}
  {\bibfnamefont {N.}~\bibnamefont {Li}}, \bibinfo {author} {\bibfnamefont
  {S.}~\bibnamefont {Wang}}, \bibinfo {author} {\bibfnamefont {Y.}~\bibnamefont
  {Zhao}}, \bibinfo {author} {\bibfnamefont {J.}~\bibnamefont {Tang}}, \bibinfo
  {author} {\bibfnamefont {J.}~\bibnamefont {Liu}}, \bibinfo {author}
  {\bibfnamefont {J.}~\bibnamefont {Tian}}, \bibinfo {author} {\bibfnamefont
  {K.}~\bibnamefont {Watanabe}}, \bibinfo {author} {\bibfnamefont
  {T.}~\bibnamefont {Taniguchi}}, \bibinfo {author} {\bibfnamefont
  {R.}~\bibnamefont {Yang}}, \bibinfo {author} {\bibfnamefont {Z.~Y.}\
  \bibnamefont {Meng}}, \bibinfo {author} {\bibfnamefont {D.}~\bibnamefont
  {Shi}}, \bibinfo {author} {\bibfnamefont {O.~V.}\ \bibnamefont {Yazyev}},\
  and\ \bibinfo {author} {\bibfnamefont {G.}~\bibnamefont {Zhang}},\ }\bibfield
   {title} {\bibinfo {title} {Correlated states in twisted double bilayer
  graphene},\ }\href {https://doi.org/10.1038/s41567-020-0825-9} {\bibfield
  {journal} {\bibinfo  {journal} {Nature Physics}\ }\textbf {\bibinfo {volume}
  {16}},\ \bibinfo {pages} {520} (\bibinfo {year} {2020})}\BibitemShut
  {NoStop}%
\bibitem [{\citenamefont {Liu}\ \emph {et~al.}(2020)\citenamefont {Liu},
  \citenamefont {Hao}, \citenamefont {Khalaf}, \citenamefont {Lee},
  \citenamefont {Ronen}, \citenamefont {Yoo}, \citenamefont {Haei~Najafabadi},
  \citenamefont {Watanabe}, \citenamefont {Taniguchi}, \citenamefont
  {Vishwanath},\ and\ \citenamefont {Kim}}]{Liu2020}%
  \BibitemOpen
  \bibfield  {author} {\bibinfo {author} {\bibfnamefont {X.}~\bibnamefont
  {Liu}}, \bibinfo {author} {\bibfnamefont {Z.}~\bibnamefont {Hao}}, \bibinfo
  {author} {\bibfnamefont {E.}~\bibnamefont {Khalaf}}, \bibinfo {author}
  {\bibfnamefont {J.~Y.}\ \bibnamefont {Lee}}, \bibinfo {author} {\bibfnamefont
  {Y.}~\bibnamefont {Ronen}}, \bibinfo {author} {\bibfnamefont
  {H.}~\bibnamefont {Yoo}}, \bibinfo {author} {\bibfnamefont {D.}~\bibnamefont
  {Haei~Najafabadi}}, \bibinfo {author} {\bibfnamefont {K.}~\bibnamefont
  {Watanabe}}, \bibinfo {author} {\bibfnamefont {T.}~\bibnamefont {Taniguchi}},
  \bibinfo {author} {\bibfnamefont {A.}~\bibnamefont {Vishwanath}},\ and\
  \bibinfo {author} {\bibfnamefont {P.}~\bibnamefont {Kim}},\ }\bibfield
  {title} {\bibinfo {title} {Tunable spin-polarized correlated states in
  twisted double bilayer graphene},\ }\href
  {https://doi.org/10.1038/s41586-020-2458-7} {\bibfield  {journal} {\bibinfo
  {journal} {Nature}\ }\textbf {\bibinfo {volume} {583}},\ \bibinfo {pages}
  {221} (\bibinfo {year} {2020})}\BibitemShut {NoStop}%
\bibitem [{\citenamefont {Chebrolu}\ \emph {et~al.}(2019)\citenamefont
  {Chebrolu}, \citenamefont {Chittari},\ and\ \citenamefont
  {Jung}}]{PhysRevB.99.235417}%
  \BibitemOpen
  \bibfield  {author} {\bibinfo {author} {\bibfnamefont {N.~R.}\ \bibnamefont
  {Chebrolu}}, \bibinfo {author} {\bibfnamefont {B.~L.}\ \bibnamefont
  {Chittari}},\ and\ \bibinfo {author} {\bibfnamefont {J.}~\bibnamefont
  {Jung}},\ }\bibfield  {title} {\bibinfo {title} {Flat bands in twisted double
  bilayer graphene},\ }\href {https://doi.org/10.1103/PhysRevB.99.235417}
  {\bibfield  {journal} {\bibinfo  {journal} {Phys. Rev. B}\ }\textbf {\bibinfo
  {volume} {99}},\ \bibinfo {pages} {235417} (\bibinfo {year}
  {2019})}\BibitemShut {NoStop}%
\bibitem [{\citenamefont {Koshino}(2019)}]{PhysRevB.99.235406}%
  \BibitemOpen
  \bibfield  {author} {\bibinfo {author} {\bibfnamefont {M.}~\bibnamefont
  {Koshino}},\ }\bibfield  {title} {\bibinfo {title} {Band structure and
  topological properties of twisted double bilayer graphene},\ }\href
  {https://doi.org/10.1103/PhysRevB.99.235406} {\bibfield  {journal} {\bibinfo
  {journal} {Phys. Rev. B}\ }\textbf {\bibinfo {volume} {99}},\ \bibinfo
  {pages} {235406} (\bibinfo {year} {2019})}\BibitemShut {NoStop}%
\bibitem [{\citenamefont {Lee}\ \emph {et~al.}(2019)\citenamefont {Lee},
  \citenamefont {Khalaf}, \citenamefont {Liu}, \citenamefont {Liu},
  \citenamefont {Hao}, \citenamefont {Kim},\ and\ \citenamefont
  {Vishwanath}}]{Lee2019}%
  \BibitemOpen
  \bibfield  {author} {\bibinfo {author} {\bibfnamefont {J.~Y.}\ \bibnamefont
  {Lee}}, \bibinfo {author} {\bibfnamefont {E.}~\bibnamefont {Khalaf}},
  \bibinfo {author} {\bibfnamefont {S.}~\bibnamefont {Liu}}, \bibinfo {author}
  {\bibfnamefont {X.}~\bibnamefont {Liu}}, \bibinfo {author} {\bibfnamefont
  {Z.}~\bibnamefont {Hao}}, \bibinfo {author} {\bibfnamefont {P.}~\bibnamefont
  {Kim}},\ and\ \bibinfo {author} {\bibfnamefont {A.}~\bibnamefont
  {Vishwanath}},\ }\bibfield  {title} {\bibinfo {title} {Theory of correlated
  insulating behaviour and spin-triplet superconductivity in twisted double
  bilayer graphene},\ }\href {https://doi.org/10.1038/s41467-019-12981-1}
  {\bibfield  {journal} {\bibinfo  {journal} {Nature Communications}\ }\textbf
  {\bibinfo {volume} {10}},\ \bibinfo {pages} {5333} (\bibinfo {year}
  {2019})}\BibitemShut {NoStop}%
\bibitem [{\citenamefont {Haddadi}\ \emph {et~al.}(2020)\citenamefont
  {Haddadi}, \citenamefont {Wu}, \citenamefont {Kruchkov},\ and\ \citenamefont
  {Yazyev}}]{doi:10.1021/acs.nanolett.9b05117}%
  \BibitemOpen
  \bibfield  {author} {\bibinfo {author} {\bibfnamefont {F.}~\bibnamefont
  {Haddadi}}, \bibinfo {author} {\bibfnamefont {Q.}~\bibnamefont {Wu}},
  \bibinfo {author} {\bibfnamefont {A.~J.}\ \bibnamefont {Kruchkov}},\ and\
  \bibinfo {author} {\bibfnamefont {O.~V.}\ \bibnamefont {Yazyev}},\ }\bibfield
   {title} {\bibinfo {title} {Moiré flat bands in twisted double bilayer
  graphene},\ }\href {https://doi.org/10.1021/acs.nanolett.9b05117} {\bibfield
  {journal} {\bibinfo  {journal} {Nano Letters}\ }\textbf {\bibinfo {volume}
  {20}},\ \bibinfo {pages} {2410} (\bibinfo {year} {2020})},\ \bibinfo {note}
  {pMID: 32097013},\ \Eprint
  {https://arxiv.org/abs/https://doi.org/10.1021/acs.nanolett.9b05117}
  {https://doi.org/10.1021/acs.nanolett.9b05117} \BibitemShut {NoStop}%
\bibitem [{\citenamefont {Culchac}\ \emph {et~al.}(2020)\citenamefont
  {Culchac}, \citenamefont {Del~Grande}, \citenamefont {Capaz}, \citenamefont
  {Chico},\ and\ \citenamefont {Morell}}]{C9NR10830K}%
  \BibitemOpen
  \bibfield  {author} {\bibinfo {author} {\bibfnamefont {F.~J.}\ \bibnamefont
  {Culchac}}, \bibinfo {author} {\bibfnamefont {R.~R.}\ \bibnamefont
  {Del~Grande}}, \bibinfo {author} {\bibfnamefont {R.~B.}\ \bibnamefont
  {Capaz}}, \bibinfo {author} {\bibfnamefont {L.}~\bibnamefont {Chico}},\ and\
  \bibinfo {author} {\bibfnamefont {E.~S.}\ \bibnamefont {Morell}},\ }\bibfield
   {title} {\bibinfo {title} {Flat bands and gaps in twisted double bilayer
  graphene},\ }\href {https://doi.org/10.1039/C9NR10830K} {\bibfield  {journal}
  {\bibinfo  {journal} {Nanoscale}\ }\textbf {\bibinfo {volume} {12}},\
  \bibinfo {pages} {5014} (\bibinfo {year} {2020})}\BibitemShut {NoStop}%
\bibitem [{\citenamefont {Zhang}\ \emph {et~al.}(2019)\citenamefont {Zhang},
  \citenamefont {Mao}, \citenamefont {Cao}, \citenamefont {Jarillo-Herrero},\
  and\ \citenamefont {Senthil}}]{PhysRevB.99.075127}%
  \BibitemOpen
  \bibfield  {author} {\bibinfo {author} {\bibfnamefont {Y.-H.}\ \bibnamefont
  {Zhang}}, \bibinfo {author} {\bibfnamefont {D.}~\bibnamefont {Mao}}, \bibinfo
  {author} {\bibfnamefont {Y.}~\bibnamefont {Cao}}, \bibinfo {author}
  {\bibfnamefont {P.}~\bibnamefont {Jarillo-Herrero}},\ and\ \bibinfo {author}
  {\bibfnamefont {T.}~\bibnamefont {Senthil}},\ }\bibfield  {title} {\bibinfo
  {title} {Nearly flat chern bands in moir\'e superlattices},\ }\href
  {https://doi.org/10.1103/PhysRevB.99.075127} {\bibfield  {journal} {\bibinfo
  {journal} {Phys. Rev. B}\ }\textbf {\bibinfo {volume} {99}},\ \bibinfo
  {pages} {075127} (\bibinfo {year} {2019})}\BibitemShut {NoStop}%
\bibitem [{\citenamefont {Kerelsky}\ \emph {et~al.}(2019)\citenamefont
  {Kerelsky}, \citenamefont {Rubio-Verdú}, \citenamefont {Xian}, \citenamefont
  {Kennes}, \citenamefont {Halbertal}, \citenamefont {Finney}, \citenamefont
  {Song}, \citenamefont {Turkel}, \citenamefont {Wang}, \citenamefont
  {Watanabe}, \citenamefont {Taniguchi}, \citenamefont {Hone}, \citenamefont
  {Dean}, \citenamefont {Basov}, \citenamefont {Rubio},\ and\ \citenamefont
  {Pasupathy}}]{kerelsky2019moireless}%
  \BibitemOpen
  \bibfield  {author} {\bibinfo {author} {\bibfnamefont {A.}~\bibnamefont
  {Kerelsky}}, \bibinfo {author} {\bibfnamefont {C.}~\bibnamefont
  {Rubio-Verdú}}, \bibinfo {author} {\bibfnamefont {L.}~\bibnamefont {Xian}},
  \bibinfo {author} {\bibfnamefont {D.~M.}\ \bibnamefont {Kennes}}, \bibinfo
  {author} {\bibfnamefont {D.}~\bibnamefont {Halbertal}}, \bibinfo {author}
  {\bibfnamefont {N.}~\bibnamefont {Finney}}, \bibinfo {author} {\bibfnamefont
  {L.}~\bibnamefont {Song}}, \bibinfo {author} {\bibfnamefont {S.}~\bibnamefont
  {Turkel}}, \bibinfo {author} {\bibfnamefont {L.}~\bibnamefont {Wang}},
  \bibinfo {author} {\bibfnamefont {K.}~\bibnamefont {Watanabe}}, \bibinfo
  {author} {\bibfnamefont {T.}~\bibnamefont {Taniguchi}}, \bibinfo {author}
  {\bibfnamefont {J.}~\bibnamefont {Hone}}, \bibinfo {author} {\bibfnamefont
  {C.}~\bibnamefont {Dean}}, \bibinfo {author} {\bibfnamefont {D.}~\bibnamefont
  {Basov}}, \bibinfo {author} {\bibfnamefont {A.}~\bibnamefont {Rubio}},\ and\
  \bibinfo {author} {\bibfnamefont {A.~N.}\ \bibnamefont {Pasupathy}},\
  }\href@noop {} {\bibinfo {title} {Moir\'e-less correlations in abca
  graphene}} (\bibinfo {year} {2019}),\ \Eprint
  {https://arxiv.org/abs/1911.00007} {arXiv:1911.00007 [cond-mat.mes-hall]}
  \BibitemShut {NoStop}%
\bibitem [{\citenamefont {Rubio-Verdú}\ \emph {et~al.}(2020)\citenamefont
  {Rubio-Verdú}, \citenamefont {Turkel}, \citenamefont {Song}, \citenamefont
  {Klebl}, \citenamefont {Samajdar}, \citenamefont {Scheurer}, \citenamefont
  {Venderbos}, \citenamefont {Watanabe}, \citenamefont {Taniguchi},
  \citenamefont {Ochoa}, \citenamefont {Xian}, \citenamefont {Kennes},
  \citenamefont {Fernandes}, \citenamefont {Ángel Rubio},\ and\ \citenamefont
  {Pasupathy}}]{rubioverdu2020universal}%
  \BibitemOpen
  \bibfield  {author} {\bibinfo {author} {\bibfnamefont {C.}~\bibnamefont
  {Rubio-Verdú}}, \bibinfo {author} {\bibfnamefont {S.}~\bibnamefont
  {Turkel}}, \bibinfo {author} {\bibfnamefont {L.}~\bibnamefont {Song}},
  \bibinfo {author} {\bibfnamefont {L.}~\bibnamefont {Klebl}}, \bibinfo
  {author} {\bibfnamefont {R.}~\bibnamefont {Samajdar}}, \bibinfo {author}
  {\bibfnamefont {M.~S.}\ \bibnamefont {Scheurer}}, \bibinfo {author}
  {\bibfnamefont {J.~W.~F.}\ \bibnamefont {Venderbos}}, \bibinfo {author}
  {\bibfnamefont {K.}~\bibnamefont {Watanabe}}, \bibinfo {author}
  {\bibfnamefont {T.}~\bibnamefont {Taniguchi}}, \bibinfo {author}
  {\bibfnamefont {H.}~\bibnamefont {Ochoa}}, \bibinfo {author} {\bibfnamefont
  {L.}~\bibnamefont {Xian}}, \bibinfo {author} {\bibfnamefont {D.}~\bibnamefont
  {Kennes}}, \bibinfo {author} {\bibfnamefont {R.~M.}\ \bibnamefont
  {Fernandes}}, \bibinfo {author} {\bibnamefont {Ángel Rubio}},\ and\ \bibinfo
  {author} {\bibfnamefont {A.~N.}\ \bibnamefont {Pasupathy}},\ }\href@noop {}
  {\bibinfo {title} {Universal moir\'e nematic phase in twisted graphitic
  systems}} (\bibinfo {year} {2020}),\ \Eprint
  {https://arxiv.org/abs/2009.11645} {arXiv:2009.11645 [cond-mat.str-el]}
  \BibitemShut {NoStop}%
\bibitem [{\citenamefont {Halbertal}\ \emph {et~al.}(2021)\citenamefont
  {Halbertal}, \citenamefont {Finney}, \citenamefont {Sunku}, \citenamefont
  {Kerelsky}, \citenamefont {Rubio-Verd{\'u}}, \citenamefont {Shabani},
  \citenamefont {Xian}, \citenamefont {Carr}, \citenamefont {Chen},
  \citenamefont {Zhang}, \citenamefont {Wang}, \citenamefont
  {Gonzalez-Acevedo}, \citenamefont {McLeod}, \citenamefont {Rhodes},
  \citenamefont {Watanabe}, \citenamefont {Taniguchi}, \citenamefont {Kaxiras},
  \citenamefont {Dean}, \citenamefont {Hone}, \citenamefont {Pasupathy},
  \citenamefont {Kennes}, \citenamefont {Rubio},\ and\ \citenamefont
  {Basov}}]{halbertal2020moire}%
  \BibitemOpen
  \bibfield  {author} {\bibinfo {author} {\bibfnamefont {D.}~\bibnamefont
  {Halbertal}}, \bibinfo {author} {\bibfnamefont {N.~R.}\ \bibnamefont
  {Finney}}, \bibinfo {author} {\bibfnamefont {S.~S.}\ \bibnamefont {Sunku}},
  \bibinfo {author} {\bibfnamefont {A.}~\bibnamefont {Kerelsky}}, \bibinfo
  {author} {\bibfnamefont {C.}~\bibnamefont {Rubio-Verd{\'u}}}, \bibinfo
  {author} {\bibfnamefont {S.}~\bibnamefont {Shabani}}, \bibinfo {author}
  {\bibfnamefont {L.}~\bibnamefont {Xian}}, \bibinfo {author} {\bibfnamefont
  {S.}~\bibnamefont {Carr}}, \bibinfo {author} {\bibfnamefont {S.}~\bibnamefont
  {Chen}}, \bibinfo {author} {\bibfnamefont {C.}~\bibnamefont {Zhang}},
  \bibinfo {author} {\bibfnamefont {L.}~\bibnamefont {Wang}}, \bibinfo {author}
  {\bibfnamefont {D.}~\bibnamefont {Gonzalez-Acevedo}}, \bibinfo {author}
  {\bibfnamefont {A.~S.}\ \bibnamefont {McLeod}}, \bibinfo {author}
  {\bibfnamefont {D.}~\bibnamefont {Rhodes}}, \bibinfo {author} {\bibfnamefont
  {K.}~\bibnamefont {Watanabe}}, \bibinfo {author} {\bibfnamefont
  {T.}~\bibnamefont {Taniguchi}}, \bibinfo {author} {\bibfnamefont
  {E.}~\bibnamefont {Kaxiras}}, \bibinfo {author} {\bibfnamefont {C.~R.}\
  \bibnamefont {Dean}}, \bibinfo {author} {\bibfnamefont {J.~C.}\ \bibnamefont
  {Hone}}, \bibinfo {author} {\bibfnamefont {A.~N.}\ \bibnamefont {Pasupathy}},
  \bibinfo {author} {\bibfnamefont {D.~M.}\ \bibnamefont {Kennes}}, \bibinfo
  {author} {\bibfnamefont {A.}~\bibnamefont {Rubio}},\ and\ \bibinfo {author}
  {\bibfnamefont {D.~N.}\ \bibnamefont {Basov}},\ }\bibfield  {title} {\bibinfo
  {title} {Moir{\'e} metrology of energy landscapes in van der waals
  heterostructures},\ }\href {https://doi.org/10.1038/s41467-020-20428-1}
  {\bibfield  {journal} {\bibinfo  {journal} {Nature Communications}\ }\textbf
  {\bibinfo {volume} {12}},\ \bibinfo {pages} {242} (\bibinfo {year}
  {2021})}\BibitemShut {NoStop}%
\bibitem [{\citenamefont {Christos}\ \emph {et~al.}(2021)\citenamefont
  {Christos}, \citenamefont {Sachdev},\ and\ \citenamefont
  {Scheurer}}]{christos2021correlated}%
  \BibitemOpen
  \bibfield  {author} {\bibinfo {author} {\bibfnamefont {M.}~\bibnamefont
  {Christos}}, \bibinfo {author} {\bibfnamefont {S.}~\bibnamefont {Sachdev}},\
  and\ \bibinfo {author} {\bibfnamefont {M.~S.}\ \bibnamefont {Scheurer}},\
  }\href@noop {} {\bibinfo {title} {Correlated insulators, semimetals, and
  superconductivity in twisted trilayer graphene}} (\bibinfo {year} {2021}),\
  \Eprint {https://arxiv.org/abs/2106.02063} {arXiv:2106.02063
  [cond-mat.str-el]} \BibitemShut {NoStop}%
\bibitem [{\citenamefont {Cao}\ \emph {et~al.}(2018{\natexlab{b}})\citenamefont
  {Cao}, \citenamefont {Fatemi}, \citenamefont {Fang}, \citenamefont
  {Watanabe}, \citenamefont {Taniguchi}, \citenamefont {Kaxiras},\ and\
  \citenamefont {Jarillo-Herrero}}]{Cao2018sc}%
  \BibitemOpen
  \bibfield  {author} {\bibinfo {author} {\bibfnamefont {Y.}~\bibnamefont
  {Cao}}, \bibinfo {author} {\bibfnamefont {V.}~\bibnamefont {Fatemi}},
  \bibinfo {author} {\bibfnamefont {S.}~\bibnamefont {Fang}}, \bibinfo {author}
  {\bibfnamefont {K.}~\bibnamefont {Watanabe}}, \bibinfo {author}
  {\bibfnamefont {T.}~\bibnamefont {Taniguchi}}, \bibinfo {author}
  {\bibfnamefont {E.}~\bibnamefont {Kaxiras}},\ and\ \bibinfo {author}
  {\bibfnamefont {P.}~\bibnamefont {Jarillo-Herrero}},\ }\bibfield  {title}
  {\bibinfo {title} {Unconventional superconductivity in magic-angle graphene
  superlattices},\ }\href {https://doi.org/10.1038/nature26160} {\bibfield
  {journal} {\bibinfo  {journal} {Nature}\ }\textbf {\bibinfo {volume} {556}},\
  \bibinfo {pages} {43} (\bibinfo {year} {2018}{\natexlab{b}})}\BibitemShut
  {NoStop}%
\bibitem [{\citenamefont {Codecido}\ \emph {et~al.}(2019)\citenamefont
  {Codecido}, \citenamefont {Wang}, \citenamefont {Koester}, \citenamefont
  {Che}, \citenamefont {Tian}, \citenamefont {Lv}, \citenamefont {Tran},
  \citenamefont {Watanabe}, \citenamefont {Taniguchi}, \citenamefont {Zhang},
  \citenamefont {Bockrath},\ and\ \citenamefont {Lau}}]{Codecidoeaaw9770}%
  \BibitemOpen
  \bibfield  {author} {\bibinfo {author} {\bibfnamefont {E.}~\bibnamefont
  {Codecido}}, \bibinfo {author} {\bibfnamefont {Q.}~\bibnamefont {Wang}},
  \bibinfo {author} {\bibfnamefont {R.}~\bibnamefont {Koester}}, \bibinfo
  {author} {\bibfnamefont {S.}~\bibnamefont {Che}}, \bibinfo {author}
  {\bibfnamefont {H.}~\bibnamefont {Tian}}, \bibinfo {author} {\bibfnamefont
  {R.}~\bibnamefont {Lv}}, \bibinfo {author} {\bibfnamefont {S.}~\bibnamefont
  {Tran}}, \bibinfo {author} {\bibfnamefont {K.}~\bibnamefont {Watanabe}},
  \bibinfo {author} {\bibfnamefont {T.}~\bibnamefont {Taniguchi}}, \bibinfo
  {author} {\bibfnamefont {F.}~\bibnamefont {Zhang}}, \bibinfo {author}
  {\bibfnamefont {M.}~\bibnamefont {Bockrath}},\ and\ \bibinfo {author}
  {\bibfnamefont {C.~N.}\ \bibnamefont {Lau}},\ }\bibfield  {title} {\bibinfo
  {title} {Correlated insulating and superconducting states in twisted bilayer
  graphene below the magic angle},\ }\bibfield  {journal} {\bibinfo  {journal}
  {Science Advances}\ }\textbf {\bibinfo {volume} {5}},\ \href
  {https://doi.org/10.1126/sciadv.aaw9770} {10.1126/sciadv.aaw9770} (\bibinfo
  {year} {2019}),\ \Eprint
  {https://arxiv.org/abs/https://advances.sciencemag.org/content/5/9/eaaw9770.full.pdf}
  {https://advances.sciencemag.org/content/5/9/eaaw9770.full.pdf} \BibitemShut
  {NoStop}%
\bibitem [{\citenamefont {Wong}\ \emph {et~al.}(2020)\citenamefont {Wong},
  \citenamefont {Nuckolls}, \citenamefont {Oh}, \citenamefont {Lian},
  \citenamefont {Xie}, \citenamefont {Jeon}, \citenamefont {Watanabe},
  \citenamefont {Taniguchi}, \citenamefont {Bernevig},\ and\ \citenamefont
  {Yazdani}}]{wong2019cascade}%
  \BibitemOpen
  \bibfield  {author} {\bibinfo {author} {\bibfnamefont {D.}~\bibnamefont
  {Wong}}, \bibinfo {author} {\bibfnamefont {K.~P.}\ \bibnamefont {Nuckolls}},
  \bibinfo {author} {\bibfnamefont {M.}~\bibnamefont {Oh}}, \bibinfo {author}
  {\bibfnamefont {B.}~\bibnamefont {Lian}}, \bibinfo {author} {\bibfnamefont
  {Y.}~\bibnamefont {Xie}}, \bibinfo {author} {\bibfnamefont {S.}~\bibnamefont
  {Jeon}}, \bibinfo {author} {\bibfnamefont {K.}~\bibnamefont {Watanabe}},
  \bibinfo {author} {\bibfnamefont {T.}~\bibnamefont {Taniguchi}}, \bibinfo
  {author} {\bibfnamefont {B.~A.}\ \bibnamefont {Bernevig}},\ and\ \bibinfo
  {author} {\bibfnamefont {A.}~\bibnamefont {Yazdani}},\ }\bibfield  {title}
  {\bibinfo {title} {Cascade of electronic transitions in magic-angle twisted
  bilayer graphene},\ }\href {https://doi.org/10.1038/s41586-020-2339-0}
  {\bibfield  {journal} {\bibinfo  {journal} {Nature}\ }\textbf {\bibinfo
  {volume} {582}},\ \bibinfo {pages} {198} (\bibinfo {year}
  {2020})}\BibitemShut {NoStop}%
\bibitem [{\citenamefont {Lu}\ \emph {et~al.}(2019)\citenamefont {Lu},
  \citenamefont {Stepanov}, \citenamefont {Yang}, \citenamefont {Xie},
  \citenamefont {Aamir}, \citenamefont {Das}, \citenamefont {Urgell},
  \citenamefont {Watanabe}, \citenamefont {Taniguchi}, \citenamefont {Zhang},
  \citenamefont {Bachtold}, \citenamefont {MacDonald},\ and\ \citenamefont
  {Efetov}}]{Lu2019efetov}%
  \BibitemOpen
  \bibfield  {author} {\bibinfo {author} {\bibfnamefont {X.}~\bibnamefont
  {Lu}}, \bibinfo {author} {\bibfnamefont {P.}~\bibnamefont {Stepanov}},
  \bibinfo {author} {\bibfnamefont {W.}~\bibnamefont {Yang}}, \bibinfo {author}
  {\bibfnamefont {M.}~\bibnamefont {Xie}}, \bibinfo {author} {\bibfnamefont
  {M.~A.}\ \bibnamefont {Aamir}}, \bibinfo {author} {\bibfnamefont
  {I.}~\bibnamefont {Das}}, \bibinfo {author} {\bibfnamefont {C.}~\bibnamefont
  {Urgell}}, \bibinfo {author} {\bibfnamefont {K.}~\bibnamefont {Watanabe}},
  \bibinfo {author} {\bibfnamefont {T.}~\bibnamefont {Taniguchi}}, \bibinfo
  {author} {\bibfnamefont {G.}~\bibnamefont {Zhang}}, \bibinfo {author}
  {\bibfnamefont {A.}~\bibnamefont {Bachtold}}, \bibinfo {author}
  {\bibfnamefont {A.~H.}\ \bibnamefont {MacDonald}},\ and\ \bibinfo {author}
  {\bibfnamefont {D.~K.}\ \bibnamefont {Efetov}},\ }\bibfield  {title}
  {\bibinfo {title} {Superconductors, orbital magnets and correlated states in
  magic-angle bilayer graphene},\ }\href
  {https://doi.org/10.1038/s41586-019-1695-0} {\bibfield  {journal} {\bibinfo
  {journal} {Nature}\ }\textbf {\bibinfo {volume} {574}},\ \bibinfo {pages}
  {653} (\bibinfo {year} {2019})}\BibitemShut {NoStop}%
\bibitem [{\citenamefont {Sharpe}\ \emph {et~al.}(2019)\citenamefont {Sharpe},
  \citenamefont {Fox}, \citenamefont {Barnard}, \citenamefont {Finney},
  \citenamefont {Watanabe}, \citenamefont {Taniguchi}, \citenamefont
  {Kastner},\ and\ \citenamefont {Goldhaber-Gordon}}]{Sharpe605}%
  \BibitemOpen
  \bibfield  {author} {\bibinfo {author} {\bibfnamefont {A.~L.}\ \bibnamefont
  {Sharpe}}, \bibinfo {author} {\bibfnamefont {E.~J.}\ \bibnamefont {Fox}},
  \bibinfo {author} {\bibfnamefont {A.~W.}\ \bibnamefont {Barnard}}, \bibinfo
  {author} {\bibfnamefont {J.}~\bibnamefont {Finney}}, \bibinfo {author}
  {\bibfnamefont {K.}~\bibnamefont {Watanabe}}, \bibinfo {author}
  {\bibfnamefont {T.}~\bibnamefont {Taniguchi}}, \bibinfo {author}
  {\bibfnamefont {M.~A.}\ \bibnamefont {Kastner}},\ and\ \bibinfo {author}
  {\bibfnamefont {D.}~\bibnamefont {Goldhaber-Gordon}},\ }\bibfield  {title}
  {\bibinfo {title} {Emergent ferromagnetism near three-quarters filling in
  twisted bilayer graphene},\ }\href {https://doi.org/10.1126/science.aaw3780}
  {\bibfield  {journal} {\bibinfo  {journal} {Science}\ }\textbf {\bibinfo
  {volume} {365}},\ \bibinfo {pages} {605} (\bibinfo {year} {2019})},\ \Eprint
  {https://arxiv.org/abs/https://science.sciencemag.org/content/365/6453/605.full.pdf}
  {https://science.sciencemag.org/content/365/6453/605.full.pdf} \BibitemShut
  {NoStop}%
\bibitem [{\citenamefont {Seo}\ \emph {et~al.}(2019)\citenamefont {Seo},
  \citenamefont {Kotov},\ and\ \citenamefont {Uchoa}}]{Seo_2019}%
  \BibitemOpen
  \bibfield  {author} {\bibinfo {author} {\bibfnamefont {K.}~\bibnamefont
  {Seo}}, \bibinfo {author} {\bibfnamefont {V.~N.}\ \bibnamefont {Kotov}},\
  and\ \bibinfo {author} {\bibfnamefont {B.}~\bibnamefont {Uchoa}},\ }\bibfield
   {title} {\bibinfo {title} {Ferromagnetic mott state in twisted graphene
  bilayers at the magic angle},\ }\bibfield  {journal} {\bibinfo  {journal}
  {Physical Review Letters}\ }\textbf {\bibinfo {volume} {122}},\ \href
  {https://doi.org/10.1103/physrevlett.122.246402}
  {10.1103/physrevlett.122.246402} (\bibinfo {year} {2019})\BibitemShut
  {NoStop}%
\bibitem [{\citenamefont {Oka}\ and\ \citenamefont
  {Kitamura}(2019)}]{Oka_2019}%
  \BibitemOpen
  \bibfield  {author} {\bibinfo {author} {\bibfnamefont {T.}~\bibnamefont
  {Oka}}\ and\ \bibinfo {author} {\bibfnamefont {S.}~\bibnamefont {Kitamura}},\
  }\bibfield  {title} {\bibinfo {title} {Floquet engineering of quantum
  materials},\ }\href
  {https://doi.org/10.1146/annurev-conmatphys-031218-013423} {\bibfield
  {journal} {\bibinfo  {journal} {Annual Review of Condensed Matter Physics}\
  }\textbf {\bibinfo {volume} {10}},\ \bibinfo {pages} {387–408} (\bibinfo
  {year} {2019})}\BibitemShut {NoStop}%
\bibitem [{\citenamefont {Rudner}\ and\ \citenamefont
  {Lindner}(2020)}]{rudner2020_review}%
  \BibitemOpen
  \bibfield  {author} {\bibinfo {author} {\bibfnamefont {M.~S.}\ \bibnamefont
  {Rudner}}\ and\ \bibinfo {author} {\bibfnamefont {N.~H.}\ \bibnamefont
  {Lindner}},\ }\bibfield  {title} {\bibinfo {title} {Band structure
  engineering and non-equilibrium dynamics in floquet topological insulators},\
  }\href {https://doi.org/10.1038/s42254-020-0170-z} {\bibfield  {journal}
  {\bibinfo  {journal} {Nature Reviews Physics}\ }\textbf {\bibinfo {volume}
  {2}},\ \bibinfo {pages} {229} (\bibinfo {year} {2020})}\BibitemShut {NoStop}%
\bibitem [{\citenamefont {Giovannini}\ and\ \citenamefont
  {Hübener}(2019)}]{Giovannini_2019}%
  \BibitemOpen
  \bibfield  {author} {\bibinfo {author} {\bibfnamefont {U.~D.}\ \bibnamefont
  {Giovannini}}\ and\ \bibinfo {author} {\bibfnamefont {H.}~\bibnamefont
  {Hübener}},\ }\bibfield  {title} {\bibinfo {title} {Floquet analysis of
  excitations in materials},\ }\href {https://doi.org/10.1088/2515-7639/ab387b}
  {\bibfield  {journal} {\bibinfo  {journal} {Journal of Physics: Materials}\
  }\textbf {\bibinfo {volume} {3}},\ \bibinfo {pages} {012001} (\bibinfo {year}
  {2019})}\BibitemShut {NoStop}%
\bibitem [{\citenamefont {McIver}\ \emph {et~al.}(2020)\citenamefont {McIver},
  \citenamefont {Schulte}, \citenamefont {Stein}, \citenamefont {Matsuyama},
  \citenamefont {Jotzu}, \citenamefont {Meier},\ and\ \citenamefont
  {Cavalleri}}]{McIver2020}%
  \BibitemOpen
  \bibfield  {author} {\bibinfo {author} {\bibfnamefont {J.~W.}\ \bibnamefont
  {McIver}}, \bibinfo {author} {\bibfnamefont {B.}~\bibnamefont {Schulte}},
  \bibinfo {author} {\bibfnamefont {F.-U.}\ \bibnamefont {Stein}}, \bibinfo
  {author} {\bibfnamefont {T.}~\bibnamefont {Matsuyama}}, \bibinfo {author}
  {\bibfnamefont {G.}~\bibnamefont {Jotzu}}, \bibinfo {author} {\bibfnamefont
  {G.}~\bibnamefont {Meier}},\ and\ \bibinfo {author} {\bibfnamefont
  {A.}~\bibnamefont {Cavalleri}},\ }\bibfield  {title} {\bibinfo {title}
  {Light-induced anomalous hall effect in graphene},\ }\href
  {https://doi.org/10.1038/s41567-019-0698-y} {\bibfield  {journal} {\bibinfo
  {journal} {Nature Physics}\ }\textbf {\bibinfo {volume} {16}},\ \bibinfo
  {pages} {38} (\bibinfo {year} {2020})}\BibitemShut {NoStop}%
\bibitem [{\citenamefont {Oka}\ and\ \citenamefont {Aoki}(2009)}]{oka2009}%
  \BibitemOpen
  \bibfield  {author} {\bibinfo {author} {\bibfnamefont {T.}~\bibnamefont
  {Oka}}\ and\ \bibinfo {author} {\bibfnamefont {H.}~\bibnamefont {Aoki}},\
  }\bibfield  {title} {\bibinfo {title} {Photovoltaic hall effect in
  graphene},\ }\href@noop {} {\bibfield  {journal} {\bibinfo  {journal} {Phys.
  Rev. B}\ }\textbf {\bibinfo {volume} {79}},\ \bibinfo {pages} {081406}
  (\bibinfo {year} {2009})}\BibitemShut {NoStop}%
\bibitem [{\citenamefont {Lindner}\ \emph {et~al.}(2011)\citenamefont
  {Lindner}, \citenamefont {Refael},\ and\ \citenamefont
  {Galitski}}]{lindner2011}%
  \BibitemOpen
  \bibfield  {author} {\bibinfo {author} {\bibfnamefont {N.~H.}\ \bibnamefont
  {Lindner}}, \bibinfo {author} {\bibfnamefont {G.}~\bibnamefont {Refael}},\
  and\ \bibinfo {author} {\bibfnamefont {V.}~\bibnamefont {Galitski}},\
  }\bibfield  {title} {\bibinfo {title} {Floquet topological insulator in
  semiconductor quantum wells},\ }\href {https://doi.org/10.1038/nphys1926}
  {\bibfield  {journal} {\bibinfo  {journal} {Nat. Phys.}\ }\textbf {\bibinfo
  {volume} {7}},\ \bibinfo {pages} {490} (\bibinfo {year} {2011})}\BibitemShut
  {NoStop}%
\bibitem [{\citenamefont {Rechtsman}\ \emph {et~al.}(2013)\citenamefont
  {Rechtsman}, \citenamefont {Zeuner}, \citenamefont {Plotnik}, \citenamefont
  {Lumer}, \citenamefont {Podolsky}, \citenamefont {Dreisow}, \citenamefont
  {Nolte}, \citenamefont {Segev},\ and\ \citenamefont
  {Szameit}}]{rechtsman2013}%
  \BibitemOpen
  \bibfield  {author} {\bibinfo {author} {\bibfnamefont {M.~C.}\ \bibnamefont
  {Rechtsman}}, \bibinfo {author} {\bibfnamefont {J.~M.}\ \bibnamefont
  {Zeuner}}, \bibinfo {author} {\bibfnamefont {Y.}~\bibnamefont {Plotnik}},
  \bibinfo {author} {\bibfnamefont {Y.}~\bibnamefont {Lumer}}, \bibinfo
  {author} {\bibfnamefont {D.}~\bibnamefont {Podolsky}}, \bibinfo {author}
  {\bibfnamefont {F.}~\bibnamefont {Dreisow}}, \bibinfo {author} {\bibfnamefont
  {S.}~\bibnamefont {Nolte}}, \bibinfo {author} {\bibfnamefont
  {M.}~\bibnamefont {Segev}},\ and\ \bibinfo {author} {\bibfnamefont
  {A.}~\bibnamefont {Szameit}},\ }\bibfield  {title} {\bibinfo {title}
  {Photonic floquet topological insulators},\ }\href@noop {} {\bibfield
  {journal} {\bibinfo  {journal} {Nature}\ }\textbf {\bibinfo {volume} {496}},\
  \bibinfo {pages} {196} (\bibinfo {year} {2013})}\BibitemShut {NoStop}%
\bibitem [{\citenamefont {Luo}(2021)}]{PhysRevB.103.195422}%
  \BibitemOpen
  \bibfield  {author} {\bibinfo {author} {\bibfnamefont {M.}~\bibnamefont
  {Luo}},\ }\bibfield  {title} {\bibinfo {title} {Tuning of a bilayer graphene
  heterostructure by horizontally incident circular polarized light},\ }\href
  {https://doi.org/10.1103/PhysRevB.103.195422} {\bibfield  {journal} {\bibinfo
   {journal} {Phys. Rev. B}\ }\textbf {\bibinfo {volume} {103}},\ \bibinfo
  {pages} {195422} (\bibinfo {year} {2021})}\BibitemShut {NoStop}%
\bibitem [{\citenamefont {Kibis}\ \emph {et~al.}(2016)\citenamefont {Kibis},
  \citenamefont {Morina}, \citenamefont {Dini},\ and\ \citenamefont
  {Shelykh}}]{PhysRevB.93.115420}%
  \BibitemOpen
  \bibfield  {author} {\bibinfo {author} {\bibfnamefont {O.~V.}\ \bibnamefont
  {Kibis}}, \bibinfo {author} {\bibfnamefont {S.}~\bibnamefont {Morina}},
  \bibinfo {author} {\bibfnamefont {K.}~\bibnamefont {Dini}},\ and\ \bibinfo
  {author} {\bibfnamefont {I.~A.}\ \bibnamefont {Shelykh}},\ }\bibfield
  {title} {\bibinfo {title} {Magnetoelectronic properties of graphene dressed
  by a high-frequency field},\ }\href
  {https://doi.org/10.1103/PhysRevB.93.115420} {\bibfield  {journal} {\bibinfo
  {journal} {Phys. Rev. B}\ }\textbf {\bibinfo {volume} {93}},\ \bibinfo
  {pages} {115420} (\bibinfo {year} {2016})}\BibitemShut {NoStop}%
\bibitem [{\citenamefont {Rodriguez-Vega}\ \emph {et~al.}(2021)\citenamefont
  {Rodriguez-Vega}, \citenamefont {Vogl},\ and\ \citenamefont
  {Fiete}}]{MoireFloquetRev}%
  \BibitemOpen
  \bibfield  {author} {\bibinfo {author} {\bibfnamefont {M.}~\bibnamefont
  {Rodriguez-Vega}}, \bibinfo {author} {\bibfnamefont {M.}~\bibnamefont
  {Vogl}},\ and\ \bibinfo {author} {\bibfnamefont {G.~A.}\ \bibnamefont
  {Fiete}},\ }\bibfield  {title} {\bibinfo {title} {Low-frequency and
  moiré–floquet engineering: A review},\ }\href
  {https://doi.org/https://doi.org/10.1016/j.aop.2021.168434} {\bibfield
  {journal} {\bibinfo  {journal} {Annals of Physics}\ ,\ \bibinfo {pages}
  {168434}} (\bibinfo {year} {2021})}\BibitemShut {NoStop}%
\bibitem [{\citenamefont {Blanes}\ \emph {et~al.}(2009)\citenamefont {Blanes},
  \citenamefont {Casas}, \citenamefont {Oteo},\ and\ \citenamefont
  {Ros}}]{blanes2009}%
  \BibitemOpen
  \bibfield  {author} {\bibinfo {author} {\bibfnamefont {S.}~\bibnamefont
  {Blanes}}, \bibinfo {author} {\bibfnamefont {F.}~\bibnamefont {Casas}},
  \bibinfo {author} {\bibfnamefont {J.}~\bibnamefont {Oteo}},\ and\ \bibinfo
  {author} {\bibfnamefont {J.}~\bibnamefont {Ros}},\ }\bibfield  {title}
  {\bibinfo {title} {The magnus expansion and some of its applications},\
  }\href@noop {} {\bibfield  {journal} {\bibinfo  {journal} {Physics Reports}\
  }\textbf {\bibinfo {volume} {470}},\ \bibinfo {pages} {151} (\bibinfo {year}
  {2009})}\BibitemShut {NoStop}%
\bibitem [{\citenamefont {Rahav}\ \emph
  {et~al.}(2003{\natexlab{a}})\citenamefont {Rahav}, \citenamefont {Gilary},\
  and\ \citenamefont {Fishman}}]{rahav2003}%
  \BibitemOpen
  \bibfield  {author} {\bibinfo {author} {\bibfnamefont {S.}~\bibnamefont
  {Rahav}}, \bibinfo {author} {\bibfnamefont {I.}~\bibnamefont {Gilary}},\ and\
  \bibinfo {author} {\bibfnamefont {S.}~\bibnamefont {Fishman}},\ }\bibfield
  {title} {\bibinfo {title} {Effective hamiltonians for periodically driven
  systems},\ }\href@noop {} {\bibfield  {journal} {\bibinfo  {journal} {Phys.
  Rev. A}\ }\textbf {\bibinfo {volume} {68}},\ \bibinfo {pages} {013820}
  (\bibinfo {year} {2003}{\natexlab{a}})}\BibitemShut {NoStop}%
\bibitem [{\citenamefont {Rahav}\ \emph
  {et~al.}(2003{\natexlab{b}})\citenamefont {Rahav}, \citenamefont {Gilary},\
  and\ \citenamefont {Fishman}}]{rahav2003b}%
  \BibitemOpen
  \bibfield  {author} {\bibinfo {author} {\bibfnamefont {S.}~\bibnamefont
  {Rahav}}, \bibinfo {author} {\bibfnamefont {I.}~\bibnamefont {Gilary}},\ and\
  \bibinfo {author} {\bibfnamefont {S.}~\bibnamefont {Fishman}},\ }\bibfield
  {title} {\bibinfo {title} {Time independent description of rapidly
  oscillating potentials},\ }\href@noop {} {\bibfield  {journal} {\bibinfo
  {journal} {Phys. Rev. Lett.}\ }\textbf {\bibinfo {volume} {91}},\ \bibinfo
  {pages} {110404} (\bibinfo {year} {2003}{\natexlab{b}})}\BibitemShut
  {NoStop}%
\bibitem [{\citenamefont {Bukov}\ \emph {et~al.}(2015)\citenamefont {Bukov},
  \citenamefont {D’Alessio},\ and\ \citenamefont {Polkovnikov}}]{Bukov_2015}%
  \BibitemOpen
  \bibfield  {author} {\bibinfo {author} {\bibfnamefont {M.}~\bibnamefont
  {Bukov}}, \bibinfo {author} {\bibfnamefont {L.}~\bibnamefont {D’Alessio}},\
  and\ \bibinfo {author} {\bibfnamefont {A.}~\bibnamefont {Polkovnikov}},\
  }\bibfield  {title} {\bibinfo {title} {Universal high-frequency behavior of
  periodically driven systems: from dynamical stabilization to floquet
  engineering},\ }\href {https://doi.org/10.1080/00018732.2015.1055918}
  {\bibfield  {journal} {\bibinfo  {journal} {Advances in Physics}\ }\textbf
  {\bibinfo {volume} {64}},\ \bibinfo {pages} {139–226} (\bibinfo {year}
  {2015})}\BibitemShut {NoStop}%
\bibitem [{\citenamefont {Eckardt}\ and\ \citenamefont
  {Anisimovas}(2015)}]{Eckardt_2015}%
  \BibitemOpen
  \bibfield  {author} {\bibinfo {author} {\bibfnamefont {A.}~\bibnamefont
  {Eckardt}}\ and\ \bibinfo {author} {\bibfnamefont {E.}~\bibnamefont
  {Anisimovas}},\ }\bibfield  {title} {\bibinfo {title} {High-frequency
  approximation for periodically driven quantum systems from a floquet-space
  perspective},\ }\href {https://doi.org/10.1088/1367-2630/17/9/093039}
  {\bibfield  {journal} {\bibinfo  {journal} {New Journal of Physics}\ }\textbf
  {\bibinfo {volume} {17}},\ \bibinfo {pages} {093039} (\bibinfo {year}
  {2015})}\BibitemShut {NoStop}%
\bibitem [{\citenamefont {Fel'dman}(1984)}]{Feldm1984}%
  \BibitemOpen
  \bibfield  {author} {\bibinfo {author} {\bibfnamefont {E.}~\bibnamefont
  {Fel'dman}},\ }\bibfield  {title} {\bibinfo {title} {On the convergence of
  the {Magnus} expansion for spin systems in periodic magnetic fields},\ }\href
  {https://doi.org/http://dx.doi.org/10.1016/0375-9601(84)90027-6} {\bibfield
  {journal} {\bibinfo  {journal} {Phys. Lett. A}\ }\textbf {\bibinfo {volume}
  {104}},\ \bibinfo {pages} {479} (\bibinfo {year} {1984})}\BibitemShut
  {NoStop}%
\bibitem [{\citenamefont {Magnus}(1954)}]{Magnus1954}%
  \BibitemOpen
  \bibfield  {author} {\bibinfo {author} {\bibfnamefont {W.}~\bibnamefont
  {Magnus}},\ }\bibfield  {title} {\bibinfo {title} {On the exponential
  solution of differential equations for a linear operator},\ }\href
  {https://doi.org/10.1002/cpa.3160070404} {\bibfield  {journal} {\bibinfo
  {journal} {Commun. Pure Appl. Math.}\ }\textbf {\bibinfo {volume} {7}},\
  \bibinfo {pages} {649} (\bibinfo {year} {1954})}\BibitemShut {NoStop}%
\bibitem [{\citenamefont {Abanin}\ \emph {et~al.}(2017)\citenamefont {Abanin},
  \citenamefont {De~Roeck}, \citenamefont {Ho},\ and\ \citenamefont
  {Huveneers}}]{PhysRevB.95.014112}%
  \BibitemOpen
  \bibfield  {author} {\bibinfo {author} {\bibfnamefont {D.~A.}\ \bibnamefont
  {Abanin}}, \bibinfo {author} {\bibfnamefont {W.}~\bibnamefont {De~Roeck}},
  \bibinfo {author} {\bibfnamefont {W.~W.}\ \bibnamefont {Ho}},\ and\ \bibinfo
  {author} {\bibfnamefont {F.~m.~c.}\ \bibnamefont {Huveneers}},\ }\bibfield
  {title} {\bibinfo {title} {Effective {Hamiltonians}, prethermalization, and
  slow energy absorption in periodically driven many-body systems},\ }\href
  {https://doi.org/10.1103/PhysRevB.95.014112} {\bibfield  {journal} {\bibinfo
  {journal} {Phys. Rev. B}\ }\textbf {\bibinfo {volume} {95}},\ \bibinfo
  {pages} {014112} (\bibinfo {year} {2017})}\BibitemShut {NoStop}%
\bibitem [{\citenamefont {Goldman}\ and\ \citenamefont
  {Dalibard}(2014)}]{PhysRevX.4.031027}%
  \BibitemOpen
  \bibfield  {author} {\bibinfo {author} {\bibfnamefont {N.}~\bibnamefont
  {Goldman}}\ and\ \bibinfo {author} {\bibfnamefont {J.}~\bibnamefont
  {Dalibard}},\ }\bibfield  {title} {\bibinfo {title} {Periodically driven
  quantum systems: Effective hamiltonians and engineered gauge fields},\ }\href
  {https://doi.org/10.1103/PhysRevX.4.031027} {\bibfield  {journal} {\bibinfo
  {journal} {Phys. Rev. X}\ }\textbf {\bibinfo {volume} {4}},\ \bibinfo {pages}
  {031027} (\bibinfo {year} {2014})}\BibitemShut {NoStop}%
\bibitem [{\citenamefont {Itin}\ and\ \citenamefont
  {Katsnelson}(2015)}]{PhysRevLett.115.075301}%
  \BibitemOpen
  \bibfield  {author} {\bibinfo {author} {\bibfnamefont {A.~P.}\ \bibnamefont
  {Itin}}\ and\ \bibinfo {author} {\bibfnamefont {M.~I.}\ \bibnamefont
  {Katsnelson}},\ }\bibfield  {title} {\bibinfo {title} {Effective
  {Hamiltonians} for rapidly driven many-body lattice systems: Induced exchange
  interactions and density-dependent hoppings},\ }\href
  {https://doi.org/10.1103/PhysRevLett.115.075301} {\bibfield  {journal}
  {\bibinfo  {journal} {Phys. Rev. Lett.}\ }\textbf {\bibinfo {volume} {115}},\
  \bibinfo {pages} {075301} (\bibinfo {year} {2015})}\BibitemShut {NoStop}%
\bibitem [{\citenamefont {Mikami}\ \emph {et~al.}(2016)\citenamefont {Mikami},
  \citenamefont {Kitamura}, \citenamefont {Yasuda}, \citenamefont {Tsuji},
  \citenamefont {Oka},\ and\ \citenamefont {Aoki}}]{PhysRevB.93.144307}%
  \BibitemOpen
  \bibfield  {author} {\bibinfo {author} {\bibfnamefont {T.}~\bibnamefont
  {Mikami}}, \bibinfo {author} {\bibfnamefont {S.}~\bibnamefont {Kitamura}},
  \bibinfo {author} {\bibfnamefont {K.}~\bibnamefont {Yasuda}}, \bibinfo
  {author} {\bibfnamefont {N.}~\bibnamefont {Tsuji}}, \bibinfo {author}
  {\bibfnamefont {T.}~\bibnamefont {Oka}},\ and\ \bibinfo {author}
  {\bibfnamefont {H.}~\bibnamefont {Aoki}},\ }\bibfield  {title} {\bibinfo
  {title} {{Brillouin-Wigner} theory for high-frequency expansion in
  periodically driven systems: Application to {Floquet} topological
  insulators},\ }\href {https://doi.org/10.1103/PhysRevB.93.144307} {\bibfield
  {journal} {\bibinfo  {journal} {Phys. Rev. B}\ }\textbf {\bibinfo {volume}
  {93}},\ \bibinfo {pages} {144307} (\bibinfo {year} {2016})}\BibitemShut
  {NoStop}%
\bibitem [{\citenamefont {Mohan}\ \emph {et~al.}(2016)\citenamefont {Mohan},
  \citenamefont {Saxena}, \citenamefont {Kundu},\ and\ \citenamefont
  {Rao}}]{PhysRevB.94.235419}%
  \BibitemOpen
  \bibfield  {author} {\bibinfo {author} {\bibfnamefont {P.}~\bibnamefont
  {Mohan}}, \bibinfo {author} {\bibfnamefont {R.}~\bibnamefont {Saxena}},
  \bibinfo {author} {\bibfnamefont {A.}~\bibnamefont {Kundu}},\ and\ \bibinfo
  {author} {\bibfnamefont {S.}~\bibnamefont {Rao}},\ }\bibfield  {title}
  {\bibinfo {title} {{Brillouin-Wigner} theory for {Floquet} topological phase
  transitions in spin-orbit-coupled materials},\ }\href
  {https://doi.org/10.1103/PhysRevB.94.235419} {\bibfield  {journal} {\bibinfo
  {journal} {Phys. Rev. B}\ }\textbf {\bibinfo {volume} {94}},\ \bibinfo
  {pages} {235419} (\bibinfo {year} {2016})}\BibitemShut {NoStop}%
\bibitem [{\citenamefont {Bukov}\ \emph {et~al.}(2016)\citenamefont {Bukov},
  \citenamefont {Kolodrubetz},\ and\ \citenamefont
  {Polkovnikov}}]{PhysRevLett.116.125301}%
  \BibitemOpen
  \bibfield  {author} {\bibinfo {author} {\bibfnamefont {M.}~\bibnamefont
  {Bukov}}, \bibinfo {author} {\bibfnamefont {M.}~\bibnamefont {Kolodrubetz}},\
  and\ \bibinfo {author} {\bibfnamefont {A.}~\bibnamefont {Polkovnikov}},\
  }\bibfield  {title} {\bibinfo {title} {{Schrieffer-Wolff} transformation for
  periodically driven systems: Strongly correlated systems with artificial
  gauge fields},\ }\href {https://doi.org/10.1103/PhysRevLett.116.125301}
  {\bibfield  {journal} {\bibinfo  {journal} {Phys. Rev. Lett.}\ }\textbf
  {\bibinfo {volume} {116}},\ \bibinfo {pages} {125301} (\bibinfo {year}
  {2016})}\BibitemShut {NoStop}%
\bibitem [{\citenamefont {Vogl}\ \emph
  {et~al.}(2019{\natexlab{a}})\citenamefont {Vogl}, \citenamefont {Laurell},
  \citenamefont {Barr},\ and\ \citenamefont {Fiete}}]{Vogl_2019AnalogHJ}%
  \BibitemOpen
  \bibfield  {author} {\bibinfo {author} {\bibfnamefont {M.}~\bibnamefont
  {Vogl}}, \bibinfo {author} {\bibfnamefont {P.}~\bibnamefont {Laurell}},
  \bibinfo {author} {\bibfnamefont {A.~D.}\ \bibnamefont {Barr}},\ and\
  \bibinfo {author} {\bibfnamefont {G.~A.}\ \bibnamefont {Fiete}},\ }\bibfield
  {title} {\bibinfo {title} {Analog of hamilton-jacobi theory for the
  time-evolution operator},\ }\bibfield  {journal} {\bibinfo  {journal}
  {Physical Review A}\ }\textbf {\bibinfo {volume} {100}},\ \href
  {https://doi.org/10.1103/physreva.100.012132} {10.1103/physreva.100.012132}
  (\bibinfo {year} {2019}{\natexlab{a}})\BibitemShut {NoStop}%
\bibitem [{\citenamefont {Verdeny}\ \emph {et~al.}(2013)\citenamefont
  {Verdeny}, \citenamefont {Mielke},\ and\ \citenamefont
  {Mintert}}]{verdeny2013}%
  \BibitemOpen
  \bibfield  {author} {\bibinfo {author} {\bibfnamefont {A.}~\bibnamefont
  {Verdeny}}, \bibinfo {author} {\bibfnamefont {A.}~\bibnamefont {Mielke}},\
  and\ \bibinfo {author} {\bibfnamefont {F.}~\bibnamefont {Mintert}},\
  }\bibfield  {title} {\bibinfo {title} {Accurate effective hamiltonians via
  unitary flow in floquet space},\ }\href
  {https://doi.org/10.1103/PhysRevLett.111.175301} {\bibfield  {journal}
  {\bibinfo  {journal} {Phys. Rev. Lett.}\ }\textbf {\bibinfo {volume} {111}},\
  \bibinfo {pages} {175301} (\bibinfo {year} {2013})}\BibitemShut {NoStop}%
\bibitem [{\citenamefont {G\'omez-Le\'on}\ and\ \citenamefont
  {Platero}(2013)}]{PhysRevLett.110.200403}%
  \BibitemOpen
  \bibfield  {author} {\bibinfo {author} {\bibfnamefont {A.}~\bibnamefont
  {G\'omez-Le\'on}}\ and\ \bibinfo {author} {\bibfnamefont {G.}~\bibnamefont
  {Platero}},\ }\bibfield  {title} {\bibinfo {title} {Floquet-bloch theory and
  topology in periodically driven lattices},\ }\href
  {https://doi.org/10.1103/PhysRevLett.110.200403} {\bibfield  {journal}
  {\bibinfo  {journal} {Phys. Rev. Lett.}\ }\textbf {\bibinfo {volume} {110}},\
  \bibinfo {pages} {200403} (\bibinfo {year} {2013})}\BibitemShut {NoStop}%
\bibitem [{\citenamefont {Vogl}\ \emph
  {et~al.}(2020{\natexlab{a}})\citenamefont {Vogl}, \citenamefont
  {Rodriguez-Vega},\ and\ \citenamefont {Fiete}}]{Vogl2020_effham}%
  \BibitemOpen
  \bibfield  {author} {\bibinfo {author} {\bibfnamefont {M.}~\bibnamefont
  {Vogl}}, \bibinfo {author} {\bibfnamefont {M.}~\bibnamefont
  {Rodriguez-Vega}},\ and\ \bibinfo {author} {\bibfnamefont {G.~A.}\
  \bibnamefont {Fiete}},\ }\bibfield  {title} {\bibinfo {title} {Effective
  floquet hamiltonian in the low-frequency regime},\ }\href
  {https://doi.org/10.1103/PhysRevB.101.024303} {\bibfield  {journal} {\bibinfo
   {journal} {Phys. Rev. B}\ }\textbf {\bibinfo {volume} {101}},\ \bibinfo
  {pages} {024303} (\bibinfo {year} {2020}{\natexlab{a}})}\BibitemShut
  {NoStop}%
\bibitem [{\citenamefont {Vogl}\ \emph
  {et~al.}(2019{\natexlab{b}})\citenamefont {Vogl}, \citenamefont {Laurell},
  \citenamefont {Barr},\ and\ \citenamefont {Fiete}}]{vogl2019}%
  \BibitemOpen
  \bibfield  {author} {\bibinfo {author} {\bibfnamefont {M.}~\bibnamefont
  {Vogl}}, \bibinfo {author} {\bibfnamefont {P.}~\bibnamefont {Laurell}},
  \bibinfo {author} {\bibfnamefont {A.~D.}\ \bibnamefont {Barr}},\ and\
  \bibinfo {author} {\bibfnamefont {G.~A.}\ \bibnamefont {Fiete}},\ }\bibfield
  {title} {\bibinfo {title} {Flow equation approach to periodically driven
  quantum systems},\ }\href {https://doi.org/10.1103/PhysRevX.9.021037}
  {\bibfield  {journal} {\bibinfo  {journal} {Phys. Rev. X}\ }\textbf {\bibinfo
  {volume} {9}},\ \bibinfo {pages} {021037} (\bibinfo {year}
  {2019}{\natexlab{b}})}\BibitemShut {NoStop}%
\bibitem [{\citenamefont {Rodriguez-Vega}\ \emph {et~al.}(2018)\citenamefont
  {Rodriguez-Vega}, \citenamefont {Lentz},\ and\ \citenamefont
  {Seradjeh}}]{Rodriguez_Vega_2018}%
  \BibitemOpen
  \bibfield  {author} {\bibinfo {author} {\bibfnamefont {M.}~\bibnamefont
  {Rodriguez-Vega}}, \bibinfo {author} {\bibfnamefont {M.}~\bibnamefont
  {Lentz}},\ and\ \bibinfo {author} {\bibfnamefont {B.}~\bibnamefont
  {Seradjeh}},\ }\bibfield  {title} {\bibinfo {title} {Floquet perturbation
  theory: formalism and application to low-frequency limit},\ }\href
  {https://doi.org/10.1088/1367-2630/aade37} {\bibfield  {journal} {\bibinfo
  {journal} {New Journal of Physics}\ }\textbf {\bibinfo {volume} {20}},\
  \bibinfo {pages} {093022} (\bibinfo {year} {2018})}\BibitemShut {NoStop}%
\bibitem [{\citenamefont {Rodriguez-Vega}\ and\ \citenamefont
  {Seradjeh}(2018)}]{PhysRevLett.121.036402}%
  \BibitemOpen
  \bibfield  {author} {\bibinfo {author} {\bibfnamefont {M.}~\bibnamefont
  {Rodriguez-Vega}}\ and\ \bibinfo {author} {\bibfnamefont {B.}~\bibnamefont
  {Seradjeh}},\ }\bibfield  {title} {\bibinfo {title} {Universal fluctuations
  of floquet topological invariants at low frequencies},\ }\href
  {https://doi.org/10.1103/PhysRevLett.121.036402} {\bibfield  {journal}
  {\bibinfo  {journal} {Phys. Rev. Lett.}\ }\textbf {\bibinfo {volume} {121}},\
  \bibinfo {pages} {036402} (\bibinfo {year} {2018})}\BibitemShut {NoStop}%
\bibitem [{\citenamefont {Martiskainen}\ and\ \citenamefont
  {Moiseyev}(2015)}]{Martiskainen2015}%
  \BibitemOpen
  \bibfield  {author} {\bibinfo {author} {\bibfnamefont {H.}~\bibnamefont
  {Martiskainen}}\ and\ \bibinfo {author} {\bibfnamefont {N.}~\bibnamefont
  {Moiseyev}},\ }\bibfield  {title} {\bibinfo {title} {Perturbation theory for
  quasienergy floquet solutions in the low-frequency regime of the oscillating
  electric field},\ }\href {https://doi.org/10.1103/PhysRevA.91.023416}
  {\bibfield  {journal} {\bibinfo  {journal} {Phys. Rev. A}\ }\textbf {\bibinfo
  {volume} {91}},\ \bibinfo {pages} {023416} (\bibinfo {year}
  {2015})}\BibitemShut {NoStop}%
\bibitem [{\citenamefont {Rigolin}\ \emph {et~al.}(2008)\citenamefont
  {Rigolin}, \citenamefont {Ortiz},\ and\ \citenamefont {Ponce}}]{rigolin2008}%
  \BibitemOpen
  \bibfield  {author} {\bibinfo {author} {\bibfnamefont {G.}~\bibnamefont
  {Rigolin}}, \bibinfo {author} {\bibfnamefont {G.}~\bibnamefont {Ortiz}},\
  and\ \bibinfo {author} {\bibfnamefont {V.~H.}\ \bibnamefont {Ponce}},\
  }\bibfield  {title} {\bibinfo {title} {Beyond the quantum adiabatic
  approximation: Adiabatic perturbation theory},\ }\href
  {https://doi.org/10.1103/PhysRevA.78.052508} {\bibfield  {journal} {\bibinfo
  {journal} {Phys. Rev. A}\ }\textbf {\bibinfo {volume} {78}},\ \bibinfo
  {pages} {052508} (\bibinfo {year} {2008})}\BibitemShut {NoStop}%
\bibitem [{\citenamefont {Weinberg}\ \emph {et~al.}(2015)\citenamefont
  {Weinberg}, \citenamefont {\"Olschl\"ager}, \citenamefont {Str\"ater},
  \citenamefont {Prelle}, \citenamefont {Eckardt}, \citenamefont {Sengstock},\
  and\ \citenamefont {Simonet}}]{weinberg2015}%
  \BibitemOpen
  \bibfield  {author} {\bibinfo {author} {\bibfnamefont {M.}~\bibnamefont
  {Weinberg}}, \bibinfo {author} {\bibfnamefont {C.}~\bibnamefont
  {\"Olschl\"ager}}, \bibinfo {author} {\bibfnamefont {C.}~\bibnamefont
  {Str\"ater}}, \bibinfo {author} {\bibfnamefont {S.}~\bibnamefont {Prelle}},
  \bibinfo {author} {\bibfnamefont {A.}~\bibnamefont {Eckardt}}, \bibinfo
  {author} {\bibfnamefont {K.}~\bibnamefont {Sengstock}},\ and\ \bibinfo
  {author} {\bibfnamefont {J.}~\bibnamefont {Simonet}},\ }\bibfield  {title}
  {\bibinfo {title} {Multiphoton interband excitations of quantum gases in
  driven optical lattices},\ }\href
  {https://doi.org/10.1103/PhysRevA.92.043621} {\bibfield  {journal} {\bibinfo
  {journal} {Phys. Rev. A}\ }\textbf {\bibinfo {volume} {92}},\ \bibinfo
  {pages} {043621} (\bibinfo {year} {2015})}\BibitemShut {NoStop}%
\bibitem [{\citenamefont {Kennes}\ \emph
  {et~al.}(2019{\natexlab{a}})\citenamefont {Kennes}, \citenamefont {M\"uller},
  \citenamefont {Pletyukhov}, \citenamefont {Weber}, \citenamefont {Bruder},
  \citenamefont {Hassler}, \citenamefont {Klinovaja}, \citenamefont {Loss},\
  and\ \citenamefont {Schoeller}}]{PhysRevB.100.041103}%
  \BibitemOpen
  \bibfield  {author} {\bibinfo {author} {\bibfnamefont {D.~M.}\ \bibnamefont
  {Kennes}}, \bibinfo {author} {\bibfnamefont {N.}~\bibnamefont {M\"uller}},
  \bibinfo {author} {\bibfnamefont {M.}~\bibnamefont {Pletyukhov}}, \bibinfo
  {author} {\bibfnamefont {C.}~\bibnamefont {Weber}}, \bibinfo {author}
  {\bibfnamefont {C.}~\bibnamefont {Bruder}}, \bibinfo {author} {\bibfnamefont
  {F.}~\bibnamefont {Hassler}}, \bibinfo {author} {\bibfnamefont
  {J.}~\bibnamefont {Klinovaja}}, \bibinfo {author} {\bibfnamefont
  {D.}~\bibnamefont {Loss}},\ and\ \bibinfo {author} {\bibfnamefont
  {H.}~\bibnamefont {Schoeller}},\ }\bibfield  {title} {\bibinfo {title}
  {Chiral one-dimensional floquet topological insulators beyond the rotating
  wave approximation},\ }\href {https://doi.org/10.1103/PhysRevB.100.041103}
  {\bibfield  {journal} {\bibinfo  {journal} {Phys. Rev. B}\ }\textbf {\bibinfo
  {volume} {100}},\ \bibinfo {pages} {041103} (\bibinfo {year}
  {2019}{\natexlab{a}})}\BibitemShut {NoStop}%
\bibitem [{\citenamefont {Li}\ \emph {et~al.}(2017)\citenamefont {Li},
  \citenamefont {Lam},\ and\ \citenamefont {You}}]{PhysRevB.96.155438}%
  \BibitemOpen
  \bibfield  {author} {\bibinfo {author} {\bibfnamefont {Z.-Z.}\ \bibnamefont
  {Li}}, \bibinfo {author} {\bibfnamefont {C.-H.}\ \bibnamefont {Lam}},\ and\
  \bibinfo {author} {\bibfnamefont {J.~Q.}\ \bibnamefont {You}},\ }\bibfield
  {title} {\bibinfo {title} {Floquet engineering of long-range $p$-wave
  superconductivity: Beyond the high-frequency limit},\ }\href
  {https://doi.org/10.1103/PhysRevB.96.155438} {\bibfield  {journal} {\bibinfo
  {journal} {Phys. Rev. B}\ }\textbf {\bibinfo {volume} {96}},\ \bibinfo
  {pages} {155438} (\bibinfo {year} {2017})}\BibitemShut {NoStop}%
\bibitem [{\citenamefont {Kennes}\ \emph
  {et~al.}(2019{\natexlab{b}})\citenamefont {Kennes}, \citenamefont {M\"uller},
  \citenamefont {Pletyukhov}, \citenamefont {Weber}, \citenamefont {Bruder},
  \citenamefont {Hassler}, \citenamefont {Klinovaja}, \citenamefont {Loss},\
  and\ \citenamefont {Schoeller}}]{Kennes-Klinovaja-2019}%
  \BibitemOpen
  \bibfield  {author} {\bibinfo {author} {\bibfnamefont {D.~M.}\ \bibnamefont
  {Kennes}}, \bibinfo {author} {\bibfnamefont {N.}~\bibnamefont {M\"uller}},
  \bibinfo {author} {\bibfnamefont {M.}~\bibnamefont {Pletyukhov}}, \bibinfo
  {author} {\bibfnamefont {C.}~\bibnamefont {Weber}}, \bibinfo {author}
  {\bibfnamefont {C.}~\bibnamefont {Bruder}}, \bibinfo {author} {\bibfnamefont
  {F.}~\bibnamefont {Hassler}}, \bibinfo {author} {\bibfnamefont
  {J.}~\bibnamefont {Klinovaja}}, \bibinfo {author} {\bibfnamefont
  {D.}~\bibnamefont {Loss}},\ and\ \bibinfo {author} {\bibfnamefont
  {H.}~\bibnamefont {Schoeller}},\ }\bibfield  {title} {\bibinfo {title}
  {Chiral one-dimensional floquet topological insulators beyond the rotating
  wave approximation},\ }\href {https://doi.org/10.1103/PhysRevB.100.041103}
  {\bibfield  {journal} {\bibinfo  {journal} {Phys. Rev. B}\ }\textbf {\bibinfo
  {volume} {100}},\ \bibinfo {pages} {041103} (\bibinfo {year}
  {2019}{\natexlab{b}})}\BibitemShut {NoStop}%
\bibitem [{\citenamefont {M\"uller}\ \emph {et~al.}(2020)\citenamefont
  {M\"uller}, \citenamefont {Kennes}, \citenamefont {Klinovaja}, \citenamefont
  {Loss},\ and\ \citenamefont {Schoeller}}]{PhysRevB.101.155417}%
  \BibitemOpen
  \bibfield  {author} {\bibinfo {author} {\bibfnamefont {N.}~\bibnamefont
  {M\"uller}}, \bibinfo {author} {\bibfnamefont {D.~M.}\ \bibnamefont
  {Kennes}}, \bibinfo {author} {\bibfnamefont {J.}~\bibnamefont {Klinovaja}},
  \bibinfo {author} {\bibfnamefont {D.}~\bibnamefont {Loss}},\ and\ \bibinfo
  {author} {\bibfnamefont {H.}~\bibnamefont {Schoeller}},\ }\bibfield  {title}
  {\bibinfo {title} {Electronic transport in one-dimensional floquet
  topological insulators via topological and nontopological edge states},\
  }\href {https://doi.org/10.1103/PhysRevB.101.155417} {\bibfield  {journal}
  {\bibinfo  {journal} {Phys. Rev. B}\ }\textbf {\bibinfo {volume} {101}},\
  \bibinfo {pages} {155417} (\bibinfo {year} {2020})}\BibitemShut {NoStop}%
\bibitem [{\citenamefont {Katz}\ \emph {et~al.}(2020)\citenamefont {Katz},
  \citenamefont {Refael},\ and\ \citenamefont {Lindner}}]{katz2019floquet}%
  \BibitemOpen
  \bibfield  {author} {\bibinfo {author} {\bibfnamefont {O.}~\bibnamefont
  {Katz}}, \bibinfo {author} {\bibfnamefont {G.}~\bibnamefont {Refael}},\ and\
  \bibinfo {author} {\bibfnamefont {N.~H.}\ \bibnamefont {Lindner}},\
  }\bibfield  {title} {\bibinfo {title} {Optically induced flat bands in
  twisted bilayer graphene},\ }\bibfield  {journal} {\bibinfo  {journal}
  {Physical Review B}\ }\textbf {\bibinfo {volume} {102}},\ \href
  {https://doi.org/10.1103/physrevb.102.155123} {10.1103/physrevb.102.155123}
  (\bibinfo {year} {2020})\BibitemShut {NoStop}%
\bibitem [{\citenamefont {Vogl}\ \emph
  {et~al.}(2020{\natexlab{b}})\citenamefont {Vogl}, \citenamefont
  {Rodriguez-Vega},\ and\ \citenamefont {Fiete}}]{PhysRevB.101.241408}%
  \BibitemOpen
  \bibfield  {author} {\bibinfo {author} {\bibfnamefont {M.}~\bibnamefont
  {Vogl}}, \bibinfo {author} {\bibfnamefont {M.}~\bibnamefont
  {Rodriguez-Vega}},\ and\ \bibinfo {author} {\bibfnamefont {G.~A.}\
  \bibnamefont {Fiete}},\ }\bibfield  {title} {\bibinfo {title} {Floquet
  engineering of interlayer couplings: Tuning the magic angle of twisted
  bilayer graphene at the exit of a waveguide},\ }\href
  {https://doi.org/10.1103/PhysRevB.101.241408} {\bibfield  {journal} {\bibinfo
   {journal} {Phys. Rev. B}\ }\textbf {\bibinfo {volume} {101}},\ \bibinfo
  {pages} {241408} (\bibinfo {year} {2020}{\natexlab{b}})}\BibitemShut
  {NoStop}%
\bibitem [{\citenamefont {Vogl}\ \emph
  {et~al.}(2020{\natexlab{c}})\citenamefont {Vogl}, \citenamefont
  {Rodriguez-Vega},\ and\ \citenamefont {Fiete}}]{vogl2020effective}%
  \BibitemOpen
  \bibfield  {author} {\bibinfo {author} {\bibfnamefont {M.}~\bibnamefont
  {Vogl}}, \bibinfo {author} {\bibfnamefont {M.}~\bibnamefont
  {Rodriguez-Vega}},\ and\ \bibinfo {author} {\bibfnamefont {G.~A.}\
  \bibnamefont {Fiete}},\ }\bibfield  {title} {\bibinfo {title} {Effective
  floquet hamiltonians for periodically driven twisted bilayer graphene},\
  }\bibfield  {journal} {\bibinfo  {journal} {Physical Review B}\ }\textbf
  {\bibinfo {volume} {101}},\ \href
  {https://doi.org/10.1103/physrevb.101.235411} {10.1103/physrevb.101.235411}
  (\bibinfo {year} {2020}{\natexlab{c}})\BibitemShut {NoStop}%
\bibitem [{\citenamefont {Topp}\ \emph {et~al.}(2019)\citenamefont {Topp},
  \citenamefont {Jotzu}, \citenamefont {McIver}, \citenamefont {Xian},
  \citenamefont {Rubio},\ and\ \citenamefont {Sentef}}]{Topp_2019}%
  \BibitemOpen
  \bibfield  {author} {\bibinfo {author} {\bibfnamefont {G.~E.}\ \bibnamefont
  {Topp}}, \bibinfo {author} {\bibfnamefont {G.}~\bibnamefont {Jotzu}},
  \bibinfo {author} {\bibfnamefont {J.~W.}\ \bibnamefont {McIver}}, \bibinfo
  {author} {\bibfnamefont {L.}~\bibnamefont {Xian}}, \bibinfo {author}
  {\bibfnamefont {A.}~\bibnamefont {Rubio}},\ and\ \bibinfo {author}
  {\bibfnamefont {M.~A.}\ \bibnamefont {Sentef}},\ }\bibfield  {title}
  {\bibinfo {title} {Topological floquet engineering of twisted bilayer
  graphene},\ }\bibfield  {journal} {\bibinfo  {journal} {Physical Review
  Research}\ }\textbf {\bibinfo {volume} {1}},\ \href
  {https://doi.org/10.1103/physrevresearch.1.023031}
  {10.1103/physrevresearch.1.023031} (\bibinfo {year} {2019})\BibitemShut
  {NoStop}%
\bibitem [{\citenamefont {Vogl}\ \emph {et~al.}(2021)\citenamefont {Vogl},
  \citenamefont {Rodriguez-Vega}, \citenamefont {Flebus}, \citenamefont
  {MacDonald},\ and\ \citenamefont {Fiete}}]{PhysRevB.103.014310}%
  \BibitemOpen
  \bibfield  {author} {\bibinfo {author} {\bibfnamefont {M.}~\bibnamefont
  {Vogl}}, \bibinfo {author} {\bibfnamefont {M.}~\bibnamefont
  {Rodriguez-Vega}}, \bibinfo {author} {\bibfnamefont {B.}~\bibnamefont
  {Flebus}}, \bibinfo {author} {\bibfnamefont {A.~H.}\ \bibnamefont
  {MacDonald}},\ and\ \bibinfo {author} {\bibfnamefont {G.~A.}\ \bibnamefont
  {Fiete}},\ }\bibfield  {title} {\bibinfo {title} {Floquet engineering of
  topological transitions in a twisted transition metal dichalcogenide
  homobilayer},\ }\href {https://doi.org/10.1103/PhysRevB.103.014310}
  {\bibfield  {journal} {\bibinfo  {journal} {Phys. Rev. B}\ }\textbf {\bibinfo
  {volume} {103}},\ \bibinfo {pages} {014310} (\bibinfo {year}
  {2021})}\BibitemShut {NoStop}%
\bibitem [{\citenamefont {Rodriguez-Vega}\ \emph {et~al.}(2020)\citenamefont
  {Rodriguez-Vega}, \citenamefont {Vogl},\ and\ \citenamefont
  {Fiete}}]{rodriguezvega_2020a}%
  \BibitemOpen
  \bibfield  {author} {\bibinfo {author} {\bibfnamefont {M.}~\bibnamefont
  {Rodriguez-Vega}}, \bibinfo {author} {\bibfnamefont {M.}~\bibnamefont
  {Vogl}},\ and\ \bibinfo {author} {\bibfnamefont {G.~A.}\ \bibnamefont
  {Fiete}},\ }\bibfield  {title} {\bibinfo {title} {Floquet engineering of
  twisted double bilayer graphene},\ }\href
  {https://doi.org/10.1103/PhysRevResearch.2.033494} {\bibfield  {journal}
  {\bibinfo  {journal} {Phys. Rev. Research}\ }\textbf {\bibinfo {volume}
  {2}},\ \bibinfo {pages} {033494} (\bibinfo {year} {2020})}\BibitemShut
  {NoStop}%
\bibitem [{\citenamefont {Lu}\ \emph {et~al.}(2021)\citenamefont {Lu},
  \citenamefont {Zeng}, \citenamefont {Liu}, \citenamefont {Gao},\ and\
  \citenamefont {Xie}}]{PhysRevB.103.195146}%
  \BibitemOpen
  \bibfield  {author} {\bibinfo {author} {\bibfnamefont {M.}~\bibnamefont
  {Lu}}, \bibinfo {author} {\bibfnamefont {J.}~\bibnamefont {Zeng}}, \bibinfo
  {author} {\bibfnamefont {H.}~\bibnamefont {Liu}}, \bibinfo {author}
  {\bibfnamefont {J.-H.}\ \bibnamefont {Gao}},\ and\ \bibinfo {author}
  {\bibfnamefont {X.~C.}\ \bibnamefont {Xie}},\ }\bibfield  {title} {\bibinfo
  {title} {Valley-selective floquet chern flat bands in twisted multilayer
  graphene},\ }\href {https://doi.org/10.1103/PhysRevB.103.195146} {\bibfield
  {journal} {\bibinfo  {journal} {Phys. Rev. B}\ }\textbf {\bibinfo {volume}
  {103}},\ \bibinfo {pages} {195146} (\bibinfo {year} {2021})}\BibitemShut
  {NoStop}%
\bibitem [{\citenamefont {Li}\ \emph {et~al.}(2020)\citenamefont {Li},
  \citenamefont {Fertig},\ and\ \citenamefont {Seradjeh}}]{Li_2020}%
  \BibitemOpen
  \bibfield  {author} {\bibinfo {author} {\bibfnamefont {Y.}~\bibnamefont
  {Li}}, \bibinfo {author} {\bibfnamefont {H.~A.}\ \bibnamefont {Fertig}},\
  and\ \bibinfo {author} {\bibfnamefont {B.}~\bibnamefont {Seradjeh}},\
  }\bibfield  {title} {\bibinfo {title} {Floquet-engineered topological flat
  bands in irradiated twisted bilayer graphene},\ }\bibfield  {journal}
  {\bibinfo  {journal} {Physical Review Research}\ }\textbf {\bibinfo {volume}
  {2}},\ \href {https://doi.org/10.1103/physrevresearch.2.043275}
  {10.1103/physrevresearch.2.043275} (\bibinfo {year} {2020})\BibitemShut
  {NoStop}%
\bibitem [{\citenamefont {Park}\ \emph {et~al.}(2021)\citenamefont {Park},
  \citenamefont {Cao}, \citenamefont {Watanabe}, \citenamefont {Taniguchi},\
  and\ \citenamefont {Jarillo-Herrero}}]{Park2021}%
  \BibitemOpen
  \bibfield  {author} {\bibinfo {author} {\bibfnamefont {J.~M.}\ \bibnamefont
  {Park}}, \bibinfo {author} {\bibfnamefont {Y.}~\bibnamefont {Cao}}, \bibinfo
  {author} {\bibfnamefont {K.}~\bibnamefont {Watanabe}}, \bibinfo {author}
  {\bibfnamefont {T.}~\bibnamefont {Taniguchi}},\ and\ \bibinfo {author}
  {\bibfnamefont {P.}~\bibnamefont {Jarillo-Herrero}},\ }\bibfield  {title}
  {\bibinfo {title} {Tunable strongly coupled superconductivity in magic-angle
  twisted trilayer graphene},\ }\href
  {https://doi.org/10.1038/s41586-021-03192-0} {\bibfield  {journal} {\bibinfo
  {journal} {Nature}\ }\textbf {\bibinfo {volume} {590}},\ \bibinfo {pages}
  {249} (\bibinfo {year} {2021})}\BibitemShut {NoStop}%
\bibitem [{\citenamefont {Lake}\ and\ \citenamefont
  {Senthil}(2021)}]{lake2021reentrant}%
  \BibitemOpen
  \bibfield  {author} {\bibinfo {author} {\bibfnamefont {E.}~\bibnamefont
  {Lake}}\ and\ \bibinfo {author} {\bibfnamefont {T.}~\bibnamefont {Senthil}},\
  }\href@noop {} {\bibinfo {title} {Re-entrant superconductivity through a
  quantum lifshitz transition in twisted trilayer graphene}} (\bibinfo {year}
  {2021}),\ \Eprint {https://arxiv.org/abs/2104.13920} {arXiv:2104.13920
  [cond-mat.supr-con]} \BibitemShut {NoStop}%
\bibitem [{\citenamefont {Ramires}\ and\ \citenamefont
  {Lado}(2021)}]{ramires2021emulating}%
  \BibitemOpen
  \bibfield  {author} {\bibinfo {author} {\bibfnamefont {A.}~\bibnamefont
  {Ramires}}\ and\ \bibinfo {author} {\bibfnamefont {J.~L.}\ \bibnamefont
  {Lado}},\ }\href@noop {} {\bibinfo {title} {Emulating heavy fermions in
  twisted trilayer graphene}} (\bibinfo {year} {2021}),\ \Eprint
  {https://arxiv.org/abs/2102.03312} {arXiv:2102.03312 [cond-mat.mes-hall]}
  \BibitemShut {NoStop}%
\bibitem [{\citenamefont {Chou}\ \emph {et~al.}(2021)\citenamefont {Chou},
  \citenamefont {Wu}, \citenamefont {Sau},\ and\ \citenamefont
  {Sarma}}]{chou2021correlationinduced}%
  \BibitemOpen
  \bibfield  {author} {\bibinfo {author} {\bibfnamefont {Y.-Z.}\ \bibnamefont
  {Chou}}, \bibinfo {author} {\bibfnamefont {F.}~\bibnamefont {Wu}}, \bibinfo
  {author} {\bibfnamefont {J.~D.}\ \bibnamefont {Sau}},\ and\ \bibinfo {author}
  {\bibfnamefont {S.~D.}\ \bibnamefont {Sarma}},\ }\href@noop {} {\bibinfo
  {title} {Correlation-induced triplet pairing superconductivity in
  graphene-based moir\'e systems}} (\bibinfo {year} {2021}),\ \Eprint
  {https://arxiv.org/abs/2105.00561} {arXiv:2105.00561 [cond-mat.supr-con]}
  \BibitemShut {NoStop}%
\bibitem [{\citenamefont {Khalaf}\ \emph {et~al.}(2019)\citenamefont {Khalaf},
  \citenamefont {Kruchkov}, \citenamefont {Tarnopolsky},\ and\ \citenamefont
  {Vishwanath}}]{multilayerTG}%
  \BibitemOpen
  \bibfield  {author} {\bibinfo {author} {\bibfnamefont {E.}~\bibnamefont
  {Khalaf}}, \bibinfo {author} {\bibfnamefont {A.~J.}\ \bibnamefont
  {Kruchkov}}, \bibinfo {author} {\bibfnamefont {G.}~\bibnamefont
  {Tarnopolsky}},\ and\ \bibinfo {author} {\bibfnamefont {A.}~\bibnamefont
  {Vishwanath}},\ }\bibfield  {title} {\bibinfo {title} {Magic angle hierarchy
  in twisted graphene multilayers},\ }\href
  {https://doi.org/10.1103/PhysRevB.100.085109} {\bibfield  {journal} {\bibinfo
   {journal} {Phys. Rev. B}\ }\textbf {\bibinfo {volume} {100}},\ \bibinfo
  {pages} {085109} (\bibinfo {year} {2019})}\BibitemShut {NoStop}%
\bibitem [{\citenamefont {Li}\ \emph {et~al.}(2019)\citenamefont {Li},
  \citenamefont {Wu},\ and\ \citenamefont {MacDonald}}]{LiWuMcDonald}%
  \BibitemOpen
  \bibfield  {author} {\bibinfo {author} {\bibfnamefont {X.}~\bibnamefont
  {Li}}, \bibinfo {author} {\bibfnamefont {F.}~\bibnamefont {Wu}},\ and\
  \bibinfo {author} {\bibfnamefont {A.~H.}\ \bibnamefont {MacDonald}},\
  }\href@noop {} {\bibinfo {title} {Electronic structure of single-twist
  trilayer graphene}} (\bibinfo {year} {2019}),\ \Eprint
  {https://arxiv.org/abs/1907.12338} {arXiv:1907.12338 [cond-mat.mtrl-sci]}
  \BibitemShut {NoStop}%
\bibitem [{\citenamefont {McCann}(2012)}]{McCann2012}%
  \BibitemOpen
  \bibfield  {author} {\bibinfo {author} {\bibfnamefont {E.}~\bibnamefont
  {McCann}},\ }\bibinfo {title} {Electronic properties of monolayer and bilayer
  graphene},\ in\ \href {https://doi.org/10.1007/978-3-642-22984-8_8} {\emph
  {\bibinfo {booktitle} {Graphene Nanoelectronics: Metrology, Synthesis,
  Properties and Applications}}},\ \bibinfo {editor} {edited by\ \bibinfo
  {editor} {\bibfnamefont {H.}~\bibnamefont {Raza}}}\ (\bibinfo  {publisher}
  {Springer Berlin Heidelberg},\ \bibinfo {address} {Berlin, Heidelberg},\
  \bibinfo {year} {2012})\ pp.\ \bibinfo {pages} {237--275}\BibitemShut
  {NoStop}%
\bibitem [{\citenamefont {Vogl}\ \emph
  {et~al.}(2020{\natexlab{d}})\citenamefont {Vogl}, \citenamefont
  {Rodriguez-Vega},\ and\ \citenamefont {Fiete}}]{TBG2020}%
  \BibitemOpen
  \bibfield  {author} {\bibinfo {author} {\bibfnamefont {M.}~\bibnamefont
  {Vogl}}, \bibinfo {author} {\bibfnamefont {M.}~\bibnamefont
  {Rodriguez-Vega}},\ and\ \bibinfo {author} {\bibfnamefont {G.~A.}\
  \bibnamefont {Fiete}},\ }\bibfield  {title} {\bibinfo {title} {Effective
  floquet hamiltonians for periodically driven twisted bilayer graphene},\
  }\href {https://doi.org/10.1103/PhysRevB.101.235411} {\bibfield  {journal}
  {\bibinfo  {journal} {Phys. Rev. B}\ }\textbf {\bibinfo {volume} {101}},\
  \bibinfo {pages} {235411} (\bibinfo {year} {2020}{\natexlab{d}})}\BibitemShut
  {NoStop}%
\bibitem [{\citenamefont {Kumar}\ \emph {et~al.}(2021)\citenamefont {Kumar},
  \citenamefont {Herath},\ and\ \citenamefont {Apalkov}}]{Kumar_2021}%
  \BibitemOpen
  \bibfield  {author} {\bibinfo {author} {\bibfnamefont {P.}~\bibnamefont
  {Kumar}}, \bibinfo {author} {\bibfnamefont {T.~M.}\ \bibnamefont {Herath}},\
  and\ \bibinfo {author} {\bibfnamefont {V.}~\bibnamefont {Apalkov}},\
  }\bibfield  {title} {\bibinfo {title} {Bilayer graphene in strong ultrafast
  laser fields},\ }\href {https://doi.org/10.1088/1361-648x/ac0b1e} {\bibfield
  {journal} {\bibinfo  {journal} {Journal of Physics: Condensed Matter}\
  }\textbf {\bibinfo {volume} {33}},\ \bibinfo {pages} {335305} (\bibinfo
  {year} {2021})}\BibitemShut {NoStop}%
\bibitem [{\citenamefont {Fukui}\ \emph {et~al.}(2005)\citenamefont {Fukui},
  \citenamefont {Hatsugai},\ and\ \citenamefont {Suzuki}}]{ChernNo1}%
  \BibitemOpen
  \bibfield  {author} {\bibinfo {author} {\bibfnamefont {T.}~\bibnamefont
  {Fukui}}, \bibinfo {author} {\bibfnamefont {Y.}~\bibnamefont {Hatsugai}},\
  and\ \bibinfo {author} {\bibfnamefont {H.}~\bibnamefont {Suzuki}},\
  }\bibfield  {title} {\bibinfo {title} {Chern numbers in discretized brillouin
  zone: Efficient method of computing (spin) hall conductances},\ }\href
  {https://doi.org/10.1143/jpsj.74.1674} {\bibfield  {journal} {\bibinfo
  {journal} {Journal of the Physical Society of Japan}\ }\textbf {\bibinfo
  {volume} {74}},\ \bibinfo {pages} {1674} (\bibinfo {year}
  {2005})}\BibitemShut {NoStop}%
\bibitem [{\citenamefont {Dehghani}\ \emph {et~al.}(2015)\citenamefont
  {Dehghani}, \citenamefont {Oka},\ and\ \citenamefont
  {Mitra}}]{Dehghani_2015}%
  \BibitemOpen
  \bibfield  {author} {\bibinfo {author} {\bibfnamefont {H.}~\bibnamefont
  {Dehghani}}, \bibinfo {author} {\bibfnamefont {T.}~\bibnamefont {Oka}},\ and\
  \bibinfo {author} {\bibfnamefont {A.}~\bibnamefont {Mitra}},\ }\bibfield
  {title} {\bibinfo {title} {Out-of-equilibrium electrons and the hall
  conductance of a floquet topological insulator},\ }\bibfield  {journal}
  {\bibinfo  {journal} {Physical Review B}\ }\textbf {\bibinfo {volume} {91}},\
  \href {https://doi.org/10.1103/physrevb.91.155422}
  {10.1103/physrevb.91.155422} (\bibinfo {year} {2015})\BibitemShut {NoStop}%
\bibitem [{\citenamefont {Kumar}\ \emph {et~al.}(2020)\citenamefont {Kumar},
  \citenamefont {Rodriguez-Vega}, \citenamefont {Pereg-Barnea},\ and\
  \citenamefont {Seradjeh}}]{Kumar_2020}%
  \BibitemOpen
  \bibfield  {author} {\bibinfo {author} {\bibfnamefont {A.}~\bibnamefont
  {Kumar}}, \bibinfo {author} {\bibfnamefont {M.}~\bibnamefont
  {Rodriguez-Vega}}, \bibinfo {author} {\bibfnamefont {T.}~\bibnamefont
  {Pereg-Barnea}},\ and\ \bibinfo {author} {\bibfnamefont {B.}~\bibnamefont
  {Seradjeh}},\ }\bibfield  {title} {\bibinfo {title} {Linear response theory
  and optical conductivity of floquet topological insulators},\ }\bibfield
  {journal} {\bibinfo  {journal} {Physical Review B}\ }\textbf {\bibinfo
  {volume} {101}},\ \href {https://doi.org/10.1103/physrevb.101.174314}
  {10.1103/physrevb.101.174314} (\bibinfo {year} {2020})\BibitemShut {NoStop}%
\bibitem [{\citenamefont {Vogl}\ \emph
  {et~al.}(2020{\natexlab{e}})\citenamefont {Vogl}, \citenamefont
  {Rodriguez-Vega},\ and\ \citenamefont {Fiete}}]{Vogl_2020_interlayer}%
  \BibitemOpen
  \bibfield  {author} {\bibinfo {author} {\bibfnamefont {M.}~\bibnamefont
  {Vogl}}, \bibinfo {author} {\bibfnamefont {M.}~\bibnamefont
  {Rodriguez-Vega}},\ and\ \bibinfo {author} {\bibfnamefont {G.~A.}\
  \bibnamefont {Fiete}},\ }\bibfield  {title} {\bibinfo {title} {Floquet
  engineering of interlayer couplings: Tuning the magic angle of twisted
  bilayer graphene at the exit of a waveguide},\ }\bibfield  {journal}
  {\bibinfo  {journal} {Physical Review B}\ }\textbf {\bibinfo {volume}
  {101}},\ \href {https://doi.org/10.1103/physrevb.101.241408}
  {10.1103/physrevb.101.241408} (\bibinfo {year}
  {2020}{\natexlab{e}})\BibitemShut {NoStop}%
\bibitem [{\citenamefont {Tarnopolsky}\ \emph {et~al.}(2019)\citenamefont
  {Tarnopolsky}, \citenamefont {Kruchkov},\ and\ \citenamefont
  {Vishwanath}}]{Tarnopolsky_2019}%
  \BibitemOpen
  \bibfield  {author} {\bibinfo {author} {\bibfnamefont {G.}~\bibnamefont
  {Tarnopolsky}}, \bibinfo {author} {\bibfnamefont {A.~J.}\ \bibnamefont
  {Kruchkov}},\ and\ \bibinfo {author} {\bibfnamefont {A.}~\bibnamefont
  {Vishwanath}},\ }\bibfield  {title} {\bibinfo {title} {Origin of magic angles
  in twisted bilayer graphene},\ }\bibfield  {journal} {\bibinfo  {journal}
  {Physical Review Letters}\ }\textbf {\bibinfo {volume} {122}},\ \href
  {https://doi.org/10.1103/physrevlett.122.106405}
  {10.1103/physrevlett.122.106405} (\bibinfo {year} {2019})\BibitemShut
  {NoStop}%
\end{thebibliography}%

\appendix
\begin{widetext}
\section{Individual Chern numbers}
In this section, we reproduce the results in Figures \ref{fig:phaseI} \& \ref{fig:phaseII} where for each stacking configuration, we plot the variations of the Chern numbers for each of the six central bands separately. \\

\begin{figure}[!htbp]
\centering
    \includegraphics[width=0.5\linewidth]{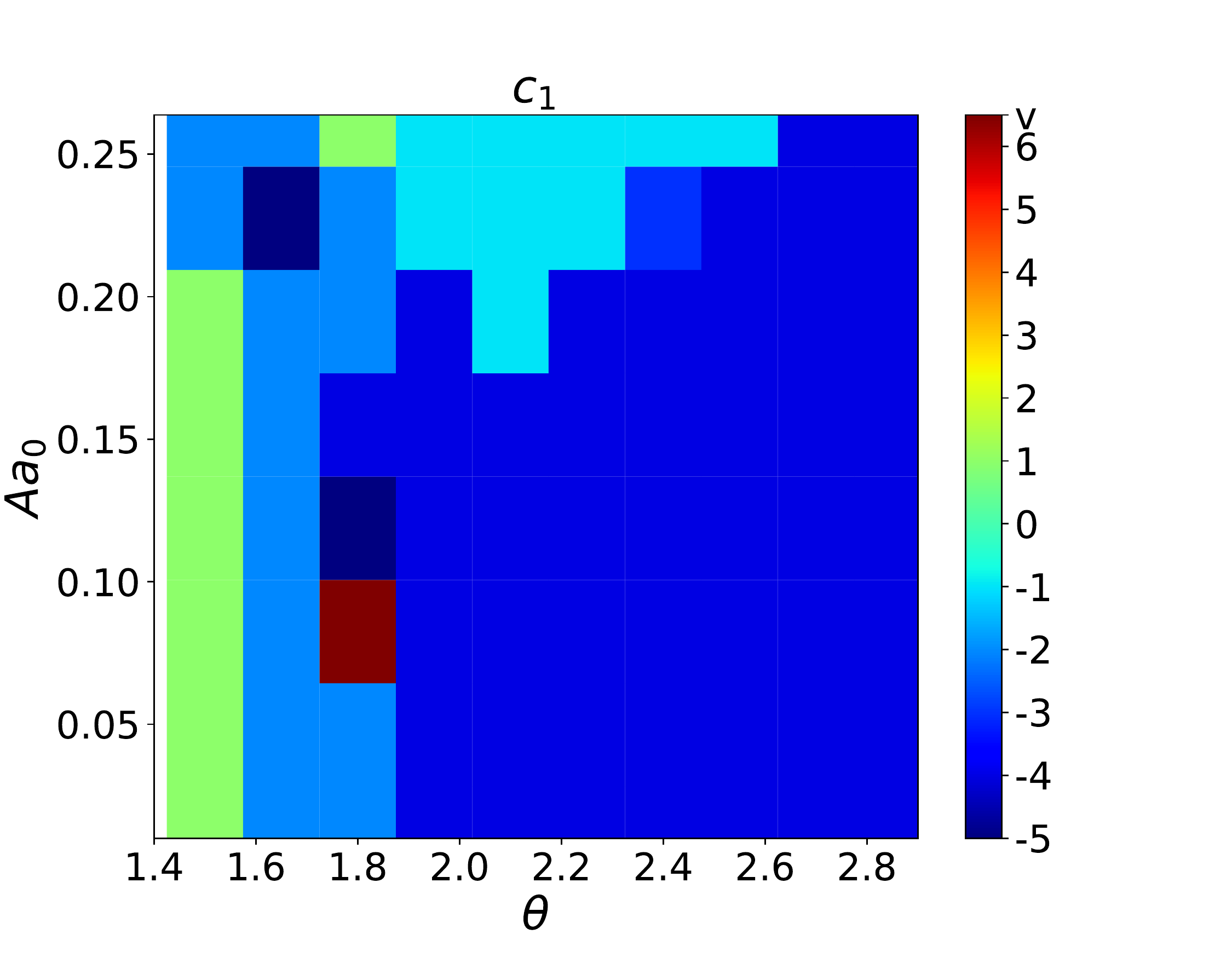}\hfill
    \includegraphics[width=0.5\linewidth]{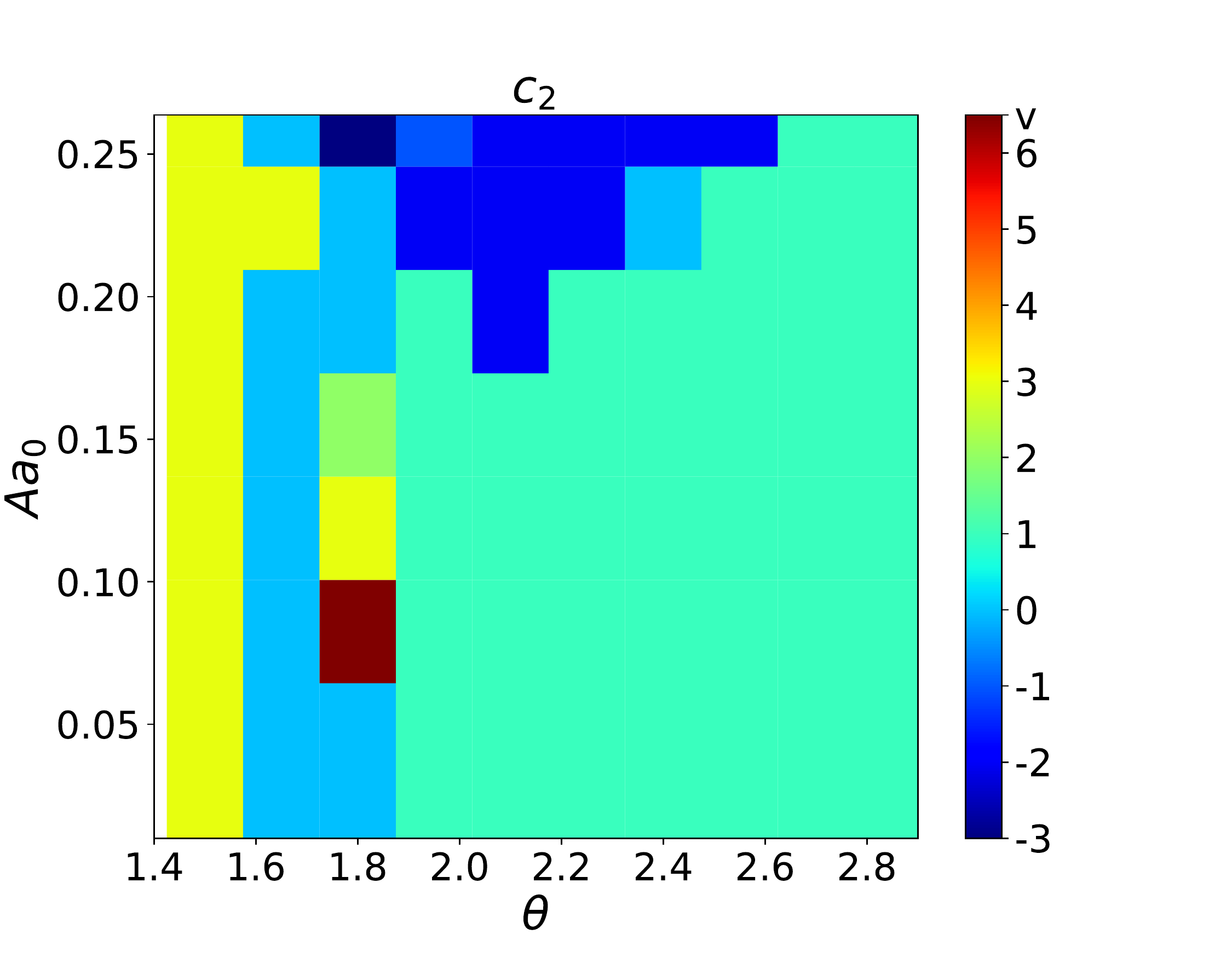}
    \hspace{0mm}
    \includegraphics[width=0.47\linewidth]{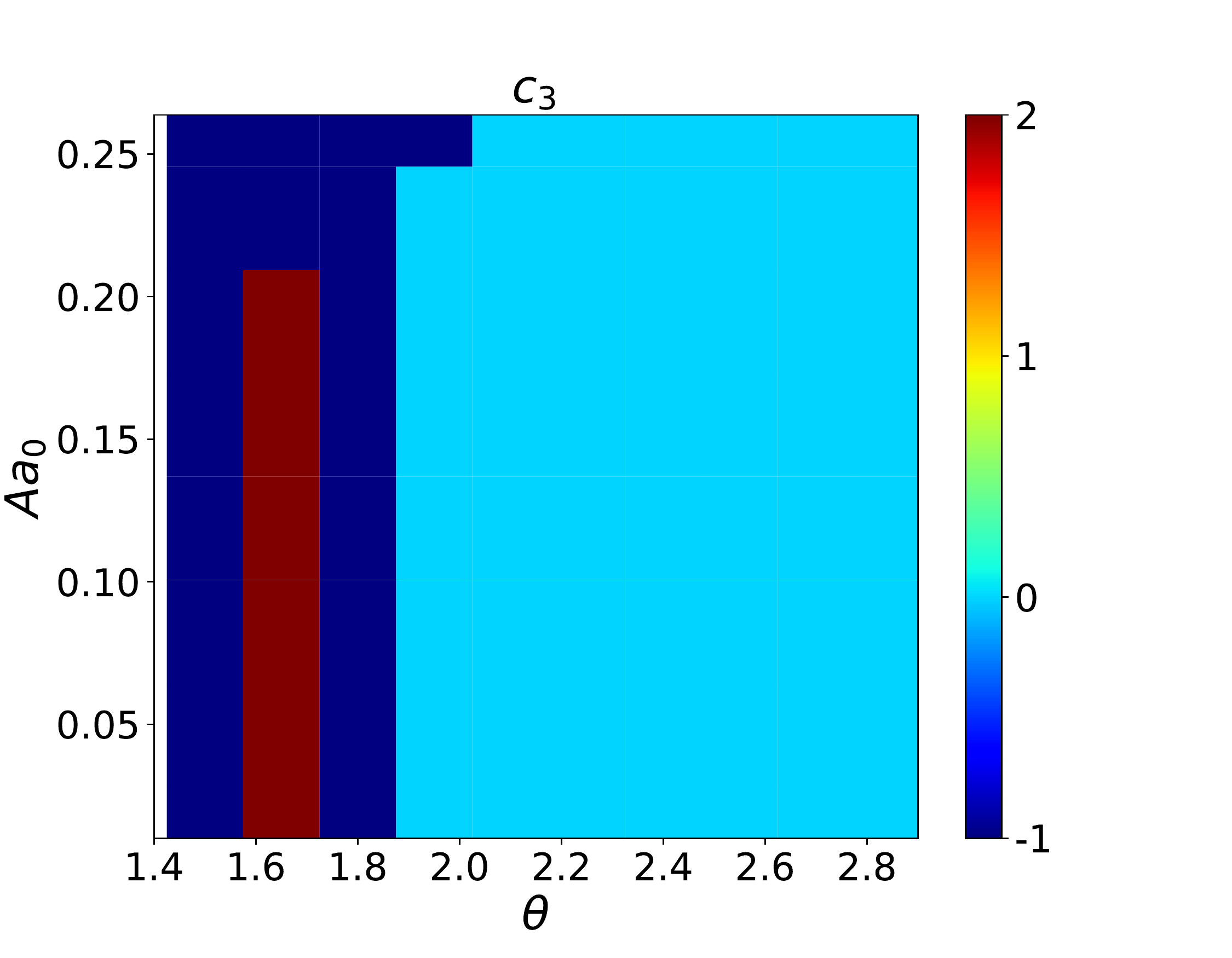}\hfill
    \includegraphics[width=0.47\linewidth]{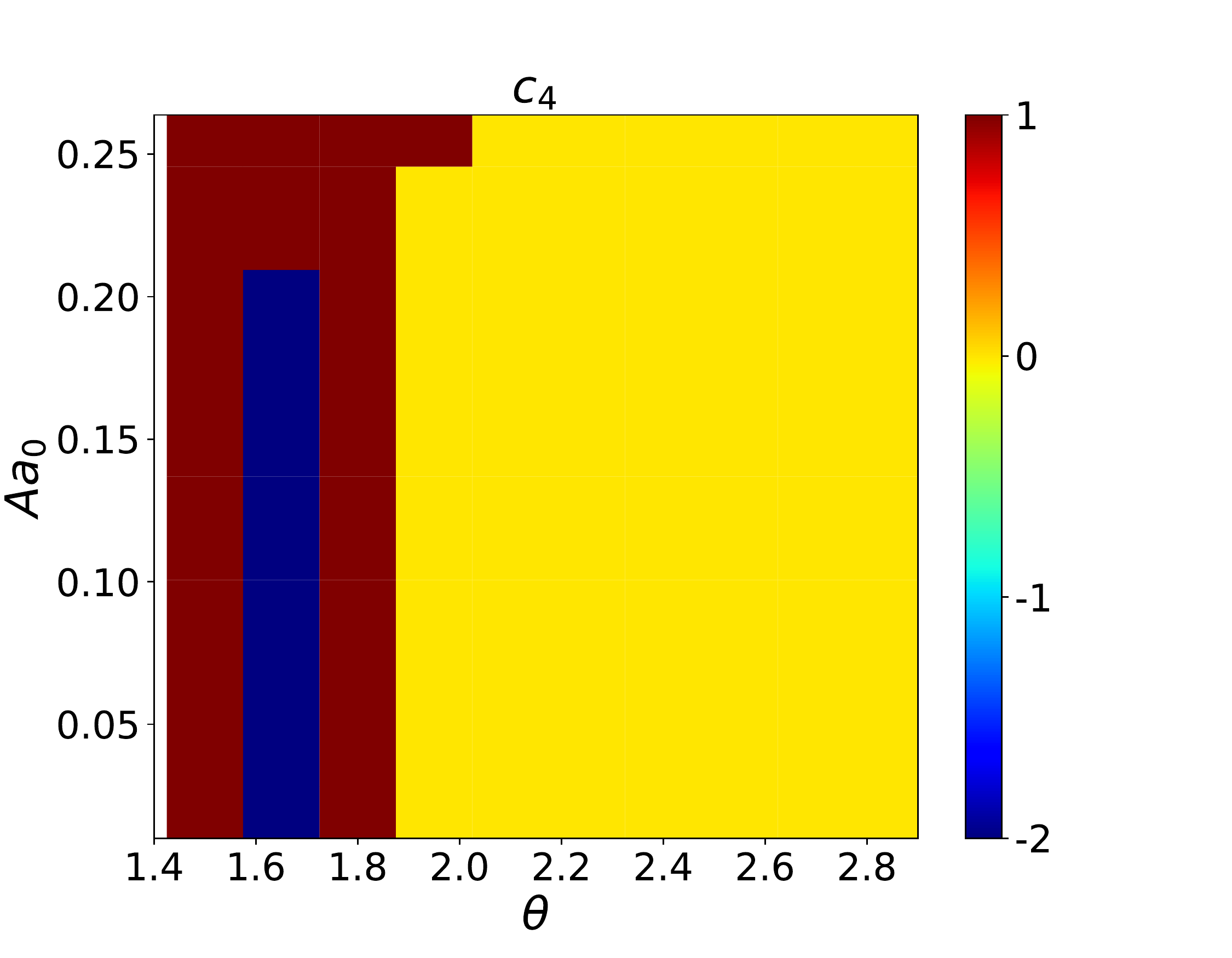}
    \hspace{0mm}
    \includegraphics[width=0.47\linewidth]{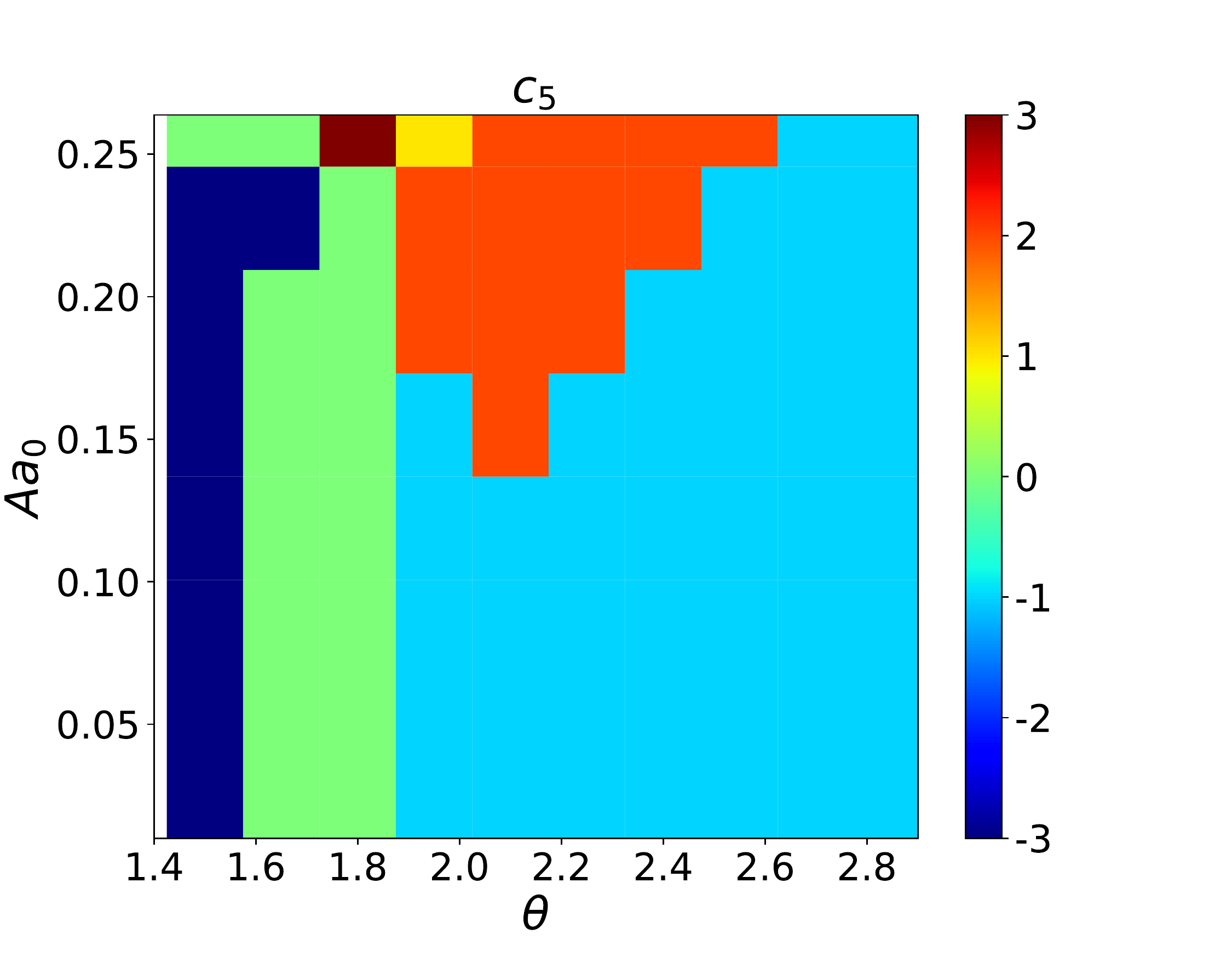}\hfill
    \includegraphics[width=0.47\linewidth]{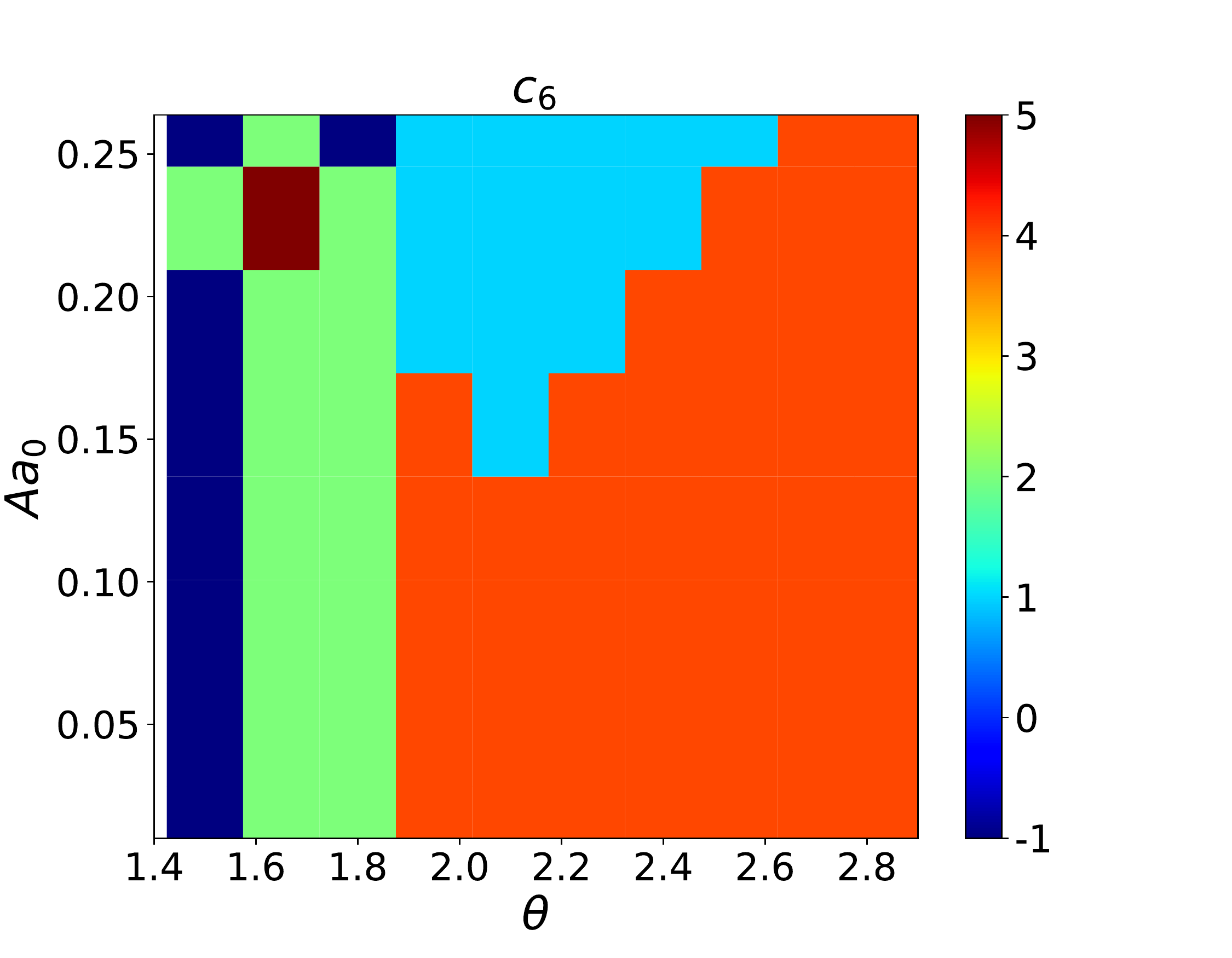}
\caption{(Color online)  The Chern number variation for the six central bands in the TTG system with the AAA stacking and top layer. This is associated with Fig. \ref{fig:phaseI} (a).}
\label{fig:chrn_AAA_TT}
\end{figure}

\begin{figure}[!htbp]
\centering
    \includegraphics[width=0.5\linewidth]{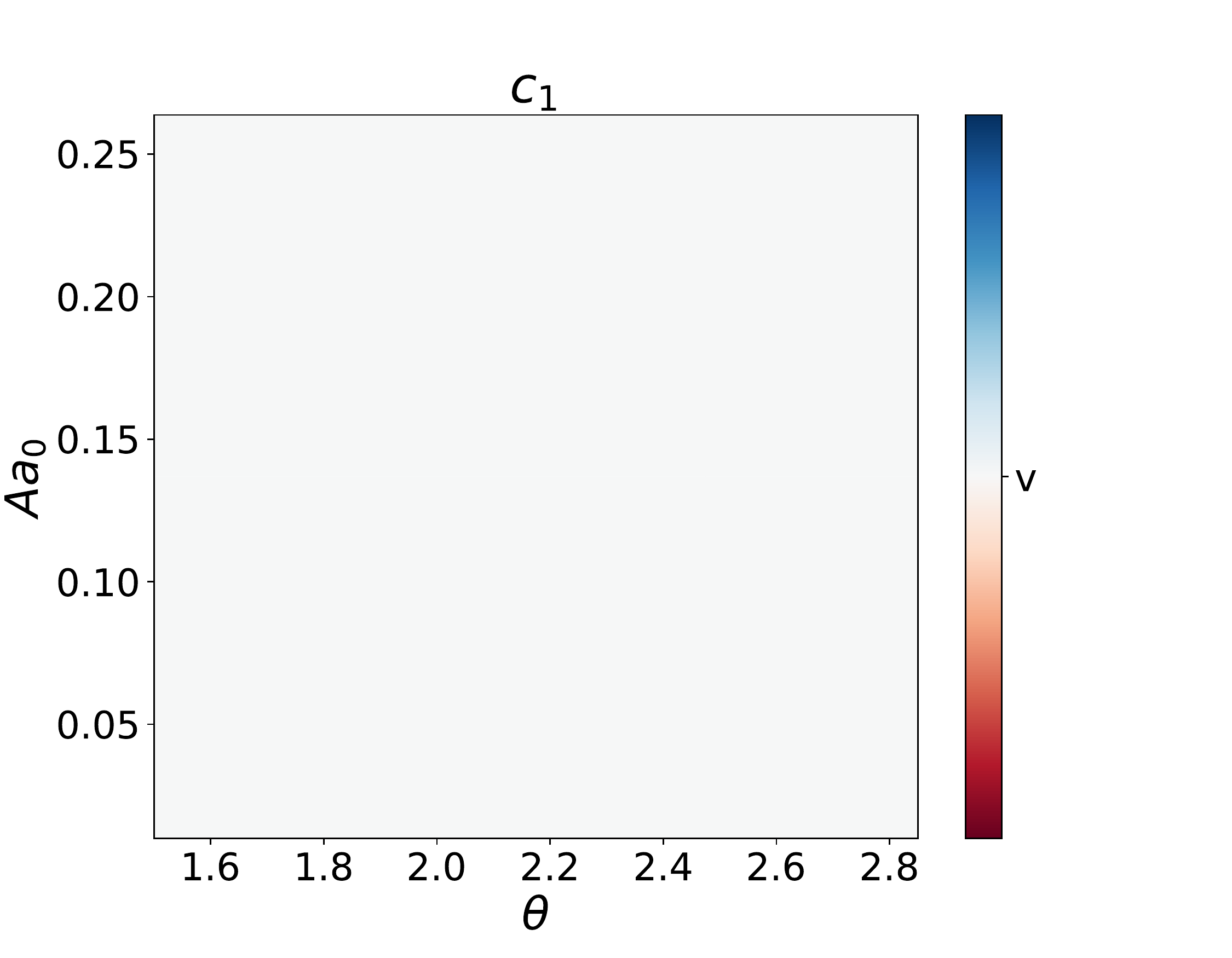}\hfill
    \includegraphics[width=0.5\linewidth]{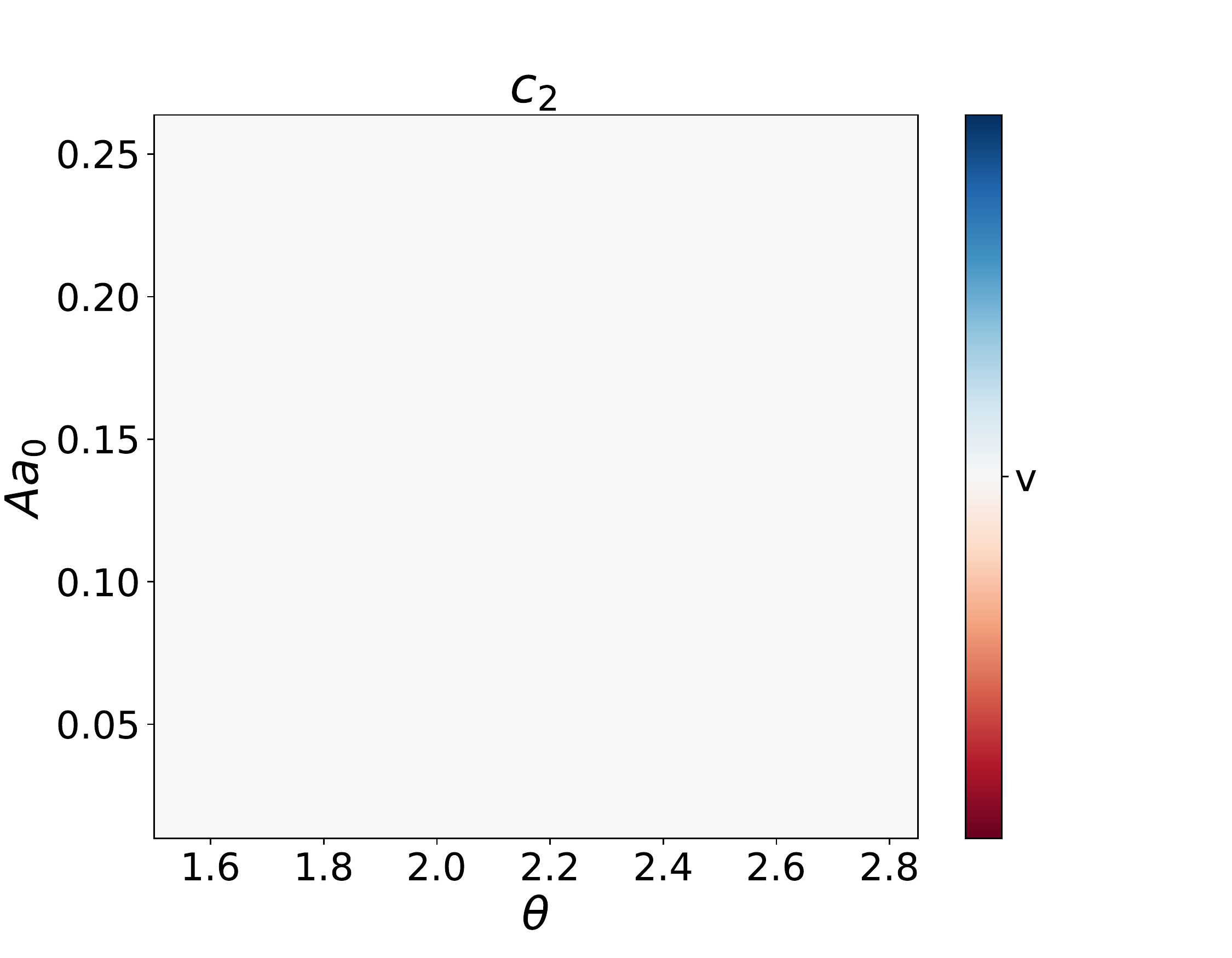}
    \hspace{0mm}
    \includegraphics[width=0.47\linewidth]{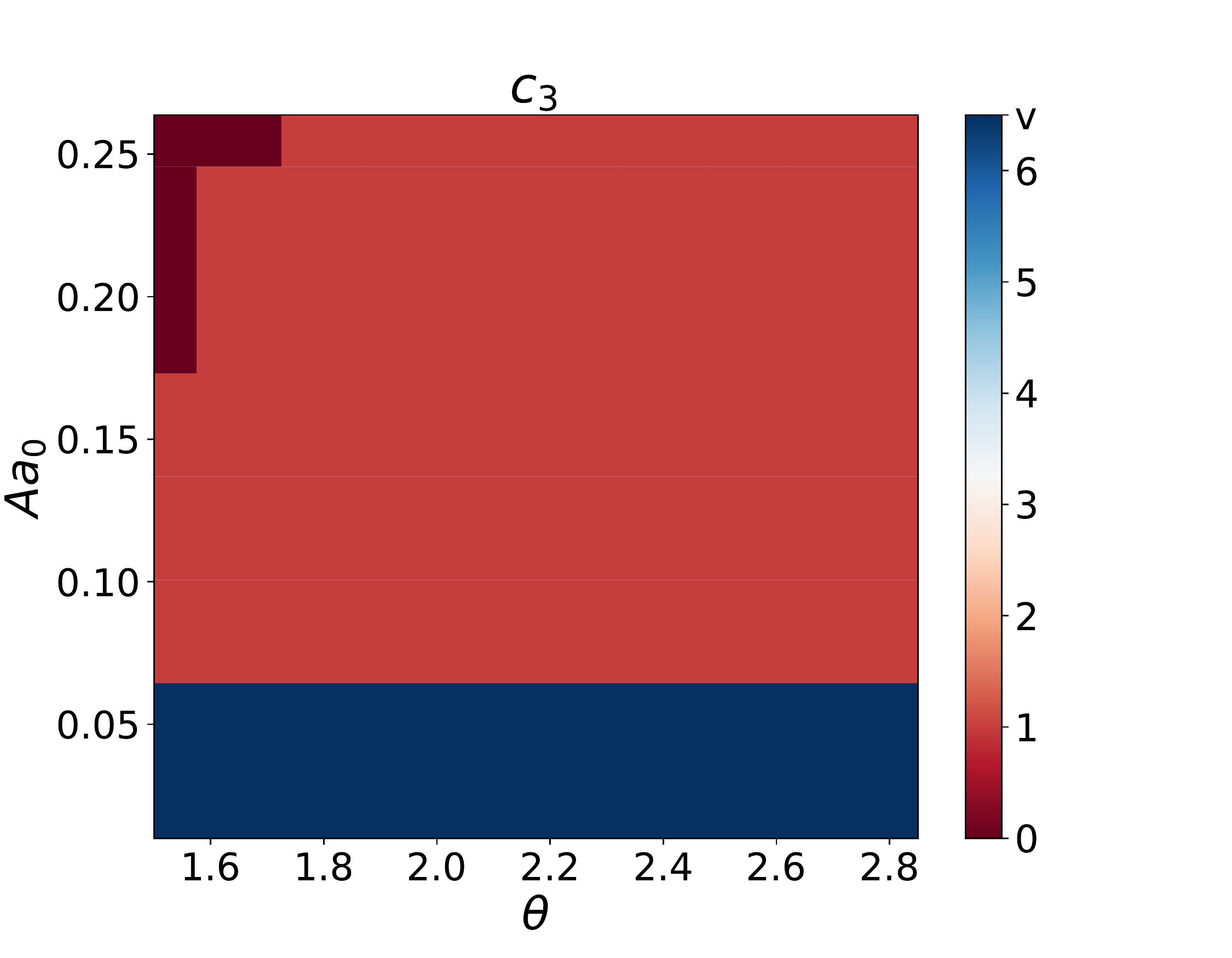}\hfill
    \includegraphics[width=0.47\linewidth]{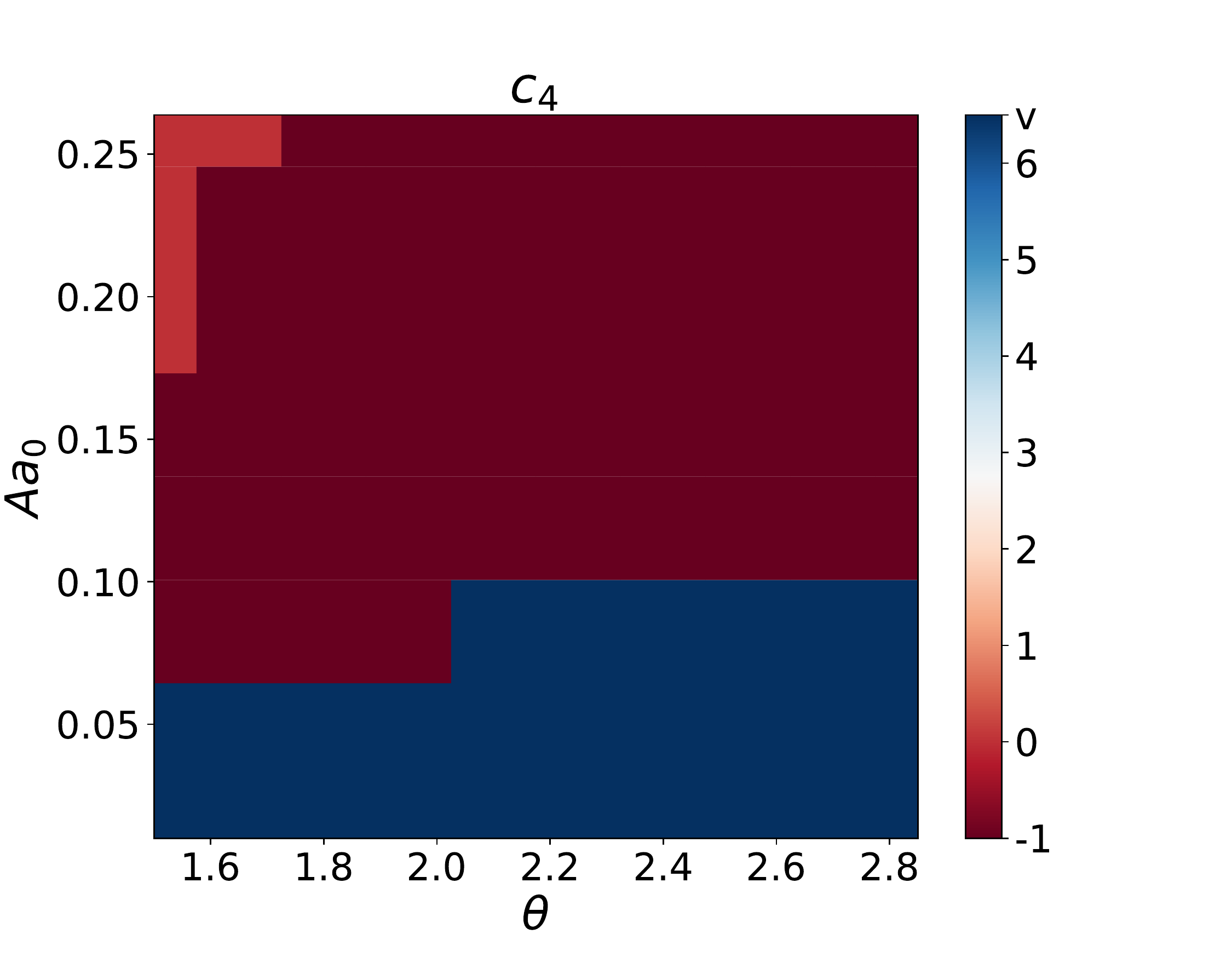}
    \hspace{0mm}
    \includegraphics[width=0.47\linewidth]{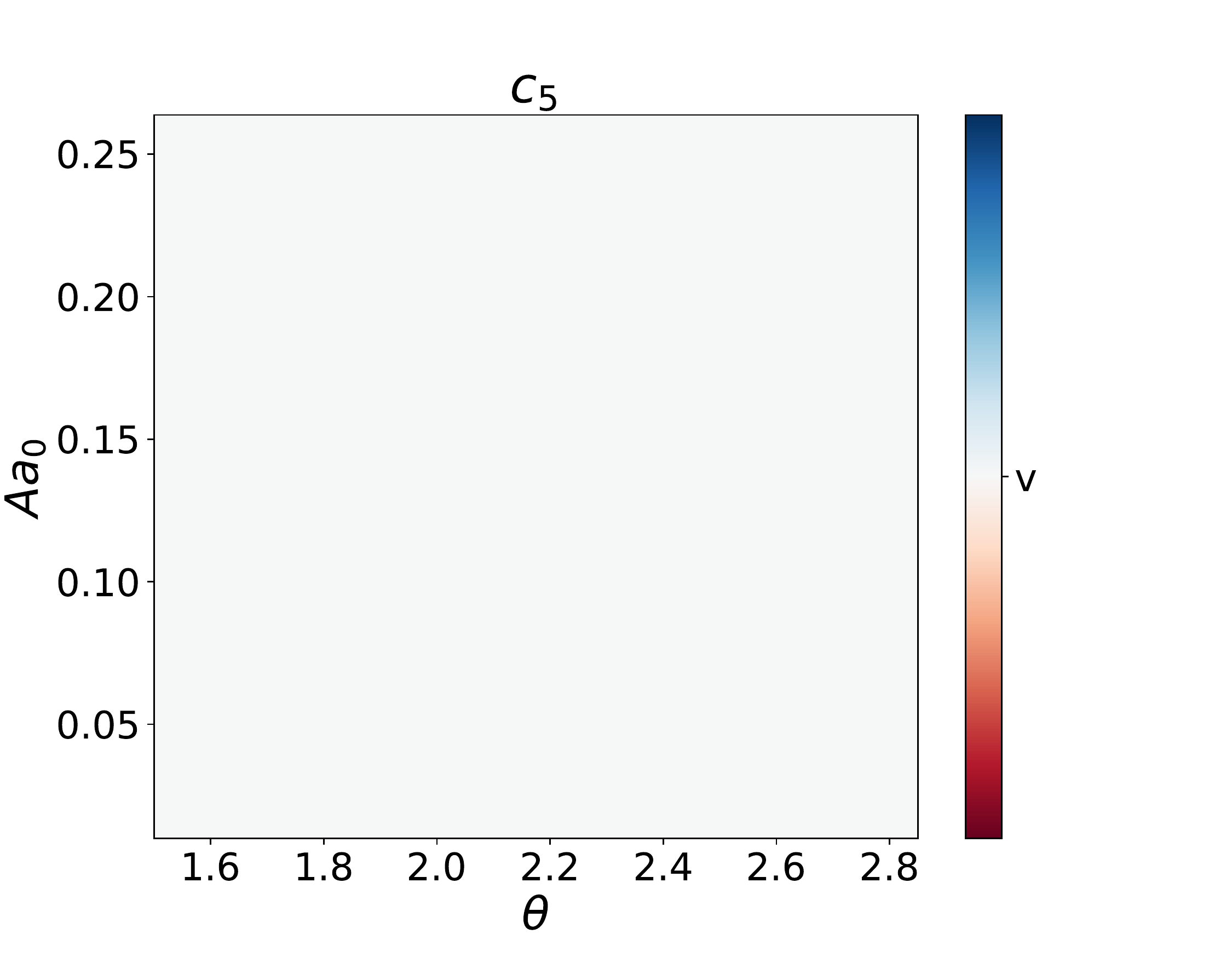}\hfill
    \includegraphics[width=0.47\linewidth]{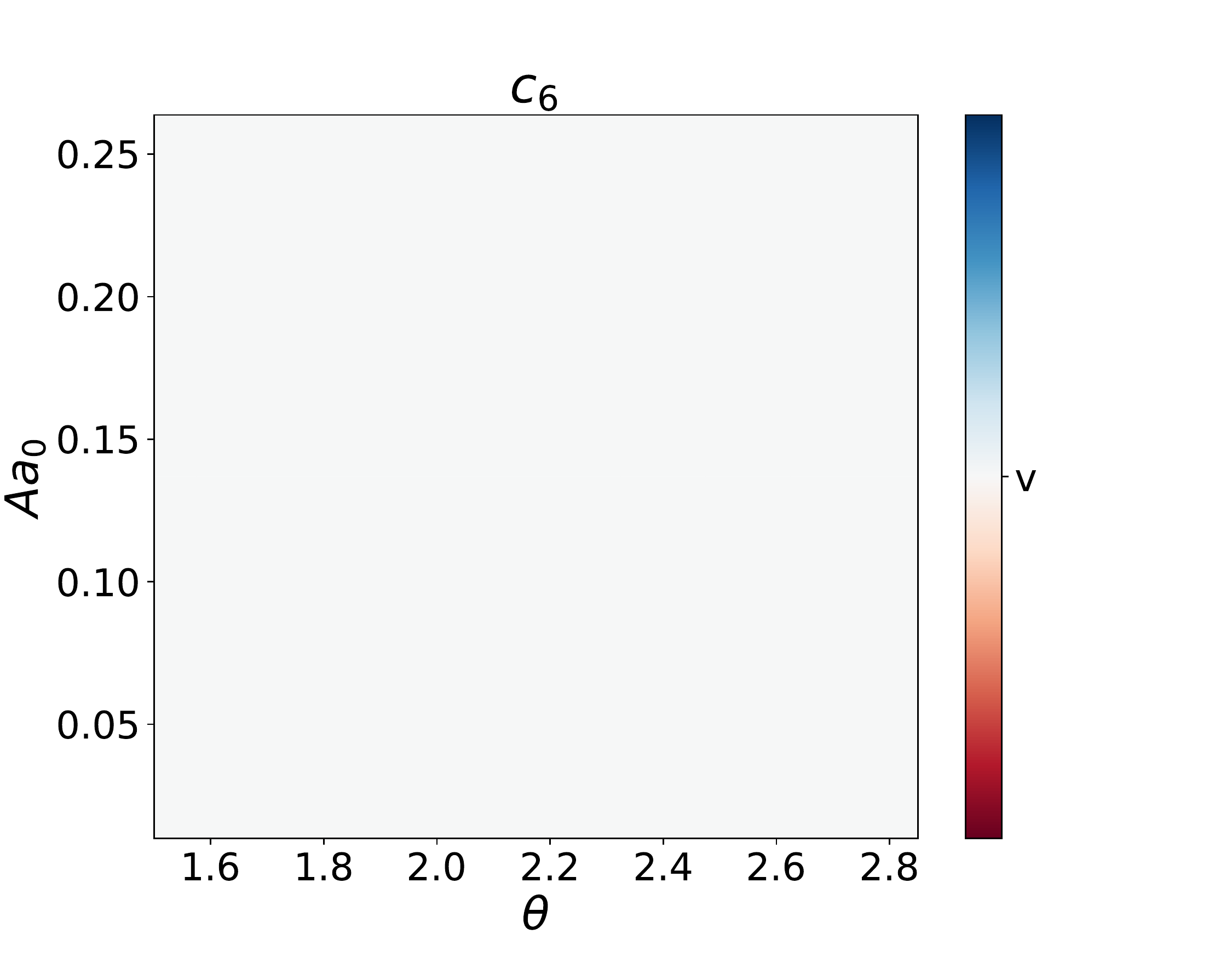}
\caption{(Color online)  The Chern number maps for the central six bands in the TTG with AAA stacking and middle layer twisted. The letter "v" stands for the case when no gap opening has been observed.}
\label{fig:chrn_AAA_MT}
\end{figure}

\begin{figure}[!htbp]
\centering
    \includegraphics[width=0.5\linewidth]{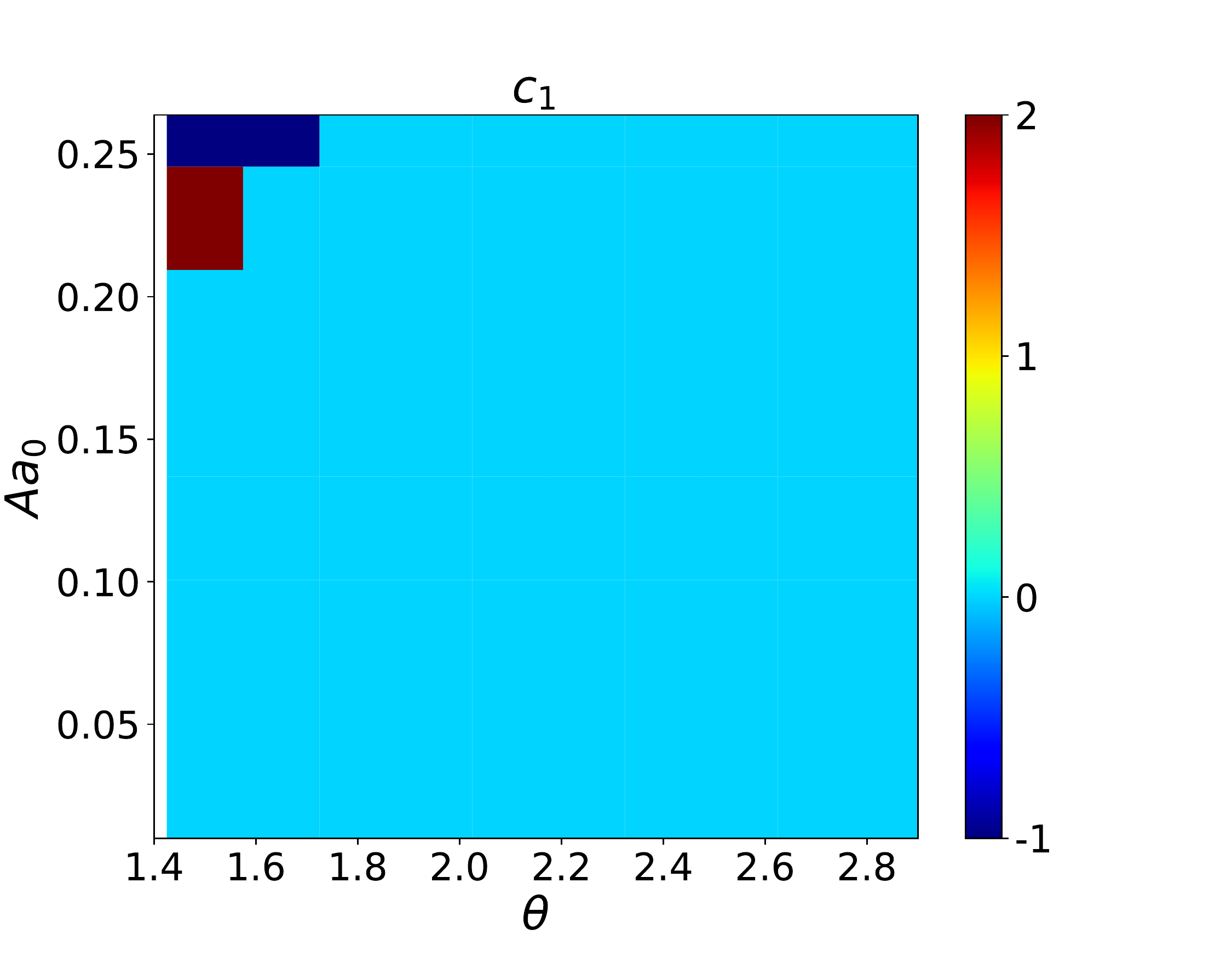}\hfill
    \includegraphics[width=0.5\linewidth]{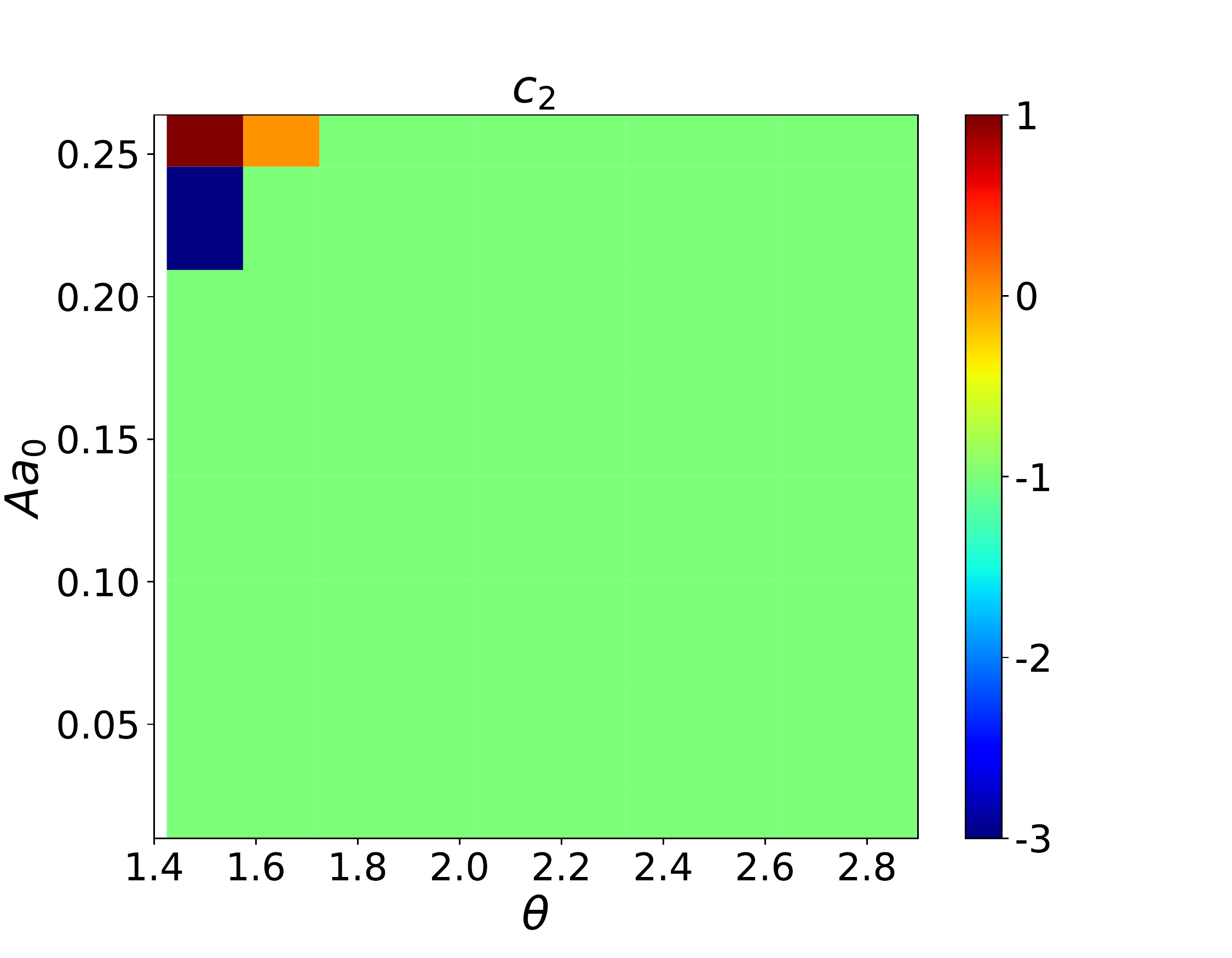}
    \hspace{0mm}
    \includegraphics[width=0.47\linewidth]{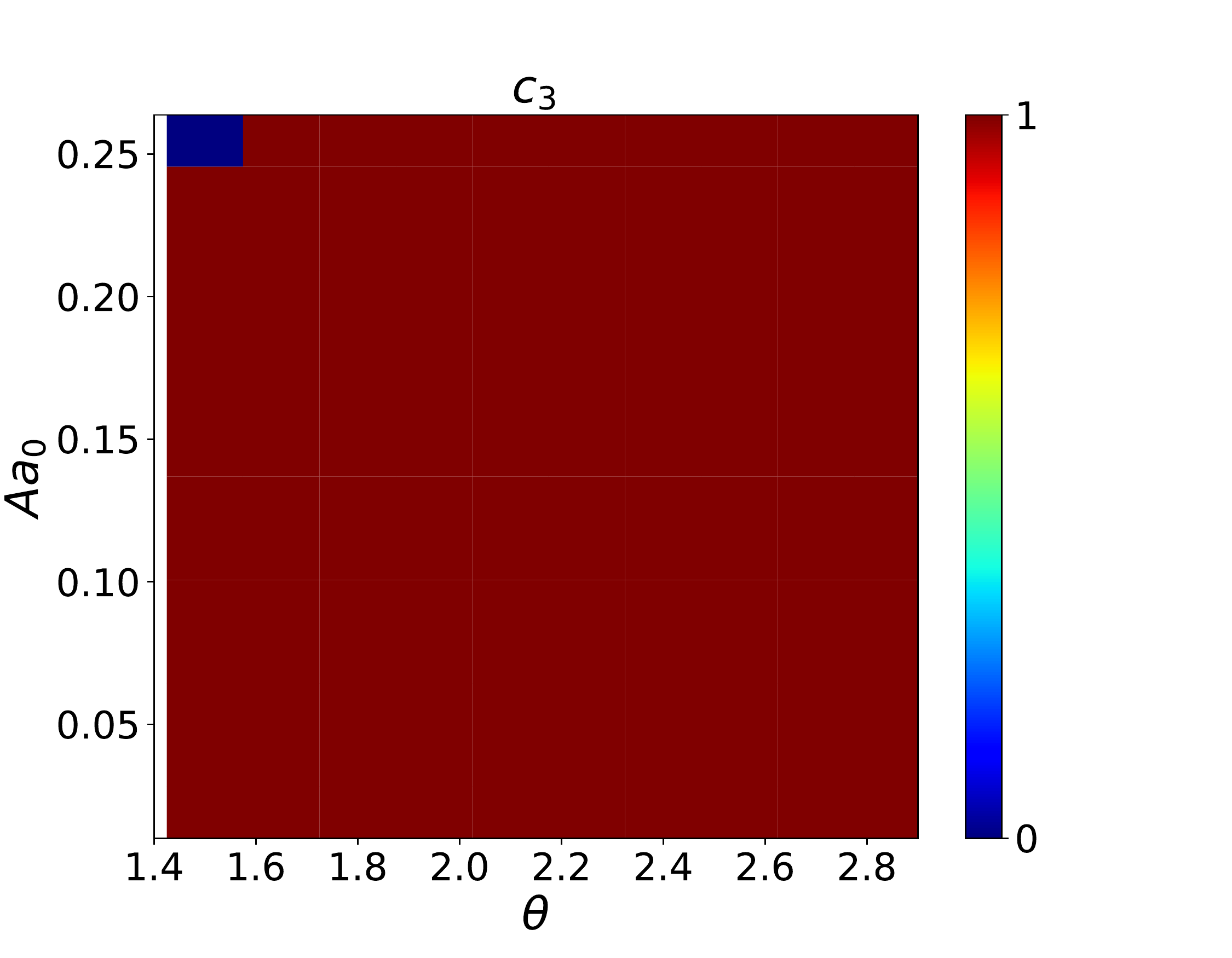}\hfill
    \includegraphics[width=0.47\linewidth]{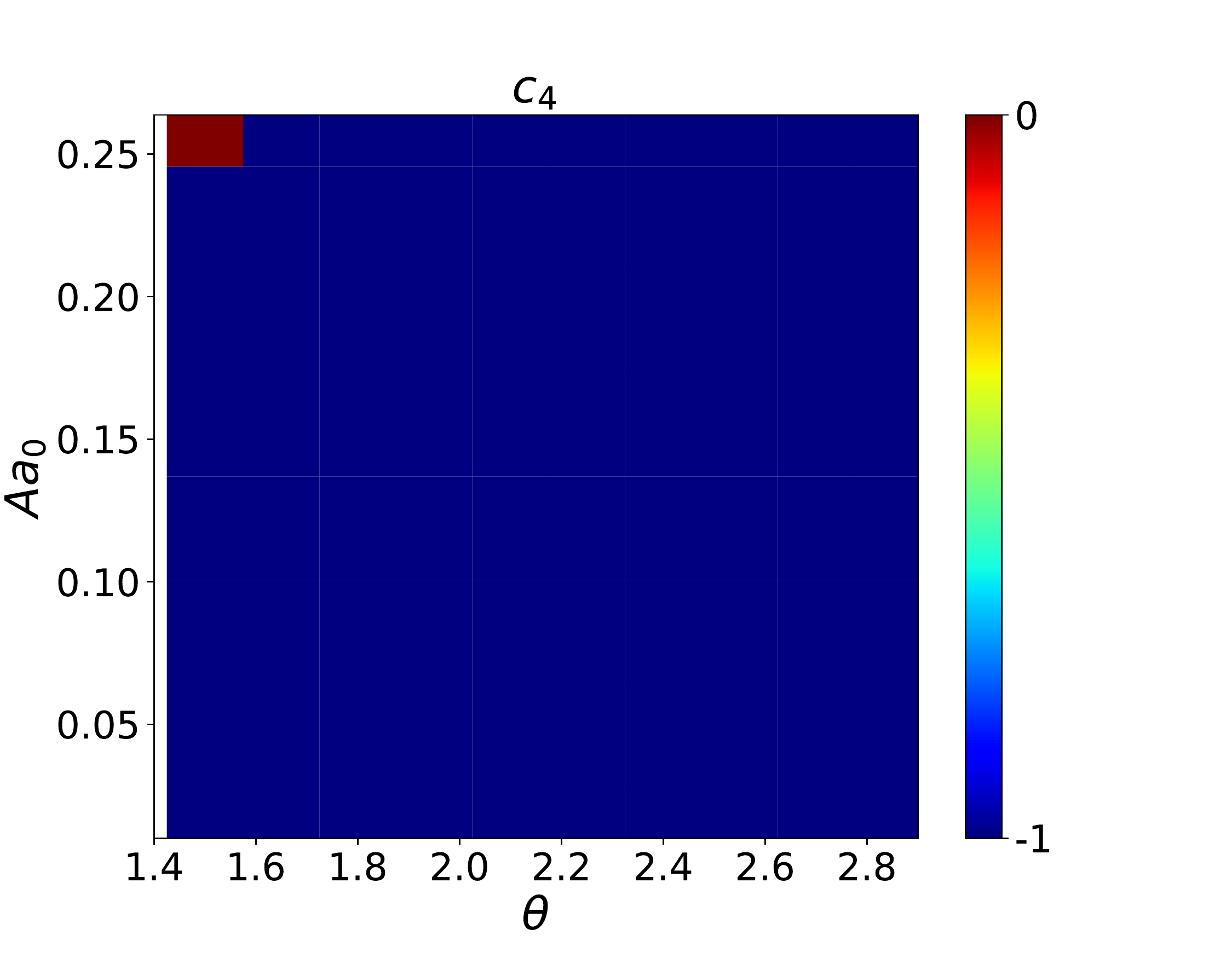}
    \hspace{0mm}
    \includegraphics[width=0.47\linewidth]{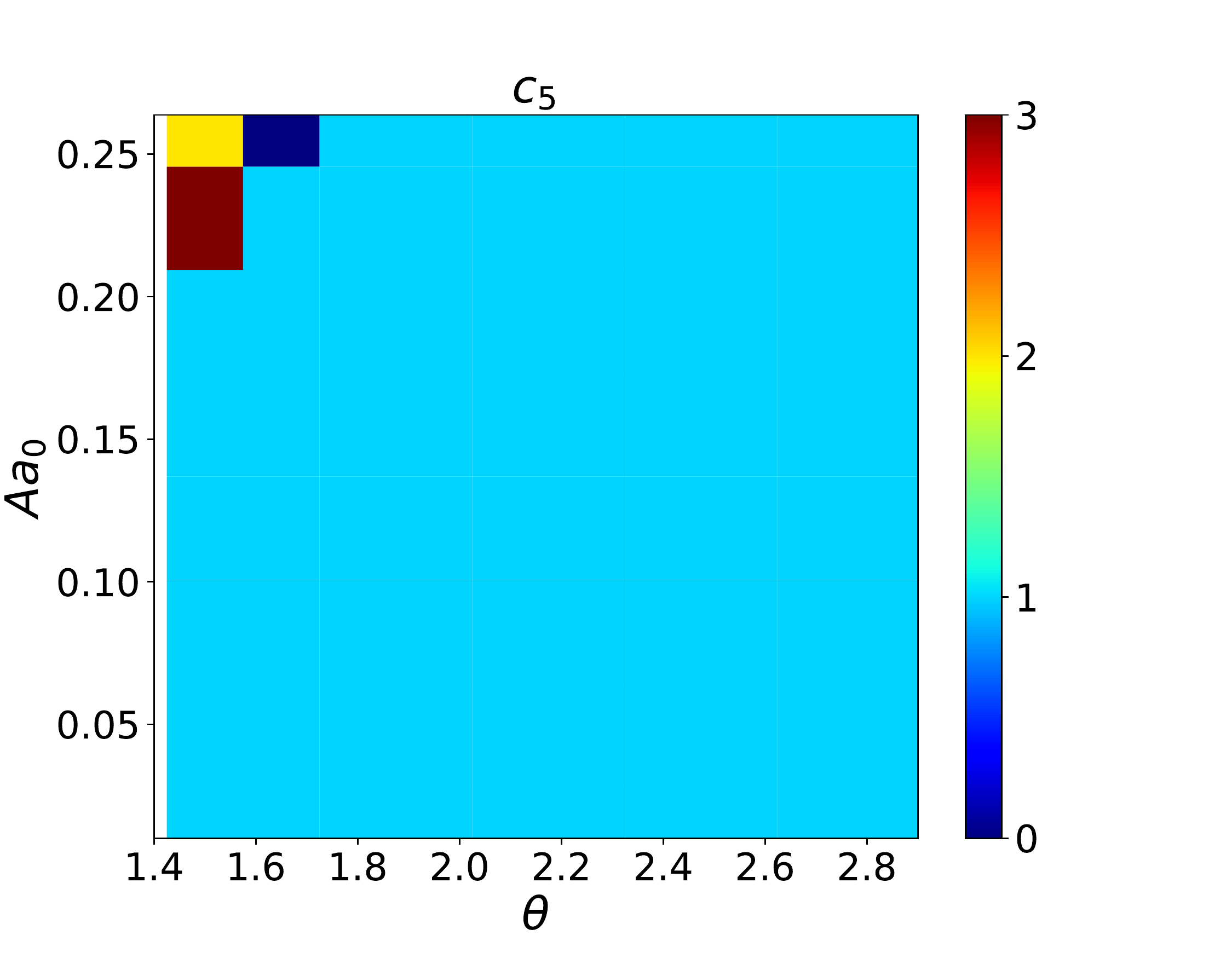}\hfill
    \includegraphics[width=0.47\linewidth]{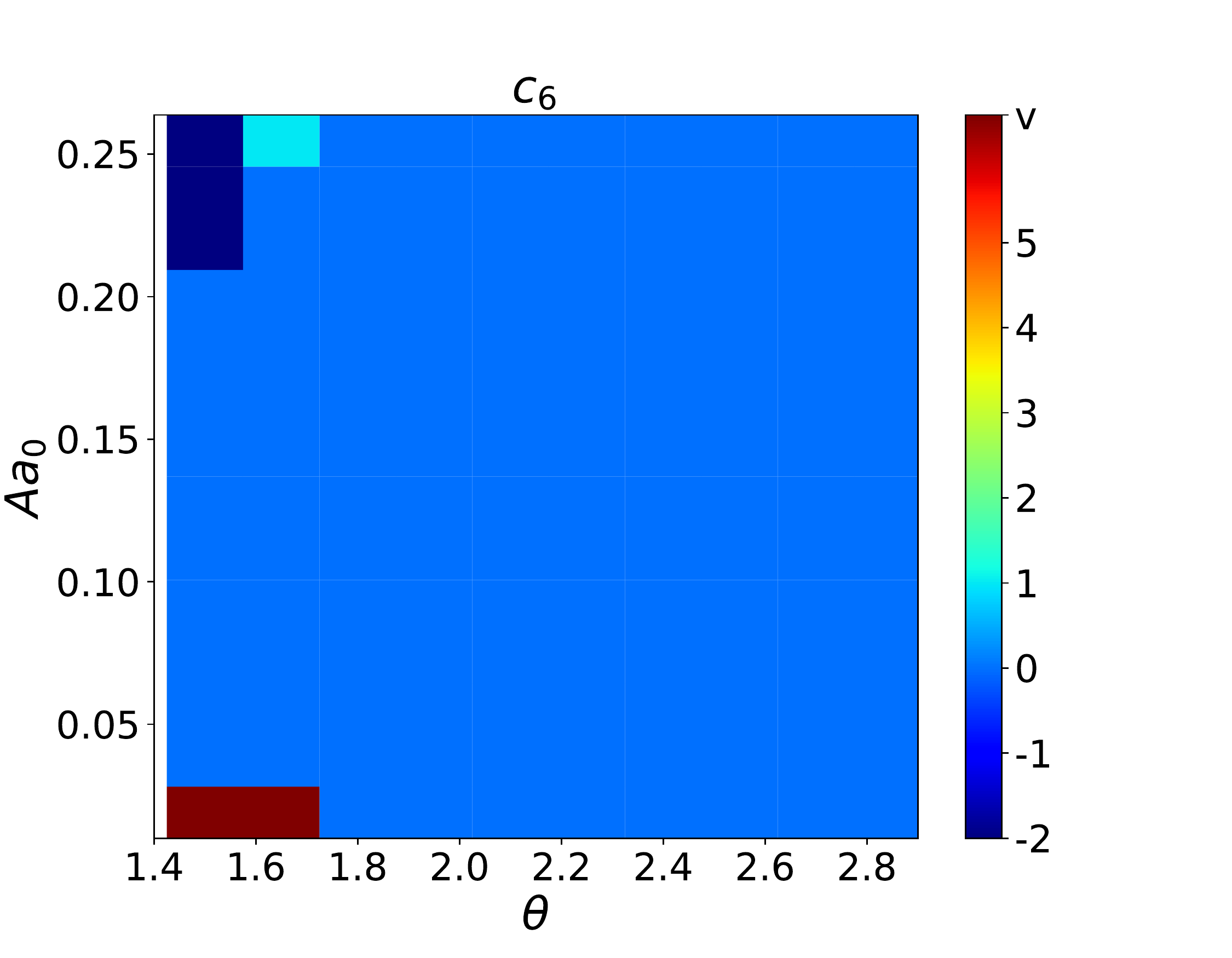}
\caption{(Color online)  Variations of the Chern numbers for each of the six central bands $c_1 \to c_6$ for the case of ABC stacking and MLT. The letter v in the topological map of the sixth band ($c_6$) corresponds to no gap opening??}
\label{fig:chrn_ABC_MT}
\end{figure}

\begin{figure}[!htbp]
\centering
    \includegraphics[width=0.5\linewidth]{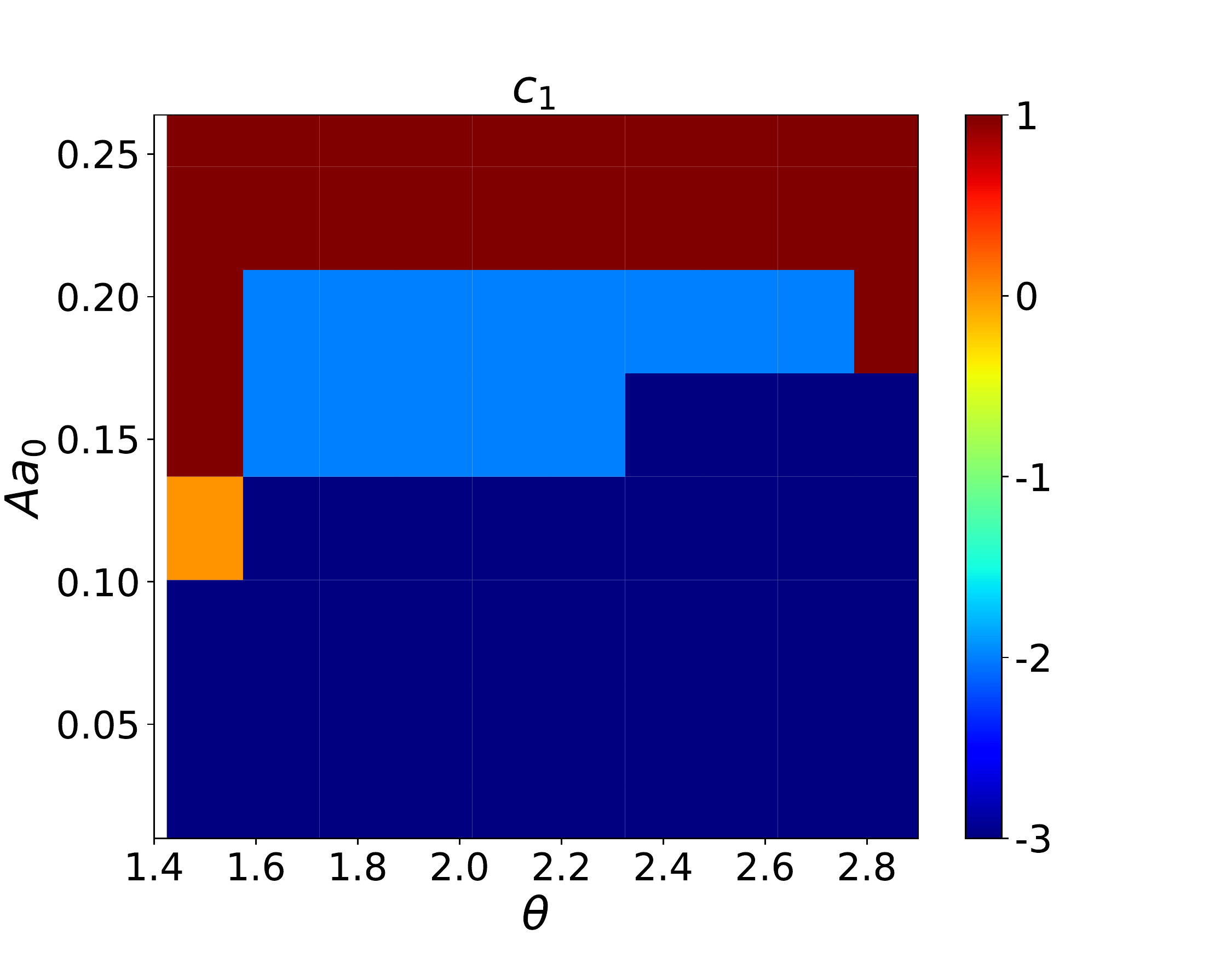}\hfill
    \includegraphics[width=0.5\linewidth]{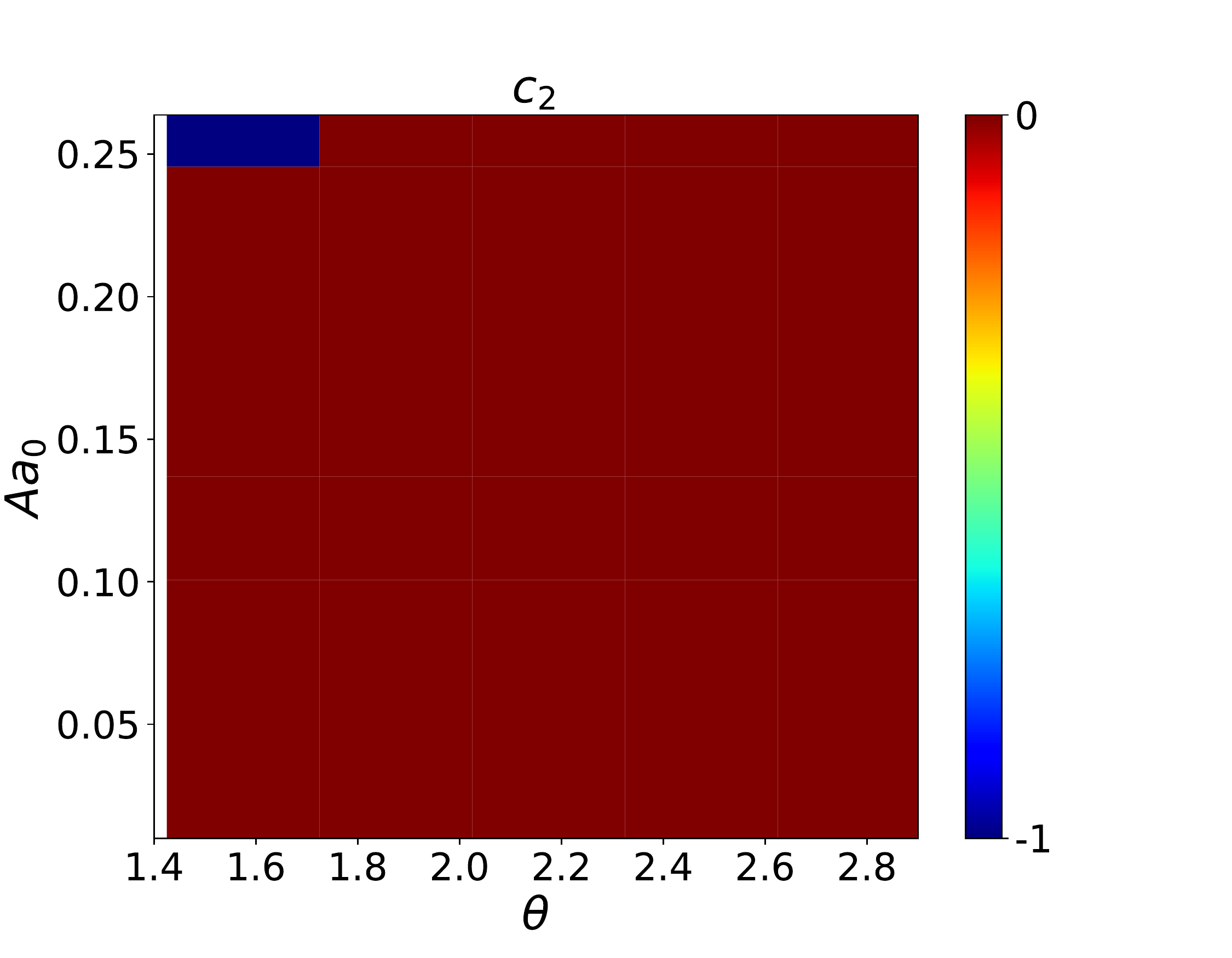}
    \hspace{0mm}
    \includegraphics[width=0.47\linewidth]{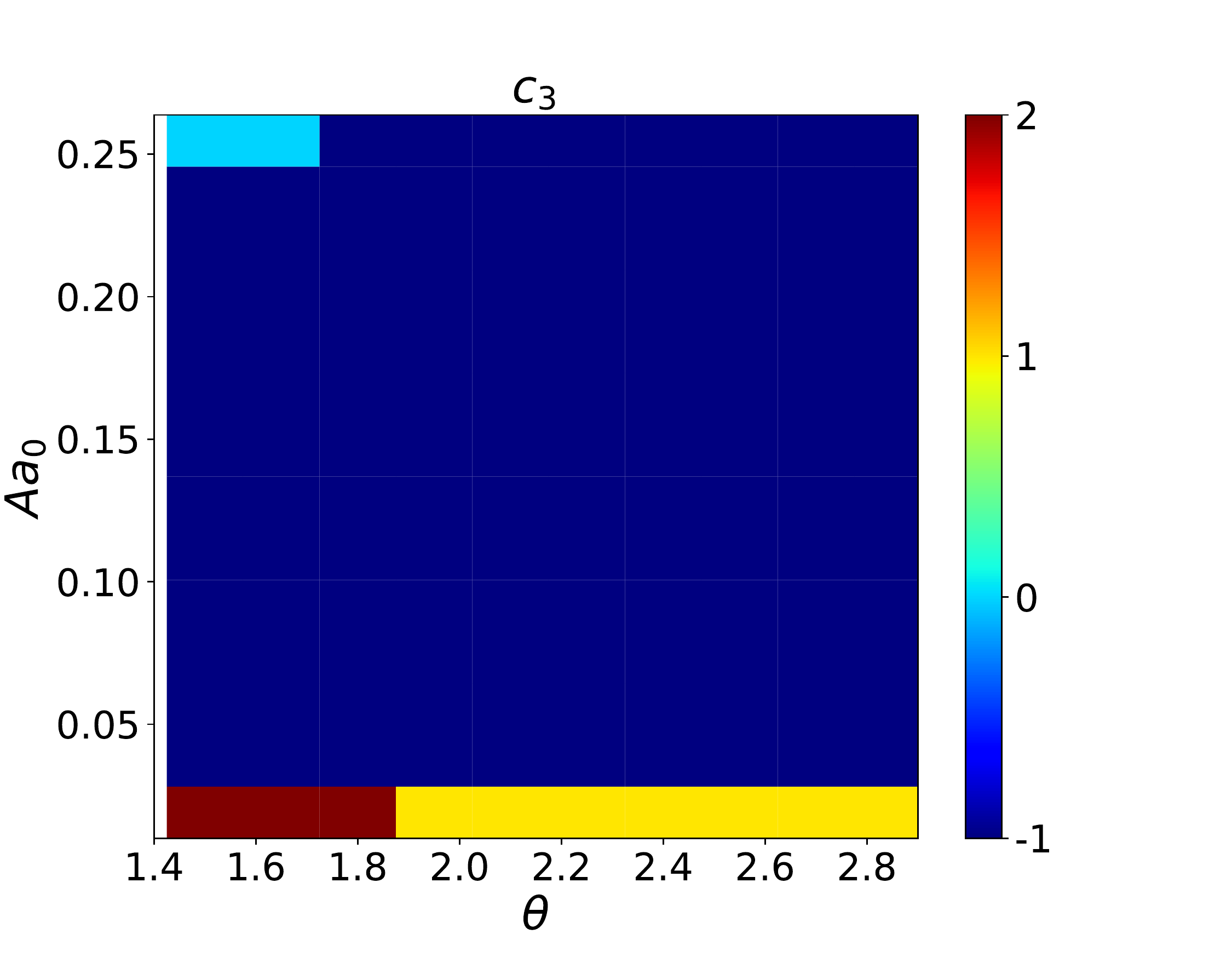}\hfill
    \includegraphics[width=0.47\linewidth]{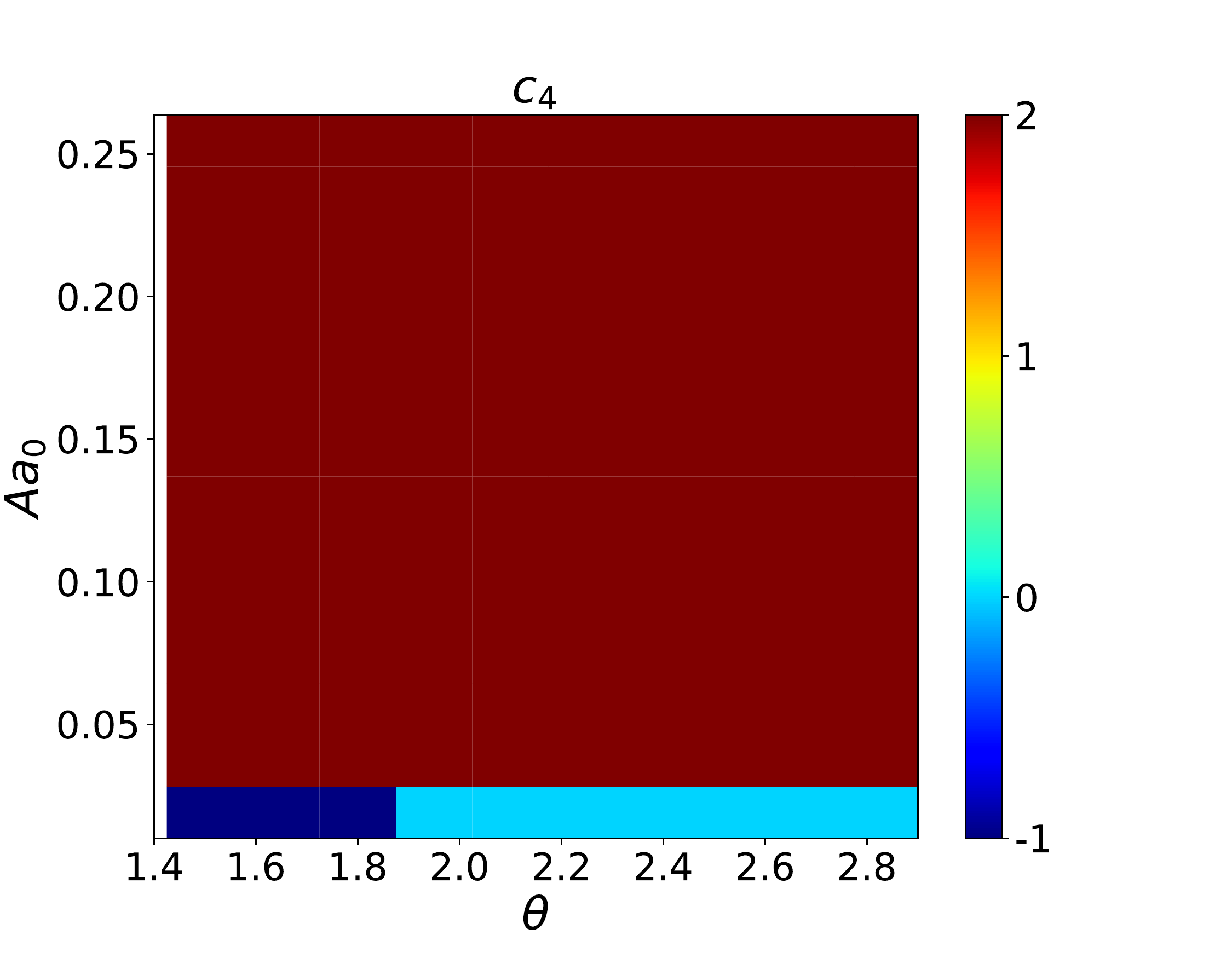}
    \hspace{0mm}
    \includegraphics[width=0.47\linewidth]{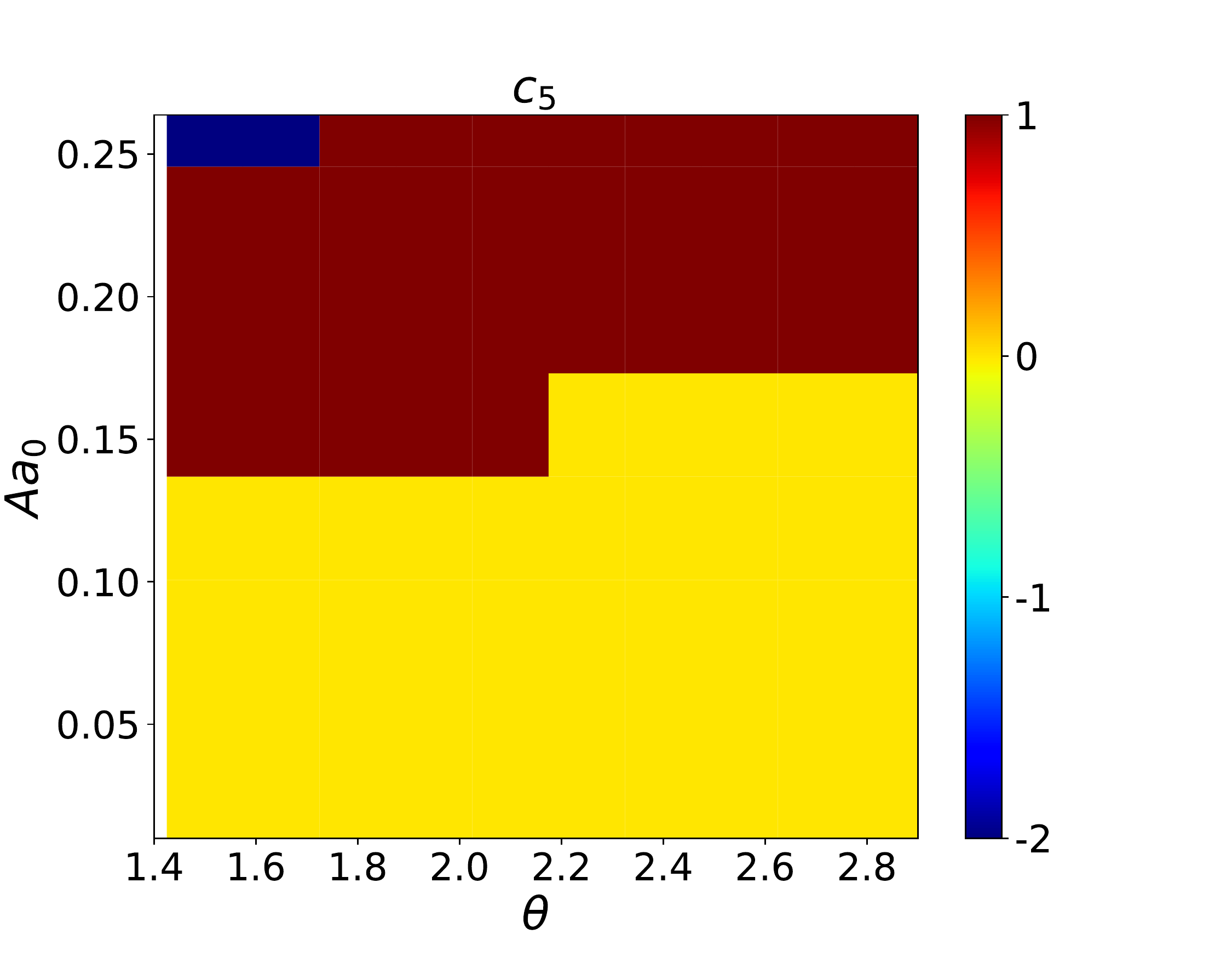}\hfill
    \includegraphics[width=0.47\linewidth]{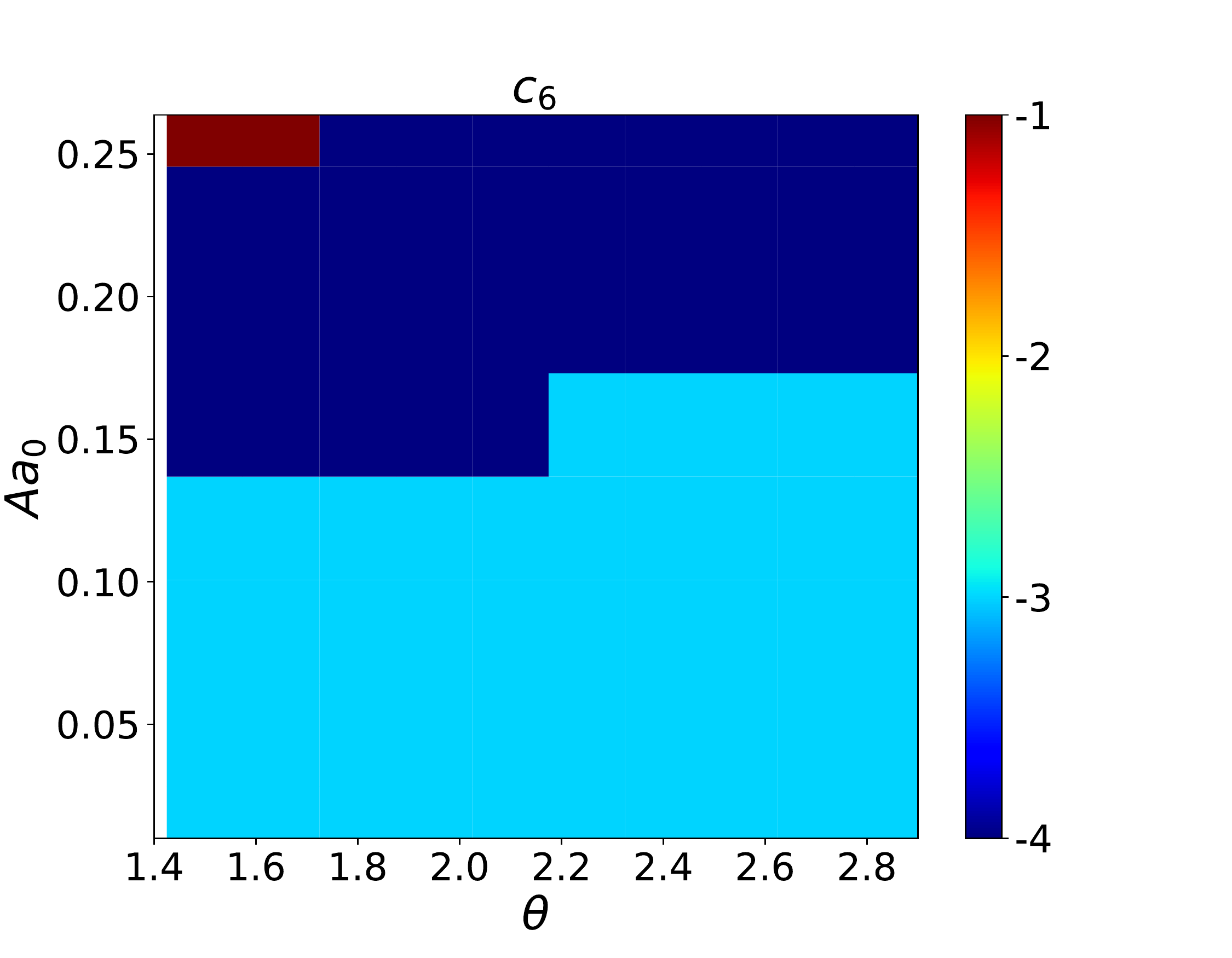}
\caption{(Color online)  Topological maps for the six bands for the ABC stacking case with top layer twisting and right-handed circularly polarized light.}
\label{fig:chrn_ABC_TT_RH}

\end{figure}

\begin{figure}[!htbp]
\centering
    \includegraphics[width=0.5\linewidth]{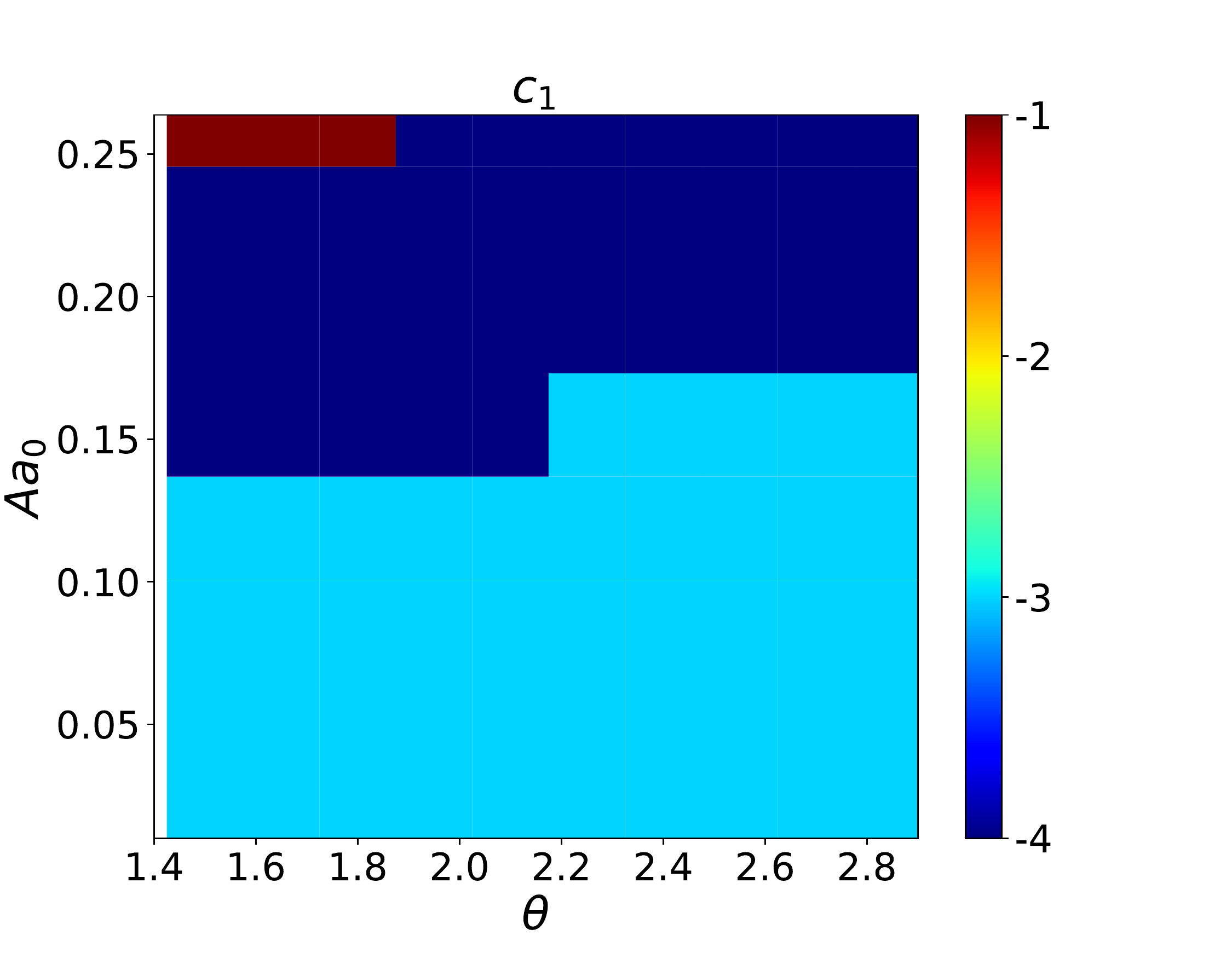}\hfill
    \includegraphics[width=0.5\linewidth]{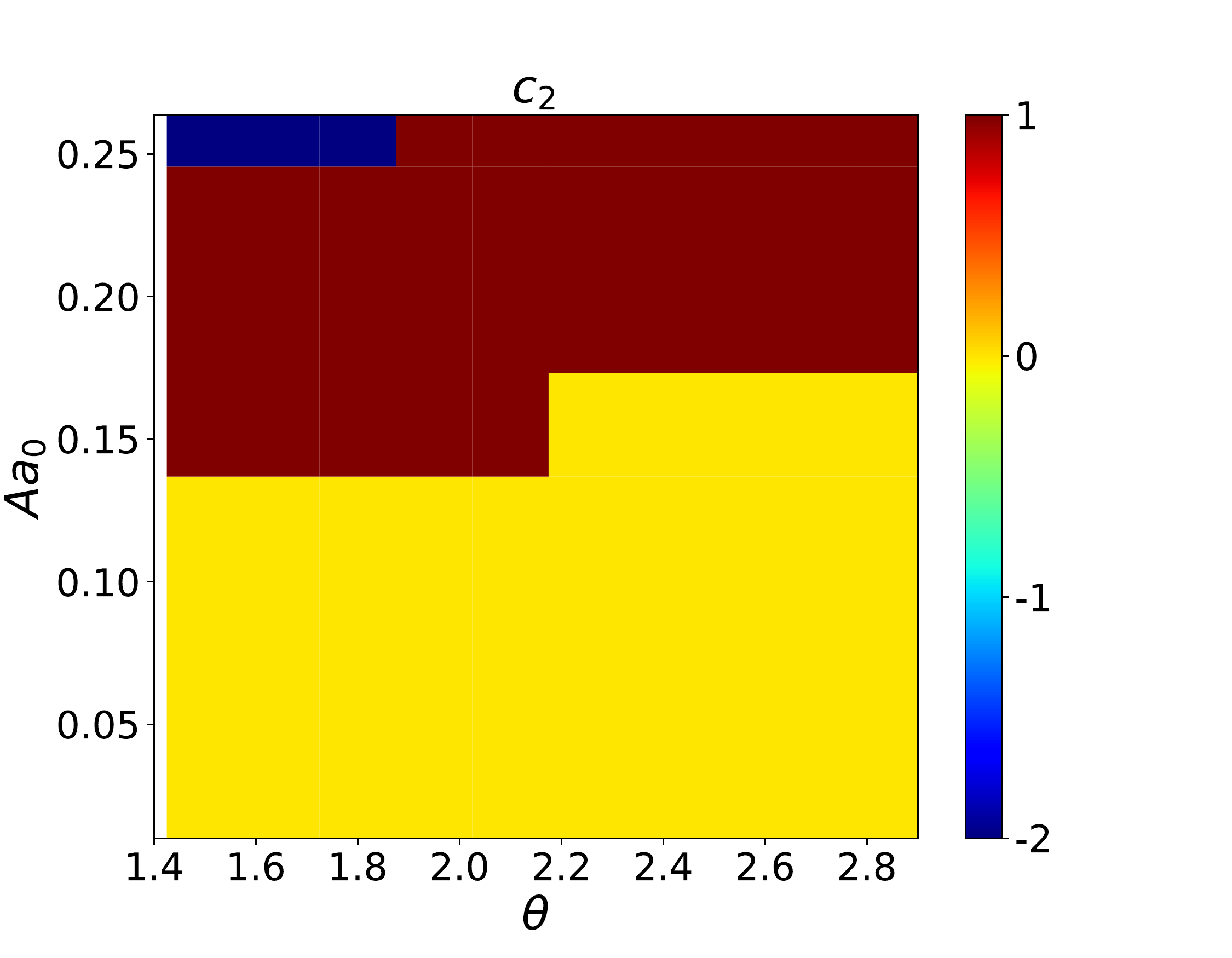}
    \hspace{0mm}
    \includegraphics[width=0.47\linewidth]{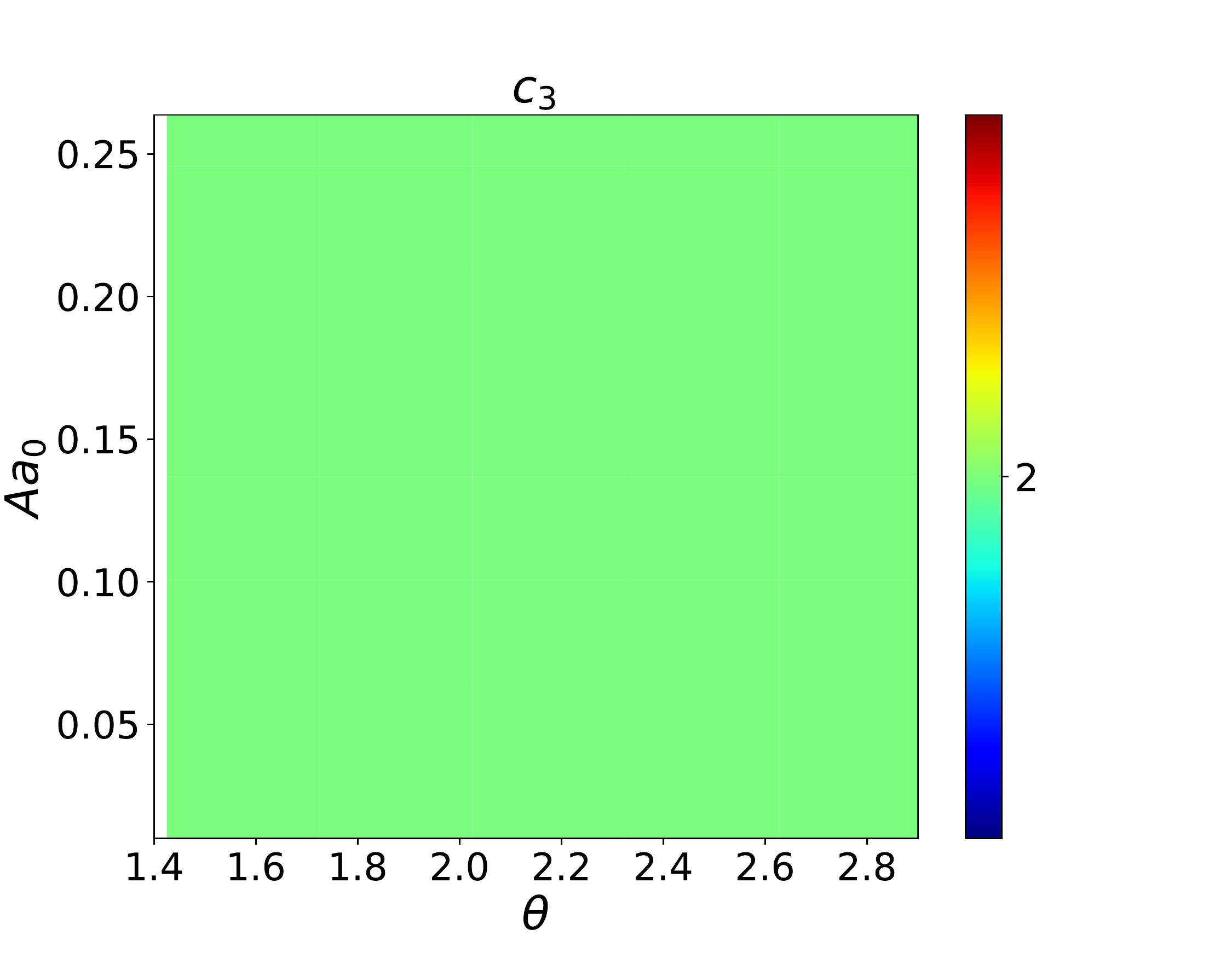}\hfill
    \includegraphics[width=0.47\linewidth]{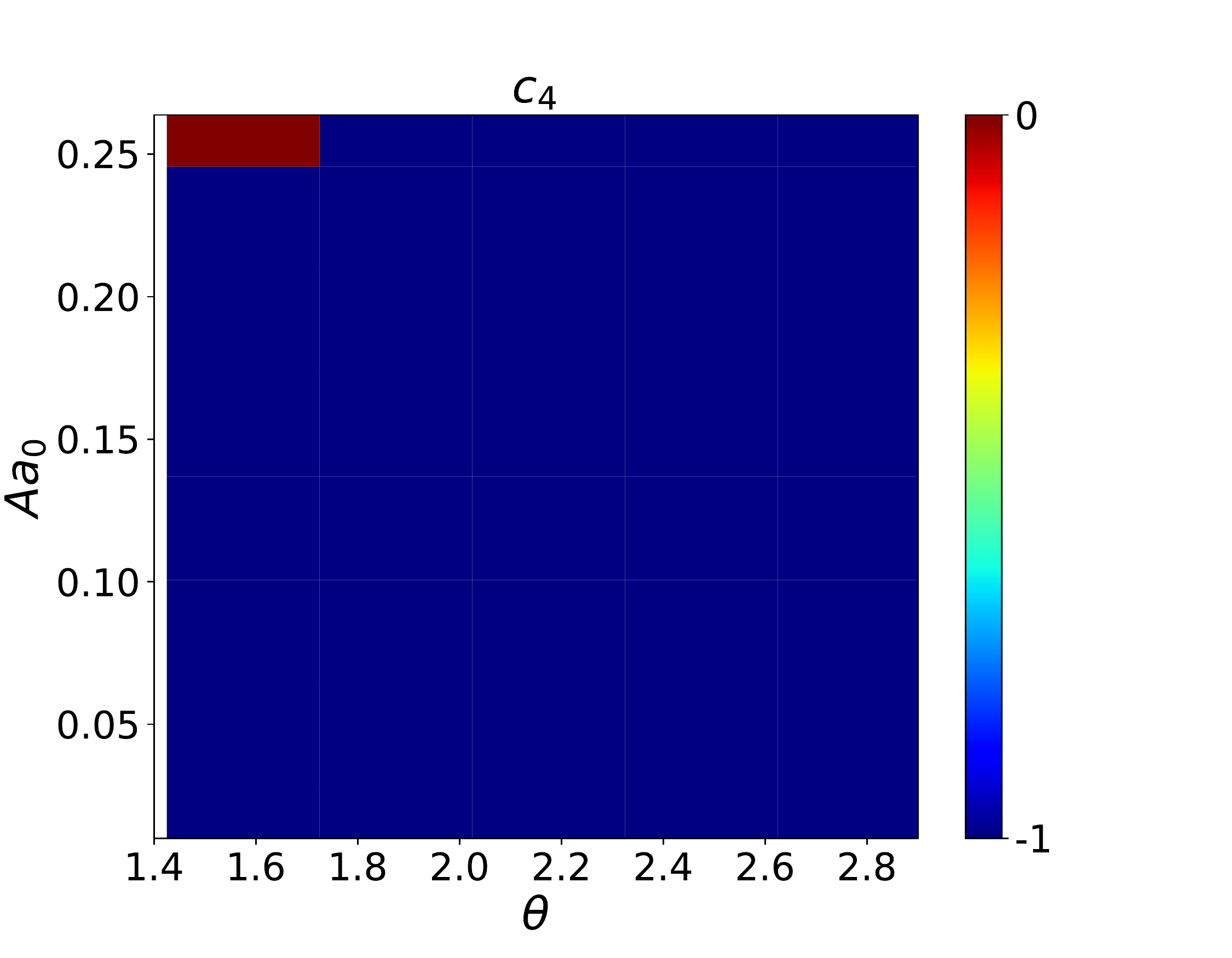}
    \hspace{0mm}
    \includegraphics[width=0.47\linewidth]{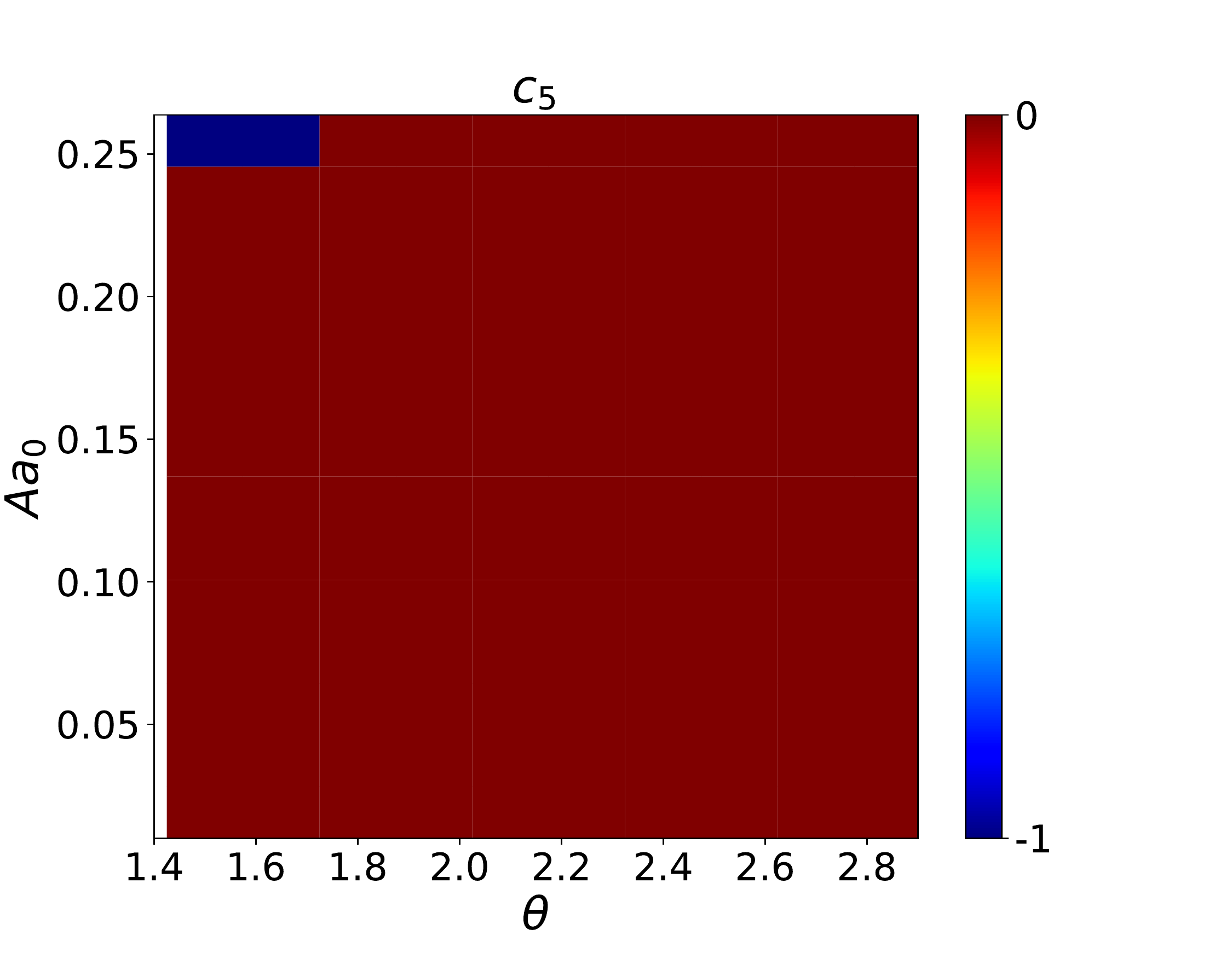}\hfill
    \includegraphics[width=0.47\linewidth]{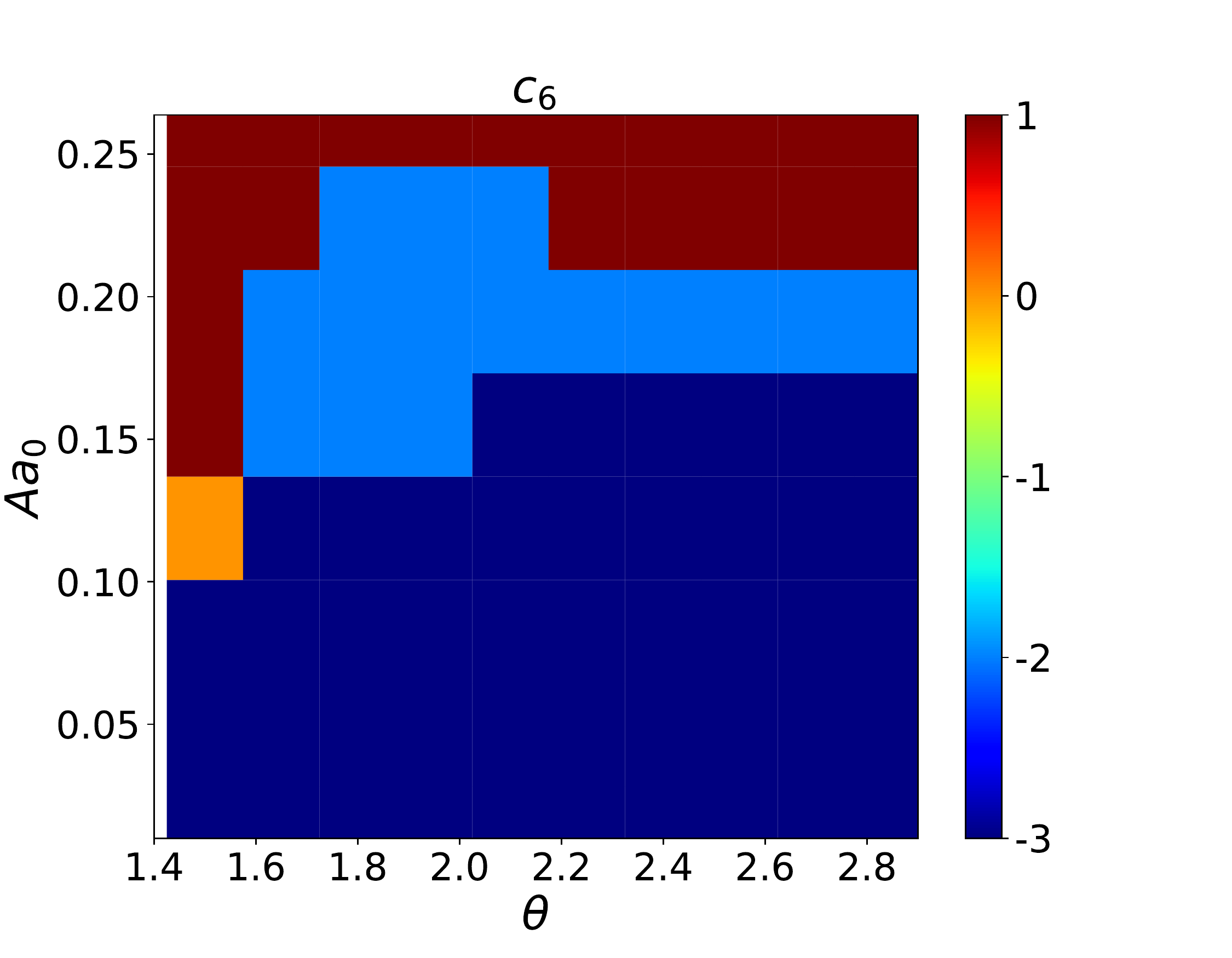}
\caption{(Color online) Reproduction of Fig. \ref{fig:chrn_ABC_TT_RH} but with incident light polarization being left-handed. }
\label{fig:chrn_ABC_TT_LH}

\end{figure}

\end{widetext}

\end{document}